\documentclass[aps,prx,twocolumn,superscriptaddress,showpacs,preprintnumbers,amsmath]{revtex4-2}
\usepackage{epsfig,amsmath,graphics,color,calc}
\usepackage{amssymb}
\usepackage{array}
\usepackage{mathtools}
\usepackage{graphicx}
\usepackage{hyperref}
\usepackage{xcolor}
\usepackage{pifont}
\usepackage[bitstream-charter, greekfamily=default]{mathdesign}
\usepackage{lipsum}
\usepackage[normalem]{ulem}
\newcommand{\tr}{\mbox{Tr}}
\newcommand{\im}{\mbox{Im}}
\newcommand{\re}{\mbox{Re}}

\newcommand{\de}{\textrm{d}}
\newcommand{\eg}{{\it e.g.}}
\newcommand{\ie}{{\it i.e.}}

\newcommand{\avg}[1]{\left\langle #1 \right\rangle}
\newcommand{\avgg}[1]{\small\langle #1 \small\rangle}
\newcommand{\abs}[1]{\left\vert #1 \right\vert}
\newcommand{\ER}{Erd\"os-R\'enyi}
\usepackage{cancel}

\newcommand{\rev}[1]{#1}
\newcommand{\revv}[1]{#1}

\begin{document}

\title{Multifractal phase in the weighted adjacency matrices of random Erd\"os-R\'enyi graphs}

\author{Leticia F. Cugliandolo}
\affiliation{Sorbonne Université, Laboratoire de Physique Théorique et Hautes Energies, CNRS-UMR 7589, 4 Place Jussieu, 75252 Paris Cedex 05, France}
\affiliation{Institut  Universitaire  de  France,  1  rue  Descartes,  75005  Paris,  France}

\author{Grégory Schehr}
\affiliation{Sorbonne Université, Laboratoire de Physique Théorique et Hautes Energies, CNRS-UMR 7589, 4 Place Jussieu, 75252 Paris Cedex 05, France}

\author{Marco Tarzia} 
\affiliation{Sorbonne Université, Laboratoire de Physique Théorique de la Matière Condensée, CNRS-UMR 7600, 4 Place Jussieu, 75252 Paris Cedex 05, France}
\affiliation{Institut  Universitaire  de  France,  1  rue  Descartes,  75005  Paris,  France}

\author{Davide Venturelli} 
\affiliation{Sorbonne Université, Laboratoire de Physique Théorique de la Matière Condensée, CNRS-UMR 7600, 4 Place Jussieu, 75252 Paris Cedex 05, France}
\affiliation{SISSA --- International School for Advanced Studies and INFN, via Bonomea 265, 34136 Trieste, Italy}

\begin{abstract}
We study the spectral properties of the adjacency matrix in the giant connected component of Erd\"os-R\'enyi random graphs, with average 
\rev{degree}
$p$ and randomly distributed hopping amplitudes.
By solving the self-consistent cavity equations satisfied by the matrix elements of the resolvent, we compute the probability distribution of the local density of states, which governs the scaling with the system size of the moments of the eigenvectors' amplitudes, as well as several other observables related to the spectral statistics. 
For small values of $p>1$ above the percolation threshold, we unveil the presence of an exotic delocalized but (weakly) multifractal phase in a broad region of the parameter space, which separates the localized phase found for $p\le1$ from the fully-delocalized and GOE-like phase expected for $p\to \infty$. We explore the fundamental physical mechanism underlying the emergence of delocalized multifractal states, rooted in the pronounced heterogeneity in the topology of the graph. This heterogeneity arises from the interplay between strong fluctuations in local degrees and hopping amplitudes, and leads to an effective fragmentation of the graph.
We further support our findings by characterizing the level statistics and the two-point spatial correlations within the multifractal phase, and address the ensuing anomalous transport and relaxation properties affecting the quantum dynamical evolution. 
\end{abstract}


\maketitle

\section{Introduction}

Ever since Anderson's groundbreaking works on localization~\cite{anderson1958absence} more than six decades ago, a significant body of research has been dedicated to exploring the transport and spectral properties of quantum particles in random environments~\cite{Thouless_1970,Lee_1985,Ferdinand_2008,Lagendijk2009}. This inquiry has profoundly influenced various realms within condensed matter physics, ranging from transport in disordered quantum systems, to random matrix theory, and to the study of quantum chaos. Despite the extensive scrutiny it has received, this field continues to reveal fresh insights and nuances.

In recent years, there has been a notable resurgence of interest in investigating Anderson localization (AL) on sparse random graphs. These tree-like structures, emblematic of the infinite-dimensional limit of the tight-binding model, offer a pathway towards exact solutions, allowing one to establish the transition point and the corresponding critical behavior. In this respect, the first --- and possibly the most celebrated --- exploration is the analysis of Abou~Chacra, Anderson, and Thouless~\cite{AbouChacra_1973} of AL on the Bethe lattice, which can be formally defined as the thermodynamic limit of a random-regular graph (RRG) of fixed connectivity. This model, amenable to analytical treatment through a supersymmetric approach~\cite{tikhonov2019statistics,efetov1985anderson,efetov1987density,efetov1987anderson,zirnbauer1986localization,zirnbauer1986anderson,verbaarschot1988graded,mirlin1991localization,mirlin1991universality,fyodorov1991localization,fyodorov1992novel,mirlin1994statistical,mirlin1997localization}, undergoes a metal-insulator transition triggered by the augmentation of the random on-site disorder strength, and constitutes one of the key elements for our current understanding of AL.

\begin{figure*}
\includegraphics[width=0.482\textwidth]{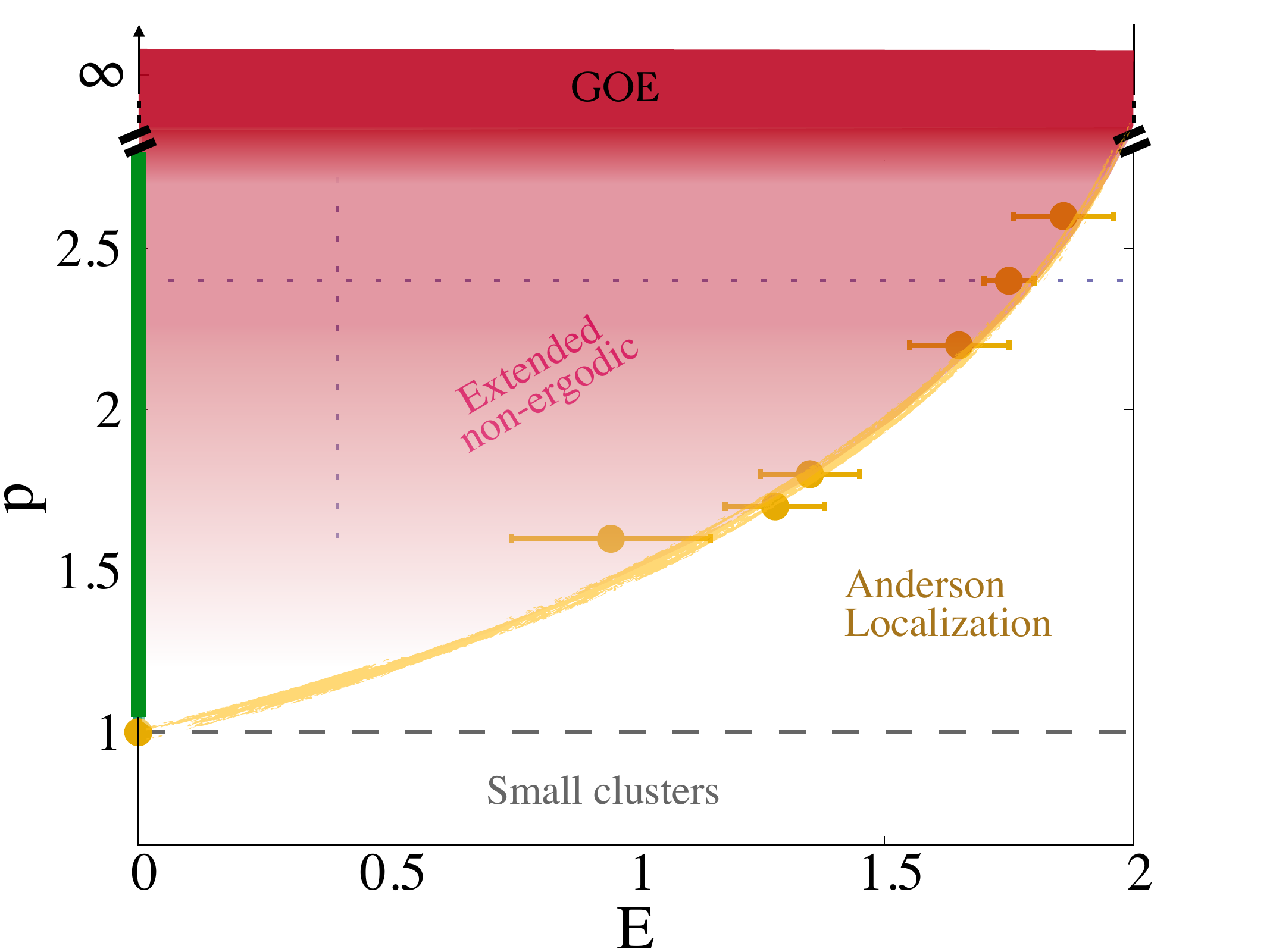}  
\hspace{0.5cm}
\includegraphics[width=0.42\textwidth]{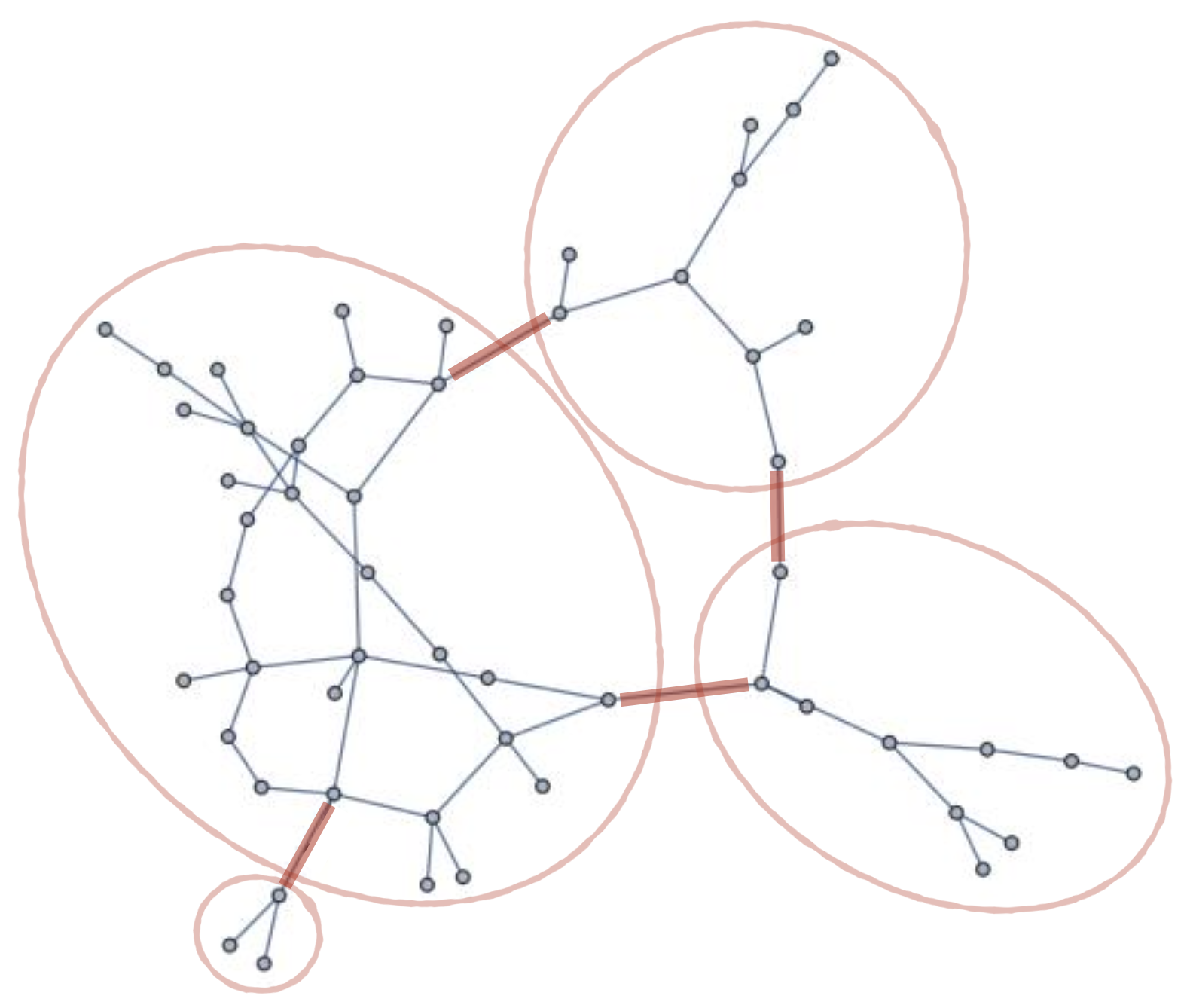}  
\put(-480,170){(a)} 
\put(-384,158){{\scriptsize (fully extended)}} 
\put(-210,170){(b)} 
 \caption{
 (a) Phase diagram of the (giant connected component of the) WER ensemble in the ($E$--$p$) plane (see Sec.~\ref{sec:phase_diagram} for details). The yellow line is a guide to the eye of the mobility edge for different values of $p$. 
 The thick green vertical line at $E=0$ represents the degenerate states at zero energy, localized on the leaves of the graph. The horizontal gray dashed line corresponds to the percolation transition at $p_c=1$. The horizontal (at $p=2.4$) and vertical (at $E=0.4$) dotted lines show the regions of the parameter space where we concentrated most of the numerical investigations. The multifractality
 of the eigenstates in the intermediate region 
 progressively weakens
 upon increasing $p$ (see Fig.~\ref{fig:Dq} below), and eventually a standard fully-\rev{extended} 
 phase, described by the GOE universality class, is reached in the $p \to \infty$ limit. \rev{The phase diagram is symmetric in the region with $E<0$ (not shown).} (b)~Physical mechanism at the origin of the multifractality in the WER ensemble. 
In the sketch, the connected component of an \ER~graph (with average 
\rev{degree}
$p=2.2$) is effectively fragmented into four weakly connected components of different sizes, for a specific realization of the random weights in which the bonds highlighted in red are very weak. 
\label{fig:PD}
 }
\end{figure*}

In general, the geometrical properties of adjacency matrices of sparse graphs encapsulate crucial structural and topological aspects of numerous physical systems~\cite{Barabasi_2002,Dorogovtsev_2003}. For example, for random walks on graphs, the eigenvalue spectrum is directly related to the relaxation time spectrum~\cite{Lovsz1973}. From the condensed-matter perspective, the spectra of such matrices have been used to characterize \eg~the gelation transition in polymers~\cite{Broderix_2000,Broderix_2002}, and the instantaneous normal modes in supercooled liquids~\cite{Cavagna2009}. 
Another relevant application is provided by Many Body Localization (MBL)~\cite{Gornyi_2005,basko2006metal}, a new kind of quantum out-of-equilibrium dynamical phase transition between an ergodic metal at low disorder, and a non-ergodic insulator at strong disorder, in which the (interacting) system is unable to self-equilibrate (see Refs.~\cite{reviewMBL,reviewMBL2,reviewMBL3,reviewMBL4,reviewMBL5,sierant2024manybody} for recent reviews). The parallelism between MBL and single-particle AL was postulated in the seminal work of Ref.~\cite{altshuler1997quasiparticle}, wherein the decay of quasi-particles in the Fock space of many-body states is transposed onto a non-interacting tight-binding model on an appropriate disordered tree-like graph (see also Refs.~\cite{basko2006metal,Gornyi_2005,Logan_1987,Logan_1990,Jacquod_1997,Bigwood1998,DeLuca2013,Tarzia_2017,Roy_2020,Tikhonov2021,Garcia-Mata_2022,biroli2023large}).

Driven by its connection to MBL, a plethora of numerical investigations have 
been conducted on the Anderson model on RRGs. Preliminary findings hinted at the possible presence of an intermediate \textit{multifractal} phase preceding the localization transition~\cite{biroli2012difference,de2014anderson,altshuler2016nonergodic,Kravtsov_2018,savitz2019anderson,pino2020scaling,pino2023correlated,zirnbauer2023wegner,arenz2023wegner}. 
In general, wavefunctions~\rev{$\psi_\alpha$} are said to be multifractal when the $q$-th moments of their amplitude scale asymptotically with the system size $N$ as 
\begin{equation}
  \avg{\sum_{i=1}^N \left | \psi_\alpha (i) \right|^{2q} } \propto N^{-(q-1)D_q} \, ,  
  \label{eq:multifractal}
\end{equation}
with distinct $q$-dependent exponents 
$0 \le D_q \le 1$ \rev{(and $D_q<1$ for at least some of the $q$'s)}. 

In the many-body setting, multifractal eigenstates that do not cover uniformly the whole accessible Hilbert space at a given intensive energy generically
violate the eigenstate thermalization hypothesis (ETH)~\cite{srednicki1994chaos,rigol2008thermalization}. For this reason, they are often called 
\textit{non-ergodic}, in contrast to the ergodic fully-delocalized eigenstates, which 
satisfy ETH. \rev{Yet,} 
in the context of single-particle non-interacting problems, such as the one considered here, 
\rev{the concept of ergodicity is not sharply defined (especially because the eigenvalues lack the extensive scaling typical of interacting systems).} 

While the possibility of this multifractal delocalized phase in the Anderson model on the RRG is undeniably fascinating, particularly due to its connection to MBL~\cite{altshuler1997quasiparticle}, it contradicts the analytical predictions based on the supersymmetric approach for the Anderson model on the 
infinite RRG~\cite{tikhonov2019statistics,efetov1985anderson,efetov1987density,efetov1987anderson,zirnbauer1986localization,zirnbauer1986anderson,verbaarschot1988graded,mirlin1991localization,mirlin1991universality,fyodorov1991localization,fyodorov1992novel,mirlin1994statistical,mirlin1997localization}. In fact, recent numerical and analytical studies have
provided compelling evidence against the existence of a truly intermediate 
multifractal extended phase in the thermodynamic limit~\cite{Garcia-Mata_2022,tikhonov2016anderson,garcia2017scaling,biroli2018delocalization,vanoni2023renormalization,baroni2023corrections,biroli2022critical}. (In contrast, it is widely accepted that the eigenstates of the Anderson model on loop-less Cayley trees are genuinely multifractal in the whole delocalized regime~\cite{Tikhonov_2016,sonner2017multifractality,monthus2011anderson,biroli2020anomalous,nosov2022statistics}.)

Yet, the existence of states exhibiting properties intermediate between full localization and full \rev{delocalization} 
has emerged as a focal point in various physical problems (including AL and MBL)~\cite{wegner1980inverse,rodriguez2011multifractal,mace2019multifractal,gornyi2017spectral,tarzia2020many,Luitz_2015,Serbyn_2017,Tikhonov_2018,Luitz_2020,Faoro_2019,baldwin2018quantum,biroli2021out,parolini2020multifractal,kechedzhi2018efficient,pino2017multifractal,pino2016nonergodic,Smelyanskiy_2020,Atland_2019,monteiro2021minimal}. Solvable (toy) dense random matrix models, exemplified by the Gaussian Rosenzweig-Porter (RP) model~\cite{RP_1960} and its extensions, showcase the appearance of fractal wavefunctions in intermediate regions of the phase diagram~\cite{Kravtsov_2015,vonSoosten_2019,Facoetti_2016,Truong_2016,Bogomolny_2018,DeTomasi_2019,amini2017spread,pino2019ergodic,berkovits2020super,us_2023,kravtsov2020localization,khaymovich2020fragile,monthus2017multifractality,biroli2021levy}.
Additionally, alternative classes of random matrix models featuring multifractal phases have surfaced across diverse physical contexts~\cite{sarkar2021mobility,roy2018multifractality,wang2016phase,nosov2019correlation,duthie2022anomalous,kutlin2021emergent,motamarri2021localization,tang2022non,tarzia2022fully,Skvortsov_2022,Cai_2013,DeGottardi_2013,Liu_2015,Das_2022,Ahmed_2022,Lee_2022}.  

In this paper we revisit a well-known sparse random matrix ensemble, revealing its surprisingly rich phase diagram, characterized by extended but multifractal eigenstates throughout a broad region of the parameter space.
Specifically, 
we focus on
the weighted adjacency matrix of random \ER~graphs (WER), which are $N \times N$ real symmetric matrices having a finite mean number $p$ of non-zero elements per row~\cite{barabasi2016book,AnttiKnowles_I,AnttiKnowles_II}. The 
latter
are drawn 
independently and identically from a certain probability distribution, which is regular in zero and has a finite variance (throughout we will focus on the Gaussian case). 
One of the interesting properties of this ensemble is the existence of a percolation transition at $p_c=1$: 
for $p<p_c$ the graph breaks into disconnected components (or finite clusters) of finite size (\ie~not scaling with $N$). 
Clearly, this should result in fully-localized eigenvectors on the small clusters, and consequently in the absence of eigenvalue correlations at distances of 
$\mathcal{O}(1/N)$.
Conversely, when $p > p_c$ a giant connected block (infinite cluster) with size of 
$\mathcal{O}(N)$
appears with probability approaching one in the large-$N$ limit, while the remaining nodes belong to an extensive number of finite clusters containing $\mathcal{O}(1)$ nodes separated from the bulk. 
Throughout this paper we will only focus on the spectral properties of the giant connected component appearing for $p>1$.
A mobility edge signaling the onset of Anderson-localized eigenvectors is expected at large energy due to large values of the random 
off-diagonal elements~\cite{Metz_2010,abou1974self,biroli2010anderson}. 
All this lends a special interest to this problem. 

In fact, the WER ensemble has already been studied in the past. The density of states (DoS) was first investigated in Ref.~\cite{Rodgers_1988} by means of the replica trick, while in Refs.~\cite{mirlin1991localization,Rodgers_1990} the spectral statistics of the model was studied using a supersymmetric formalism. 
Besides rederiving the integral equation for the DoS already found in Ref.~\cite{Rodgers_1988},
in Ref.~\cite{mirlin1991localization} Mirlin and Fyodorov
computed the density-density correlation function. The resulting supersymmetric action is similar (but not identical) to that of the Anderson model on the Bethe lattice~\cite{tikhonov2019statistics}. These authors investigated the analytical properties of the saddle point solution for $p \ll 1$ using an expansion in powers of $p$, which leads to the vanishing of the connected part of the level correlator, in agreement with the fact that below the percolation threshold the eigenvectors are all localized on the small clusters. 
Conversely, in the opposite limit 
of large connectivity
$p \gg 1$, the authors argued that the model becomes essentially equivalent to the large-connectivity limit of the Anderson model on the Bethe lattice in the presence of weak disorder in the hopping amplitudes.
Based on the close similarity of these two models, it is natural to expect that 
an Anderson transition should take place in the WER ensemble upon increasing $p$, and that the 
spectral statistics 
should be of the Dyson type in the delocalized phase. 

Below we show that the phase diagram of the model (which we anticipate here in Fig.~\ref{fig:PD}(a)) is 
more complicated than this expectation, and contains a non-trivial delocalized but multifractal phase in a broad region
of the ($E$--$p$) plane, where $p$ is the average 
\rev{degree}
and $E$ the eigenstate energy. 
The underlying physical mechanism behind the emergence of multifractal states is schematically illustrated in Fig.~\ref{fig:PD}(b) and can be 
rationalized
as follows: when the average degree is low, the percolating connected component becomes fragile,
as it contains numerous edges that, if removed, would cause it to split into disconnected components.
When the hopping amplitudes on these links are sufficiently small, this 
results in
effective fragmentation. We argue that incorporating these effects 
arising
from the pronounced heterogeneity of the graph's topology into the supersymmetric approach, which treats all nodes uniformly, 
poses
significant challenges. This difficulty explains why the multifractal states could not be observed in the analysis conducted in Ref.~\cite{mirlin1991localization}.
\revv{The decade-long debate on multifractality in the Anderson model on the RRG~\cite{biroli2012difference,de2014anderson,altshuler2016nonergodic,Kravtsov_2018,savitz2019anderson,pino2020scaling,pino2023correlated,zirnbauer2023wegner,arenz2023wegner,Garcia-Mata_2022,tikhonov2016anderson,garcia2017scaling,biroli2018delocalization,vanoni2023renormalization,baroni2023corrections,biroli2022critical} highlights the importance of handling finite-size effects with extreme caution. While we cannot entirely rule out their influence in this study, we will support our claims by showing that both semi-analytic (cavity) calculations and exact diagonalizations consistently converge to the same results.
Moreover, the physical mechanism we propose is reminiscent of the fragmentation of the Hilbert space, which has been recently argued to play a crucial role in non-ergodic many-body systems,} 
such as MBL~\cite{pietracaprina2021hilbert} and quantum kinetically constrained models~\cite{bernien2017probing,serbyn2021quantum,pancotti2020quantum}. 

In more recent years, subsequent works have refined the study of the average spectral density in ER graphs by using several approximation schemes~\cite{Monasson_1999_SDA,Semerjian_2002} and replicas~\cite{Kuhn_2008}. Other works have analyzed the spectrum of related models such as locally tree-like graphs~\cite{Rogers_2008,Slanina_2011,arras2023existence}, 
or explored the effect of row constraints (using supersymmetry~\cite{Akarapipattana2023}). 
Several papers have focused on the origin of the discrete part of the spectrum (see below for a precise definition),
first discussed in Ref.~\cite{Kirkpatrick_1972}, and later 
elucidated in Refs.~\cite{Bauer2001,golinelli2003statistics} for the adjacency matrix of random labeled trees with constant weights.
More recent studies have addressed the eigenvalue index distribution of ER graphs~\cite{Metz_2016}, and their spectral and localization properties in the large-connectivity limit~\cite{Metz_2010,Metz_2022}.
The main purpose of the present work is to push the latter studies towards smaller (and generic) values of the 
\rev{average degree}
$p$, and for a generic distribution of the hopping amplitudes,
with special focus on the statistical properties of wavefunction's amplitudes and the correlation among energy levels.

The rest of the paper is organized as follows.
In Sec.~\ref{sec:model} we introduce the model, while in Sec.~\ref{sec:detecting} we delineate the signatures of delocalized multifractal states in 
the distribution of the diagonal elements $G_{ii}$ 
of the resolvent matrix (see Eq.~\eqref{eq:resolvent} below).
In Sec.~\ref{sec:cavity} we write the self-consistent equations satisfied by $G_{ii}$, 
which we use in Sec.~\ref{sec:phase_diagram} to characterize the phase diagram anticipated here in Fig.~\ref{fig:PD}(a). 
In Sec.~\ref{sec:mechanism} we illustrate the physical mechanism that gives rise to the multifractal behavior.
In Sec.~\ref{sec:level} we explore several indicators of the level statistics in the 
extended \rev{multifractal} phase, which support the scenario put forward in Sec.~\ref{sec:phase_diagram}. In Sec.~\ref{sec:return_probability} we show that the presence of delocalized but multifractal eigenfunctions has deep consequences on the anomalous 
behavior of the return probability of a quantum particle undergoing unitary Hamiltonian dynamics, starting from a random node of the graph. 
In Sec.~\ref{sec:conclusions} we finally summarize our results, and elaborate on their possible future extensions.

\section{The model} \label{sec:model}
We consider real, symmetric $N \times N$ matrices ${\cal H}$, whose elements ${H}_{ij}$ are (up to the symmetry ${H}_{ij} = {H}_{ji}$) independent and identically-distributed random variables:
\begin{equation}
    {H}_{ij} = \frac{1}{\sqrt{p}}\sigma_{ij} h_{ij}, \quad \sigma_{ij}=\begin{dcases}
        1, & \text{with probability }p/N,\\
        0, & \text{otherwise.}
    \end{dcases}
    \label{eq:H}
\end{equation}
The Bernoulli random variables $\sigma_{ij}$ are the elements of the adjacency matrix of an \ER~graph with average degree $p\sim \mathcal{O}(1)$, while the weights (\ie~the hopping amplitudes) $h_{ij}$ are independent and identically-distributed random variables drawn from a normalized probability distribution $\pi(h)$.
Henceforth we specifically restrict to the Gaussian case, 
so that the matrix with elements $h_{ij}$ belongs to the Gaussian Orthogonal Ensemble (GOE), with $\avgg{h_{ij}}=0$, $\avgg{h_{ij}^2} = 1/2$ for $i \neq j$, and $\avgg{h_{ii}^2}=1$. 
However, we argue below that our results hold for any choice of the distribution $\pi(h)$, provided that 
it is regular \rev{and symmetric}, has a finite variance, and $\pi(0)>0$ (but we will discuss to which extent this last requirement can be relaxed).
With the definition in Eq.~\eqref{eq:H},
the eigenvalues of $\mathcal H$ are of $\mathcal O(1)$ for $N\gg 1 $. Additionally, the matrix elements 
$H_{ij}$ have been scaled by $\sqrt{p}$,
and
\begin{equation}
    \Big\langle\sum_{ij} {H}_{ij}^2 \Big\rangle = \avg{\tr {\cal H}^2} = \Big\langle \sum_\alpha \lambda_\alpha^2 \Big\rangle \longrightarrow \frac{N}{2} 
\end{equation}
in the large-$N$ limit.
Taking $N\to\infty$ first and $p\to\infty$ after, 
the average spectral density converges to a Wigner semicircle between $\left[-2,2\right]$.
Note that in the large-$N$ limit and for finite $p$
most of the diagonal elements of ${\cal H}$ are equal to zero, 
and only a few of them ($p$ on average) are Gaussian random variables with zero mean and variance $1$. 

Hereafter we will only focus on the case $p>1$, \ie~above the percolation 
threshold, where a unique giant connected component of the graph exists (with probability tending to $1$ for $N \to \infty$). 
In the thermodynamic limit, the degree of a given node $z_i$ \rev{(\textit{i.e.}~the number of incident links)} is a random variable that follows a Poisson distribution $P(z) = e^{-p} p^z/z!$ of average $\langle z \rangle = p$ and variance $\langle z^2 \rangle - \langle z \rangle^2 = p$.  
As anticipated above, 
the graph exhibits
an extensive number of small clusters containing ${\cal O}(1)$ nodes disconnected from the bulk. For instance, the probability of finding a single isolated node disconnected from the bulk 
is equal to $P(z=0)=e^{-p}$, implying that there is an extensive number $N e^{-p}$ of them. The number of small clusters decreases exponentially with $p$, and eventually vanishes in the large-connectivity limit. These small clusters are somewhat trivial, in the sense that they produce an extensive number of localized eigenvectors with eigenvalues completely uncorrelated from those of the connected component \cite{golinelli2003statistics,Bauer2001}, and no correlations on the scale of the mean level spacing. Yet, the presence of these states can hinder the numerical analysis of the spectral properties of the connected part. For this reason, in the following we will discard the small clusters, and only consider the nodes belonging to the giant cluster. 

\begin{figure}
\includegraphics[width=0.482\textwidth]{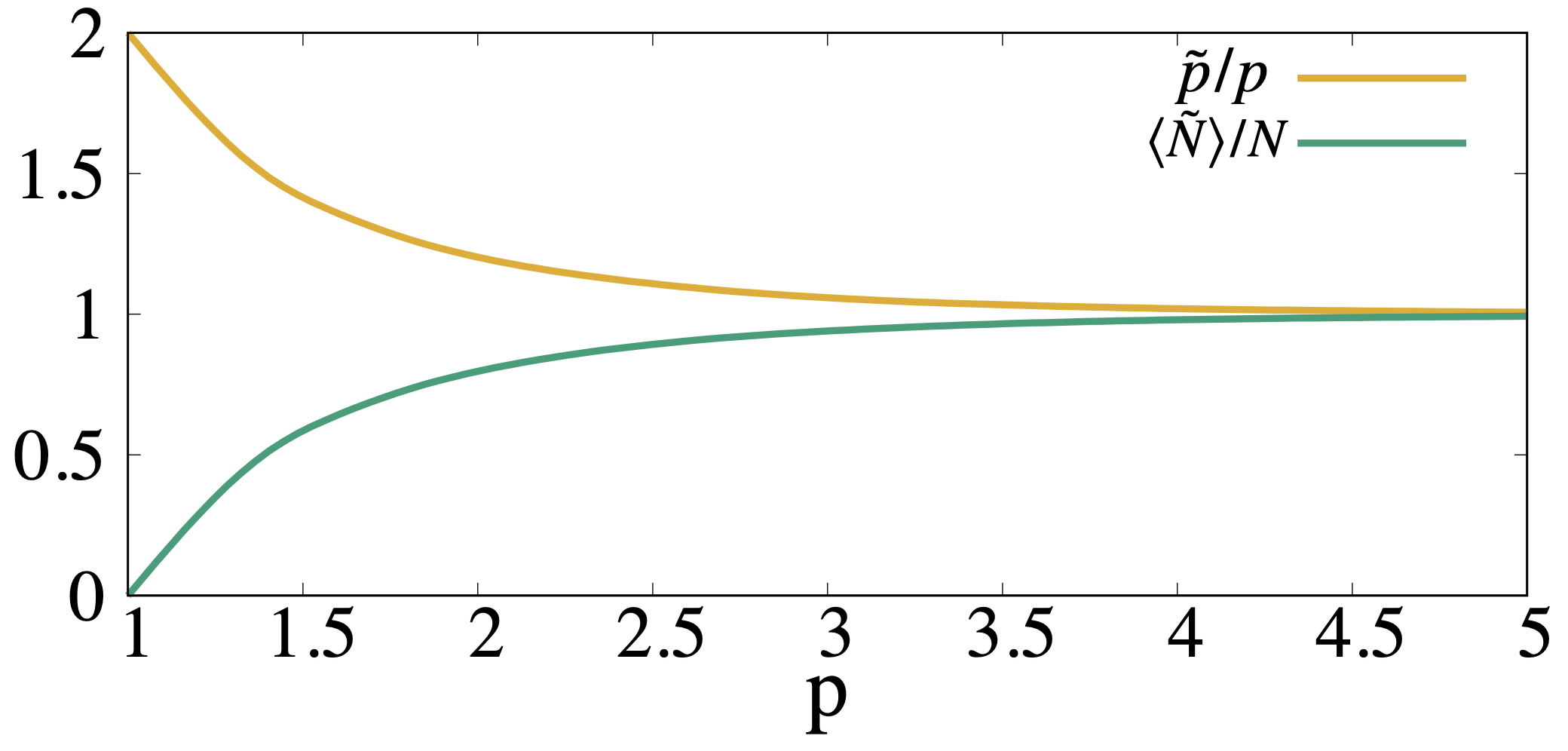}  
\put(-90,27){\includegraphics[scale=0.06]{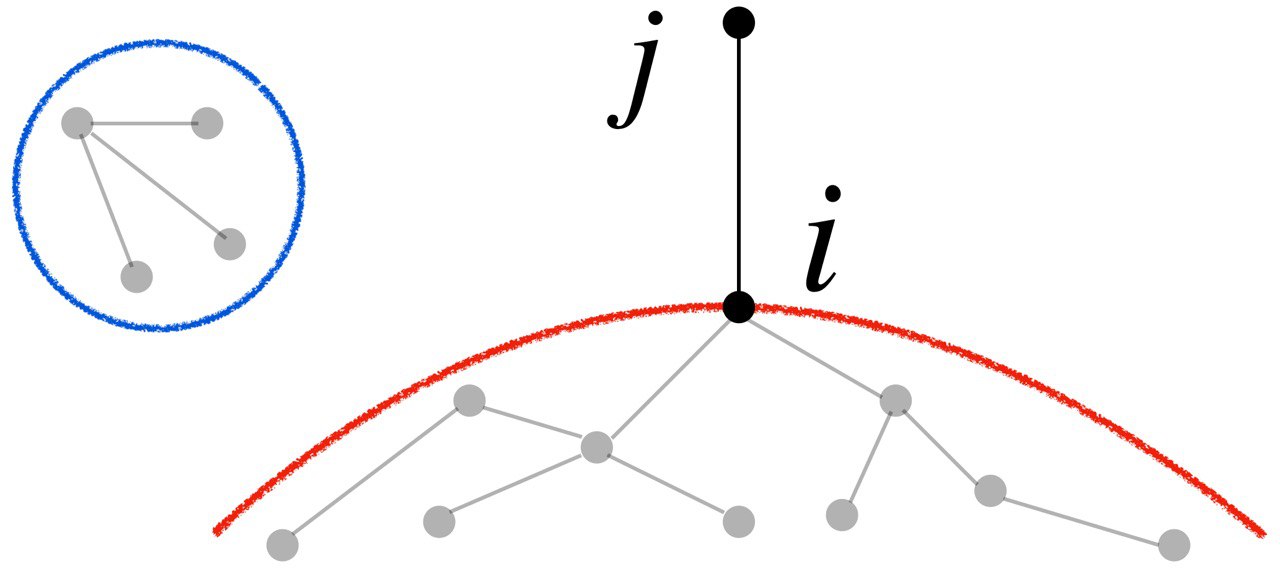}}
 \caption{Fraction of nodes belonging to the giant connected component, $\avg{\gamma} = \avgg{\tilde{N}}/N$, and renormalized effective average degree, $\tilde{p}/p$, as a function of the ``bare'' average degree $p$. 
 Inset: node $j$ is a \textit{leaf}, because it has degree 1, and is thus connected by a unique link ($i$--$j$) to the rest of the graph. 
 To obtain $\tilde p$ and $\tilde N$, we removed the disconnected components such as the one formed by the nodes in the blue circle (but not the leaves).}
 \label{fig:peff}
\end{figure}

Note that, after removing all small isolated clusters from ${\cal H}$, the number of nodes of the samples is not fixed, but slightly fluctuates from one realization to another.
Specifically, the number of vertices in the giant connected component (for $p>1$) is given by $\tilde{N} = \gamma(p) N$, where $\gamma(p)$ is a random variable whose average can be easily computed analytically as follows. We define $Q$ as the probability that a given node of the original graph does {\it not} belong to the connected component, \ie~$Q=1-\avg{\gamma}$. Since all its neighbors must also belong to small isolated clusters, it is straightforward to show that $Q$ verifies the 
self-consistent relation
\begin{equation}
    Q = \sum_{z=0}^\infty P(z) Q^z = e^{p(Q-1)} \, ,
\end{equation}
with $P(z)$ being the Poisson distribution, 
from which one immediately obtains 
$\avg{\gamma(p)}=1-\exp[-p\avg{\gamma(p)}]$. From the nontrivial solution of this equation one derives $\avg{\gamma(p)} = \avgg{\tilde{N}}/N$, plotted in Fig.~\ref{fig:peff}: this ratio is an increasing function of $p$, which decays linearly to zero for $p \to 1$, 
and saturates to $1$ for large~$p$.

The removal of the disconnected clusters from the graph has also the effect of renormalizing the effective average degree. In fact, the nodes that are removed have typically a smaller degree compared to those in the bulk, 
thus leading to
a larger effective average degree of the nodes belonging to the giant connected component. The ratio of the renormalized average degree $\tilde{p}$ 
divided by the ``bare'' degree $p$ is also plotted in Fig.~\ref{fig:peff} as a function of $p$. One finds that $\tilde{p} \to 2$ for $p \to 1$, corresponding to the fact that at the percolation threshold the connected component is essentially a one-dimensional chain, while $\tilde{p}/p \to 1$ for $p \gg 1$.

Importantly, even after removing the small disconnected components, the graph still features an extensive number $N_L = N p/e^p$ of leaves (\ie~nodes with degree $1$, see the inset of Fig.~\ref{fig:peff}), as well as groups of nodes that are connected by a unique edge to the rest of the graph (also called \textit{grafted trees}~\cite{Bauer2001,golinelli2003statistics}, see \eg~Fig.~\ref{fig:PD}(b)). As we will see below in Sec.~\ref{sec:mechanism},
these structures will play a crucial role in determining the spectral properties of the WER ensemble.

\section{Detecting multifractal 
extended states} \label{sec:detecting}

Possibly the most natural set of observables that can be used to discriminate between extended, localized, and fractal eigenstates is provided by the so-called fractal dimensions $D_q$. The fractal dimensions are introduced via the asymptotic scaling behavior with the system size of the generalized inverse 
participation ratios (IPRs),
defined as
\begin{equation} \label{eq:Iq}
I_q(E) = \frac{\left \langle \sum\limits_\alpha \sum\limits_i \; 
|\psi_\alpha (i)|^{2 q} \; \delta (E - \lambda_\alpha) \right \rangle }
    {\left \langle \sum\limits_\alpha \delta (E - \lambda_\alpha) \right \rangle }\, . 
\end{equation}
Here $\psi_\alpha (i)$ is the coefficient on node $i$ of the $\alpha$-th eigenvector of ${\cal H}$ of energy $\lambda_\alpha$, and the average is performed over the disorder realizations.
The amplitudes of fully-delocalized eigenstates are essentially Gaussian random variables with zero mean and variance equal to $1/N$ (to ensure normalization), and hence $I_q \propto N^{1-q}$. Conversely, Anderson-localized eigenstates have a non-vanishing amplitude on a finite set of nodes, and the $I_q$ are of $\mathcal O(1)$ (except for very small $q$~\cite{AbouChacra_1973,de2014anderson,Ferdinand_2008,garcia2020two}, at least $q<1/2$).
Delocalized but multifractal states correspond to an intermediate situation between these two opposite limits. 
In general, one can thus define the fractal dimensions $D_q$
as
\begin{equation} \label{eq:Dq}
I_q \propto N^{D_q (1 - q)} \,\, \Rightarrow \,\,
D_q = \frac{1}{1-q} \, \frac{\partial \ln I_q}{\partial \ln N} \, .
\end{equation}
Note that for $q=1$ the expression above should be modified: indeed, by taking the $q \to 1$ limit one obtains
\begin{equation} \label{eq:D1}
    D_1 = - \frac{\partial}{\partial \ln N} 
 \frac{\left \langle \sum\limits_\alpha \sum\limits_i |\psi_\alpha (i)|^{2} \, \ln |\psi_\alpha (i)|^{2} \, \delta (E - \lambda_\alpha) \right \rangle }
    {\left \langle \sum\limits_\alpha \delta (E - \lambda_\alpha) \right \rangle } \, .
\end{equation}
Another special case is $q \to \infty$: using the definition~\eqref{eq:Iq} and taking the limit $q \to \infty$ in Eq.~\eqref{eq:Dq}, one finds
\begin{equation} \label{eq:Dinf}
D_\infty =  \frac{\partial}{\partial \ln N} \, \ln \frac{ \left \langle \sum\limits_\alpha |\psi_\alpha|^{2}_{\rm max} \, \delta (E - \lambda_\alpha) \right \rangle }
    {\left \langle \sum\limits_\alpha \delta (E - \lambda_\alpha) \right \rangle } \, ,
\end{equation}
where $ |\psi_\alpha|^{2}_{\rm max} $ is the maximum amplitude of the eigenvector $\psi_\alpha$.
We thus have 
$D_q=1$ identically for fully-extended states, $D_q=0$ for Anderson-localized states (except for very small $q$~\cite{AbouChacra_1973,de2014anderson,Ferdinand_2008,garcia2020two}), and $0<D_q<1$ for intermediate delocalized but \rev{multifractal} 
states (at least for some values of~$q>1/2$). 

All information about the 
statistics of the wavefunctions and
of the energy levels is encoded in the statistical properties of the elements of the resolvent matrix, defined as 
\begin{equation}
    {\cal G}(z) = \left({\cal H} - z {\cal I}_N \right)^{-1} \, ,
    \label{eq:resolvent}
\end{equation}
where ${\cal I}_N$ is the $N \times N$ identity matrix. 
Here $z = E + {\rm i} \eta$, where $E$ is the energy at which we 
inspect the spectral statistics of the model, and ${\rm i} \eta$ is an imaginary regulator
that 
softens the poles in the denominator of $\mathcal G$ (and that should be sent to zero in the thermodynamic limit at the end of the calculation, 
see below). The key physical observable in this context is the local density of states (LDoS)
\begin{equation} \label{eq:ldos}
\rho_i(E) = \sum_{\alpha = 1}^N \abs{\psi_\alpha (i)}^{2} \, \delta (E - \lambda_\alpha) \, ,
\end{equation}
which is intimately related to the inverse lifetime of a particle of energy $E$ sitting on a given node $i$ of the graph.
Summing over all nodes
one immediately obtains the average density of states as $\rho_{\rm av} (E) = \avg{ (1/N) \sum_i \rho_i(E)}$. By using the spectral decomposition of ${\cal G}$ over the basis of the eigenstates of ${\cal H}$, 
the LDoS can be expressed in terms of the 
diagonal elements $G_{ii}$ of the resolvent 
as
\begin{equation} \label{eq:rhoiG}
\rho_i (E) = \frac{1}{\pi} \lim_{\eta \to 0^+} \im G_{ii} (z) \, .
\end{equation}
The LDoS is a random variable, 
whose
probability distribution
\begin{equation} \label{eq:Prho}
    P(\rho; z) = \avg{\frac{1}{N} \sum_i \delta \left( \rho - \pi^{-1} {\rm Im} G_{ii} (z) \right) } 
\end{equation}
provides the order parameter distribution function for AL.
Although strictly speaking the LDoS is only obtained in the $\eta\to 0$ limit, in the definition~\eqref{eq:Prho} we will keep $\eta$ small but finite. Indeed,
as explained below, the behavior of $P(\rho)$ with small $\eta$ allows one to distinguish between fully-extended, localized, and fractal regimes. 
Note that the probability distribution of the LDoS depends on the energy $E$ and on the regulator $\eta$ (through $z = E + i \eta$); in order to 
lighten
the notations, we will omit this explicit dependence throughout the paper, and simply write $P(\rho)$. 

It is easy to show (again using 
their
spectral decomposition 
over the basis of the eigenstates of ${\cal H}$) that the generalized IPRs introduced in Eq.~\eqref{eq:Iq} can be expressed
as
\begin{equation} \label{eq:IqG}
    I_q (E)= \frac{\sqrt{\pi} \, \Gamma \left ( \frac{q}{2} \right)}{\Gamma \left( \frac{q-1}{2} \right)} \lim_{\eta \to 0^+} \frac{\eta^{q-1} \left \langle \frac{1}{N} \sum\limits_i \, |G_{ii} (z)|^q \right \rangle}{\left \langle \frac{1}{N} \sum\limits_i \, {\rm Im} G_{ii}(z) \right \rangle } \, .
\end{equation}
The denominator in Eq.~\eqref{eq:IqG}
is proportional to the average DoS $\rho_{\rm av} (E)$, and accounts for the number of states at a given value $E$ of the energy. However, the average DoS is completely insensitive to the nature of the eigenstates, and does not allow one to discriminate between the different regions of the phase diagram. Conversely, the numerator is related to the $q$-th moment of the LDoS, and exhibits different behaviors for extended, localized, and fractal eigenstates. To understand this, we need first to discuss the shape of the probability distribution of the LDoS.

\begin{figure}
\includegraphics[width=0.482\textwidth]{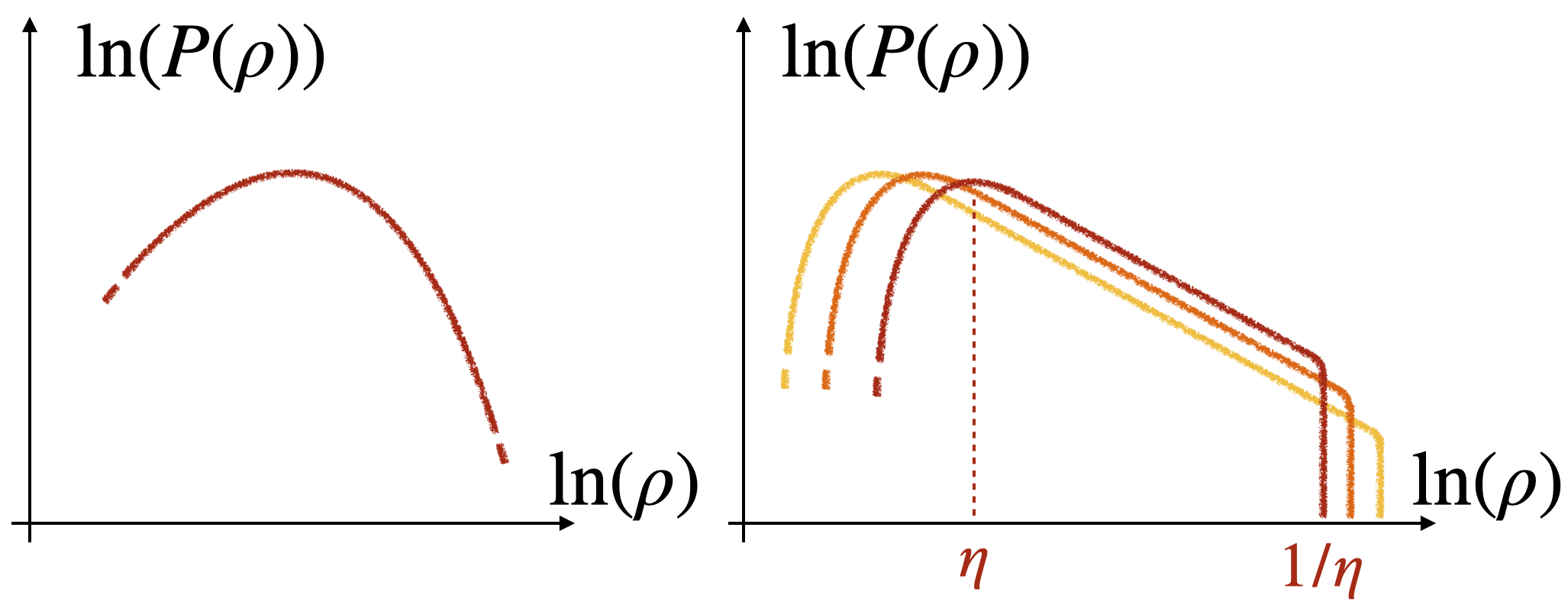}
\put(-160,83){(a)} 
\put(-25,83){(b)} 
 \caption{Schematic representation of the LDoS distribution $P(\rho)$, (a) in the fully-delocalized phase, and (b) in the Anderson-localized phase. In the localized case, colors from dark to light correspond to decreasing values of $\eta$, while in the delocalized case the distribution becomes $\eta$-independent for sufficiently small $\eta$.}
 \label{fig:sketch_dist}
\end{figure}

\begin{itemize}
\item[(a)] In the fully-delocalized metallic phase, $P(\rho)$ converges to a non-singular, $\eta$-independent distribution for $\eta \to 0^+$ in the thermodynamic limit, with tails that decay 
rapidly
to zero at large $\rho$ (as in 
Fig.~\ref{fig:sketch_dist}(a)). All the moments $\avg{\rho^q}$ are finite
and, since the main contribution to the numerator of Eq.~\eqref{eq:IqG} generically comes from the imaginary part,
this implies that
the generalized IPRs (for $q>1$)
vanish identically 
for $\eta \to 0^+$: indeed, one has
\begin{eqnarray} \label{eq:rhoqD}
    \qquad \avg{\frac{1}{N} \sum_i |G_{ii} |^q} \simeq \avg{\rho^q} \sim \mathcal O(1) \, \Rightarrow \, I_q \propto \eta^{q-1} \, .
\end{eqnarray}
\item[(b)] In the Anderson-localized phase, $P(\rho)$ is singular in the $\eta \to 0$ limit: it has a maximum in the region $\rho \sim \eta$, and power-law tails $P(\rho) \sim \sqrt{\eta}/\rho^{3/2}$ with a cutoff at $\eta^{-1}$~\cite{mirlin1994statistical} (as in 
Fig.~\ref{fig:sketch_dist}(b)). 
The main contribution to the moments 
\begin{equation}
    \qquad\avg{\rho^q} =
    \int \! \mathrm{d}\rho\,P(\rho) \,\rho^q 
    \simeq 
    \sqrt{\eta} \int_{\rho_0}^{\eta^{-1}} \!\!\!\! \mathrm{d}\rho\, \rho^{q-\frac32} 
    \propto
    \eta^{1-q} 
\end{equation}
comes from the upper cutoff at $\rho \sim \eta^{-1}$ (for $q \ge 1/2$), while 
the typical value of $\rho$ is of 
$\mathcal{O}(\eta)$.
This behavior reflects the fact that in the localized phase the wavefunctions are exponentially localized on a few $\mathcal O(1)$ sites, where $\rho_i$ takes very large values, while the typical value of the LDoS is exponentially small and vanishes in the thermodynamic limit for $\eta \to 0^+$. 
From Eq.~\eqref{eq:IqG} we thus get (for $q>1/2$)
\begin{equation} \label{eq:rhoqL}
    \qquad \avg{\frac{1}{N} \sum_i |G_{ii} |^q} \simeq \avg{\rho^q} \propto \eta^{1-q} \, \Rightarrow \, I_q \sim \mathcal O(1) \, .
\end{equation}
\item[(c)] As we will show below, for some values of the parameters $E$ and $p$ we find an intermediate situation corresponding to a specific kind of multifractal eigenstates. The probability distribution $P(\rho)$ is given by a bulk part which shares most of the properties of the distribution found in the fully-delocalized case, namely it is regular and $\eta$-independent for sufficiently small $\eta$. Conversely, at large values of $\rho$ the distribution exhibits power-law tails of the form $\rho^{-(1+\beta)}$ 
(similarly to the localized case, although with no $\eta$-dependent prefactor), which extend up to a cut-off at the scale $1/\eta$.
As it will become clear later (see the discussion after Eq.~\eqref{eq:ImGii}), the $\eta$-dependence of the cut-off on the right tails of the probability distribution
necessarily causes
a residual $\eta$-dependence on the left tails,
which describe the behavior at very small $\rho$. 
We 
find that the exponent $\beta$ depends on the parameters $p$ and $E$, and is larger than $1$ (see Fig.~\ref{fig:Dq}(a) below): this implies 
that the tails do not contribute to the normalization integral, and that small enough moments, and in particular the average DoS, do not depend on the cut-off and thus on $\eta$. Conversely, large moments are dominated by the cut-off. In particular, one has
\begin{eqnarray} 
\label{eq:rhoqI}
\qquad \;\; \avg{\rho^q} &\propto&
\left \{
\begin{array}{ll}
\eta^{\beta - q}, & {\rm if~} \;\;\; q>\beta,  \\
\mathcal O(1), & {\rm if~} \;\;\; q \leq \beta,  
\end{array}
\right .
\nonumber\\
[5pt]
\Rightarrow \,\,\,\,\,\, I_q &\propto& \left \{
\begin{array}{ll}
\eta^{\beta - 1}, & {\rm if~} \;\;\; q>\beta,   \\
\eta^{q - 1}, & {\rm if~} \;\;\; q \leq \beta . 
\end{array}
\right .
\end{eqnarray}
\end{itemize}

In the following we will compute the probability distribution of the LDoS for instances of the WER ensemble of large but {\it finite} sizes, using a self-consistent approximation for the diagonal elements of the Green's function that becomes asymptotically exact in the $N \to \infty$ limit~\cite{Bordenave2010} (see Sec.~\ref{sec:cavity}). 
In this respect, the following subtle point requires a thorough discussion.
We first recall that in the delocalized 
regime 
the spectrum is absolutely continuous, meaning that the energy levels tend to 
form a continuum set
for $N \to \infty$  giving rise to 
a smooth
function, and the sum over the eigenvalues in Eq.~\eqref{eq:ldos} converges to an integral. 
Conversely,
in the Anderson-localized phase the spectrum is pure point, meaning that it is the result of a sum of uncorrelated $\delta$-peaks even in the thermodynamic limit. 
Of course,
the spectrum of any finite system is, by definition, point-like, independently of the nature of the eigenstates: indeed 
at finite $N$, even deep into the metallic regime, for any value of the energy $E$ there are no eigenvalues 
$\lambda_\alpha$ strictly equal to $E$, and thus the LDoS vanishes. 
Hence, in order to compute the spectral observables, one has to replace the $\delta$-function in the definitions of $\rho_i$ and $I_q$ (see Eqs.~\eqref{eq:Iq} and~\eqref{eq:ldos}) with a smooth function, such as
$\delta(x) \rightarrow \eta (x^2+\eta^2)^{-1} \pi^{-1}$, 
depending on a broadening parameter $\eta$. 
The fact that the system is in the delocalized phase manifests itself in the emergence of a broad interval of $\eta$ over which the probability distribution $P(\rho)$ of the LDoS 
becomes essentially independent of $\eta$. 
This interval extends over a broader and broader range of $\eta$ as the system size is increased, and eventually diverges in the thermodynamic limit~\cite{baroni2023corrections,biroli2018delocalization}. 
In other words, using the terminology of spontaneous symmetry breaking of conventional phase transitions, the limits $N \to \infty$ and $\eta \to 0$ do not commute in the metallic phase, \ie
\begin{equation}
    \lim_{\eta \to 0} \lim_{N \to \infty} P(\rho) \neq \lim_{N \to \infty} \lim_{\eta \to 0} P(\rho) .
\end{equation}
In principle, one should take the thermodynamic limit first, and the limit $\eta \to 0$ only at the end. Yet, since below we work at finite $N$, we will take these two limits concomitantly. 
Hence, the evaluation of the generalized IPRs at finite $N$ must be performed at small but finite $\eta$. Since the regulator $\eta$ serves as a broadening of the energy levels, whose mean distance is $\delta = 1/(N \rho_{\rm av}(E))$, 
its natural scale is $\eta=\delta \propto 1/N$. With this choice, 
using the definition~\eqref{eq:Dq} of $D_q$ in terms of $I_q$,
we immediately find that 
$D_q = 1$ in the fully-extended phase (see Eq.~\eqref{eq:rhoqD}),
$D_q = 0$ in the localized phase (except for very small $q$~\cite{AbouChacra_1973,de2014anderson,Ferdinand_2008,garcia2020two}, see Eq.~\eqref{eq:rhoqL}), and
\begin{equation} \label{eq:Dqbeta}
D_q = 
\begin{dcases}
\dfrac{\beta-1}{q-1} & {\rm if~} \;\;\; q>\beta  \, , \\
1 & {\rm if~} \;\;\; q \leq \beta  \, ,
\end{dcases} 
\end{equation}
in the intermediate delocalized but \rev{multifractal} 
phase with the properties described in the item (c) above (see Eq.~\eqref{eq:rhoqI}). 
Since $\beta$ is found to be larger than one throughout the extended multifractal phase (as we anticipated, see also Fig.~\ref{fig:Dq}(a) below), Eq.~\eqref{eq:Dqbeta} implies
$D_1=1$. 
Since $D_1$ is the scaling dimension of the support set of the eigenfunctions~\cite{de2014anderson,altshuler2016nonergodic}, $D_1=1$ implies that the support set of the multifractal eigenstates is
\rev{extensive in $N$},
and occupies a finite fraction of all the nodes of the graph. 
Multifractality manifests itself only in the higher moments of the wavefunctions' coefficients, which corresponds to the fact that the amplitudes are anomalously large on a vanishingly small fraction of nodes. 
This situation 
corresponds to a peculiar and rather weak form of multifractality, 
very different from the multifractality observed at the AL critical point in $d$ dimensions~\cite{rodriguez2011multifractal,Evers_2014,Mirlin_2000}, and from the fractal regime observed in the intermediate phase of the RP model and its generalizations~\cite{Kravtsov_2015,vonSoosten_2019,Facoetti_2016,Truong_2016,Bogomolny_2018,DeTomasi_2019,amini2017spread,pino2019ergodic,berkovits2020super,us_2023,kravtsov2020localization,khaymovich2020fragile,monthus2017multifractality,biroli2021levy,buijsman2022circular,khaymovich2021dynamical,Kutlin_2024}. This kind of multifractality is instead similar to the one found in the generalized $\beta$-ensemble for $\beta>2$~\cite{Breuer2007,Das_2023,Das_2024}.

\section{Self-consistent recursion relation for the diagonal elements of the resolvent} \label{sec:cavity}

Our approach to investigate the WER ensemble involves deriving the probability distribution of the LDoS for samples of significant yet finite size through the resolution of the so-called \textit{cavity} recursion equations \cite{Mezard_1987}. To derive these equations, we need first to introduce the resolvent matrices of the modified Hamiltonians ${\cal H}^{(i)}$, where the node $i$ has been removed from the graph: ${\cal H}^{(i)}$ is a $(N-1) \times (N-1)$ real symmetric matrix 
obtained from ${\cal H}$ by removing the $i$-th row and column. 
The key insight here is that, due to the sparse nature of the graph, short loops are rare. Specifically, one can show that for \ER~graphs the typical length of the loops scales as $\ln N/\ln (p-1)$~\cite{Marinari2004}. 
Hence, in the large-$N$ limit the graph locally resembles a tree~\cite{bordenave_notes}.
As a result, removing one node causes each of its neighbors to become essentially uncorrelated from the others. Consequently, assuming that in the absence of the site $i$ the LDoS on its neighbors factorizes, one can obtain (\eg~through direct Gaussian integration~\cite{Dean_2002,Rogers_2008,biroli2010anderson,Susca_2021}, or by using the block matrix inversion formula, also called the Schur complement formula~\cite{arous2008spectrum}) the recursion relation
\begin{equation} \label{eq:cavity}
    G_{i \to j} (z) =\left( {H}_{ii} - z - \!\!\! \sum_{m \in \partial i \setminus j} {H}_{mi}^2 G_{m \to i} (z) \right)^{-1} \, .
\end{equation}
Here again $z = E + {\rm i} \eta$, while 
\begin{eqnarray}
    G_{i \to j}(z)=({\cal H}^{(j)} - z {\cal I}_{N-1} )^{-1}_{ii}
\end{eqnarray}
is the so-called cavity Green's function, \ie~the diagonal element on node $i$ of the resolvent of the Hamiltonian 
$\mathcal H^{(j)}$ obtained by removing one of its neighbors $j$ --- with ${\cal I}_{N-1}$ being the $(N-1) \times (N-1)$ identity matrix. 
In Eq.~\eqref{eq:cavity}, the notation $\partial i \setminus j$ denotes the set of all $z_i - 1$ neighbors of $i$ except $j$. The ${H}_{im}=h_{im}/\sqrt{p}$ are the off-diagonal elements connecting two neighboring sites of the graph, where the $h_{im}$ are independent random numbers drawn from a Gaussian distribution with zero mean and variance $1/2$.  
Since ${H}_{ii}$ is the diagonal element of ${\cal H}$ on node $i$, which is $0$ on almost all  nodes ($N-p$ on average), and equal to a Gaussian random number of zero mean and unit variance on the few remaining nodes ($p$ on average), we will set it to zero throughout.

Henceforth we focus on the nodes belonging to the connected component of the graph, so that $N$ in the expressions above has to be replaced by $\tilde N$ (see Sec.~\ref{sec:model} and Fig.~\ref{fig:peff}).
For a node $i$ with degree $z_i$, one can define $z_i$ cavity Green’s functions, each one satisfying a recursion relation of this kind when one of the $z_i$ neighbors of the node has been removed. These relations represent in fact a set of $2 \tilde{N} \tilde{p}$ (on average) coupled non-linear equations, which can be solved relatively easily by iteration for graphs up to $\tilde{N} \sim 2^{24}$ nodes. Most importantly, the time required to find the solution scales linearly with $\tilde{N}$. 
We remark that population dynamics methods \cite{Mezard_2001}, which have been widely employed in the literature to solve cavity-like equations, 
appear unable to fully capture the spectral statistics
in the present context. Indeed, their underlying assumption of 
homogeneity of the graph
is not fulfilled in \ER~graphs, where the number of connections fluctuates from site to site --- and where the poorly-connected sites (or groups of sites) will play a pivotal role, which we will elucidate in Sec.~\ref{sec:mechanism}. To circumvent this limitation, we construct explicitly several instances of the graph, solve the cavity equations for each particular realization of $\mathcal H$, and finally average our measurements over all distinct realizations.

From the solution of these equations one can finally obtain the diagonal elements of the resolvent of the original problem as
\begin{equation} \label{eq:green}
G_{ii} (z) =\left( - \, z - \frac{1}{p} \sum_{m \in \partial i} h_{mi}^2 G_{m \to i} (z) \right)^{-1} \, ,
\end{equation}
where we already set $H_{ii} =0$ (see the discussion above) and  $z=E +{\rm i} \eta $.
It is possible to show that for \ER~graphs in the $N~\to~\infty$ limit these equations become asymptotically exact~\cite{Bordenave2010}, as the typical size of the loops diverges~\cite{Marinari2004}. However, at finite $N$ 
one should expect finite-size corrections
due to the presence of loops. Generally, these corrections go to zero 
as the system size is increased (at least on most of the nodes), yet they are expected to become very large close to the localization transition~\cite{baroni2023corrections}. 

We stress that, since the cavity approximation is obtained by discarding the effect of the loops, 
the finite-size effects due to the loops 
cannot be captured by our approach
even though we consider instances of finite size. In a certain sense, the cavity approximation corresponds to studying the system already in the thermodynamic limit, even though we solve the equations on finite (but very large) graphs. Yet, a small residual $N$-dependence 
still affects
the cavity calculations:
taking larger and larger graphs allows us to sample more accurately rare realizations of the disorder distributions, such as nodes with anomalously large degree, clusters of nodes connected to the bulk
by a single edge (\textit{grafted trees}~\cite{Bauer2001,golinelli2003statistics}), or edges with anomalously large values of the hopping amplitudes. 
(In this respect, here the size of the graphs plays a similar role as the size of the pool in the standard population dynamics approach.) 

From the knowledge of the diagonal elements $G_{ii}$ of the Green's function one 
immediately obtains, 
via Eq.~\eqref{eq:rhoiG}, the probability distribution of the LDoS, 
as well as 
the generalized IPRs
given in Eq.~\eqref{eq:IqG}.
These observables allow 
us to directly inspect the statistics of the wavefunctions' coefficients and 
that of the energy levels. 
In particular, from Eq.~\eqref{eq:green} one 
can express
the imaginary part of the Green's functions (which is proportional to the LDoS)
as
\begin{equation}
\im G_{ii} 
= 
\frac{\eta + 
\sum\limits_m 
\frac{h_{mi}^2}{p} \im G_{m \to i} 
}{\left(E + 
\sum\limits_m 
\frac{h_{mi}^2}{p} \re G_{m \to i} 
\right)^2  \!\! + \left(\eta + 
\sum\limits_m 
\frac{h_{mi}^2}{p} \im G_{m \to i} 
\right)^2 } \, ,
\label{eq:ImGii}
\end{equation}
which is bounded by $\eta^{-1}$.
Note that large (respectively small) values of $\im G_{ii}$ are obtained from small (respectively large) values of the Green's functions on the neighboring nodes. Specifically, values of the imaginary part of the Green's functions on one of the neighbors of $i$ close to the cut-off, $\im G_{m \to i} \sim \eta^{-1}$, produce values of $\im G_{ii}$ of 
$\mathcal{O}(\eta)$~\cite{abou1974self}.

\section{Phase diagram} \label{sec:phase_diagram}

In this section we present the phase diagram found by solving the self-consistent equations~\eqref{eq:cavity}~and~\eqref{eq:green} on the giant connected component of many instances of the WER ensemble, of large but finite size. The largest systems 
we could analyze
contain about 
$2^{24}$ nodes. 
Recall that, after removing all small isolated clusters from ${\cal H}$, the number of nodes $\tilde N$ of the samples is not fixed, but slightly fluctuates from one realization to another,
and the average effective connectivity is slightly renormalized (as discussed in Sec.~\ref{sec:model}). 
We vary the average degree $p$ in the range $p \in [1.6,5]$, and the energy $E$ across the bandwidth in the range $E \in [0,2]$. 
For each value of $p$ and $E$, we solve the recursion equations for many values of the imaginary regulator $\eta \in [2^{-25},2^{-4}]$, and many system sizes in the range $N \in [2^{14},2^{24}]= [16384,16777216]$, making sure that $N$ is large enough so that the probability distribution of the LDoS found within the cavity approximation becomes independent of $N$. 

\begin{figure}
\includegraphics[width=0.482\textwidth]{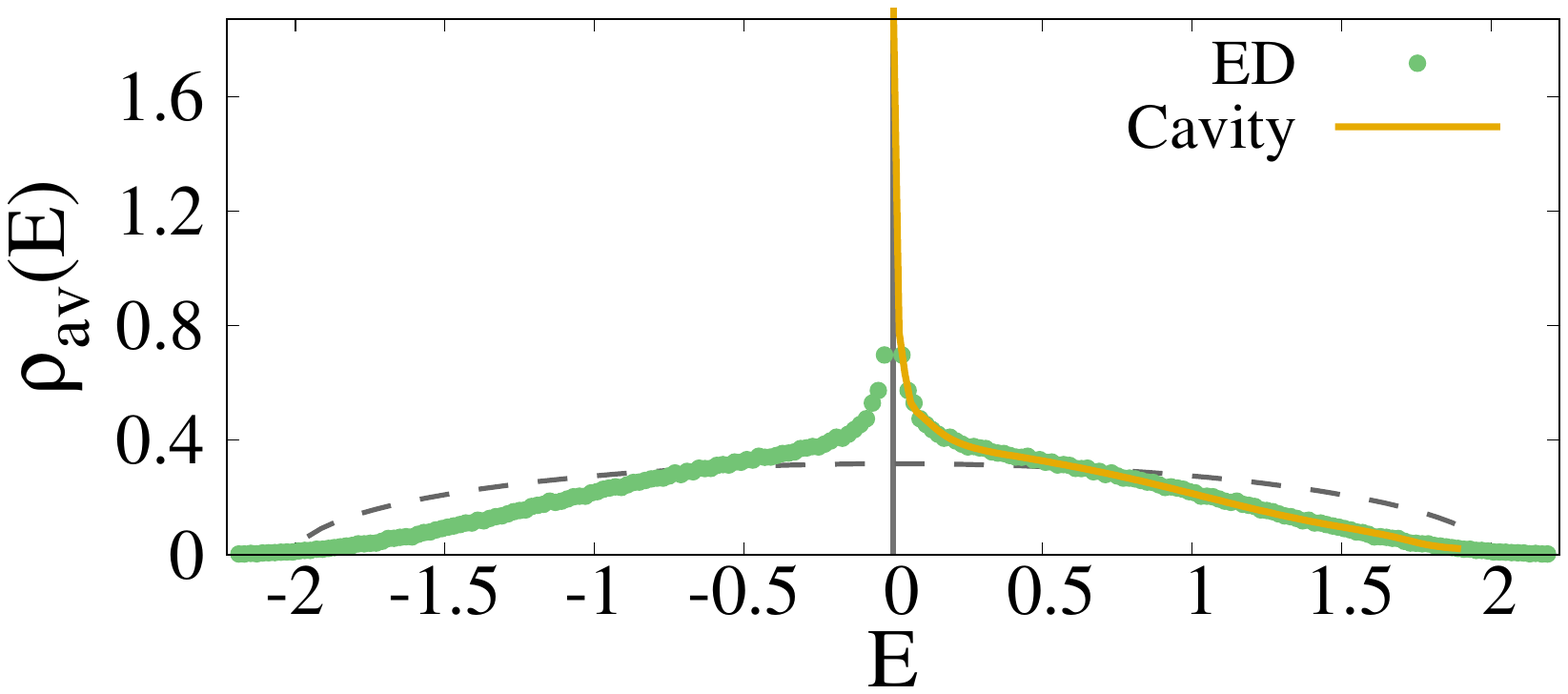}  
\vspace{-2.7cm}
 \caption{Average DoS for $p=2.4$. The yellow curve shows the results obtained by solving the cavity recursion relations~\eqref{eq:cavity} and~\eqref{eq:green} for systems of $N \simeq 2^{22}$ nodes. Green circles are the results of exact diagonalization (ED) of instances of $2^{12}$ nodes. We have checked that, for such values of $N$, the average DoS has converged to its thermodynamic limiting value within our numerical accuracy. The gray dashed line shows 
 for comparison
 the Wigner semicircle, which is recovered in the $p\to \infty$ limit.}
 \label{fig:DoS}
\end{figure}

\begin{figure*}
\includegraphics[width=0.338\textwidth]{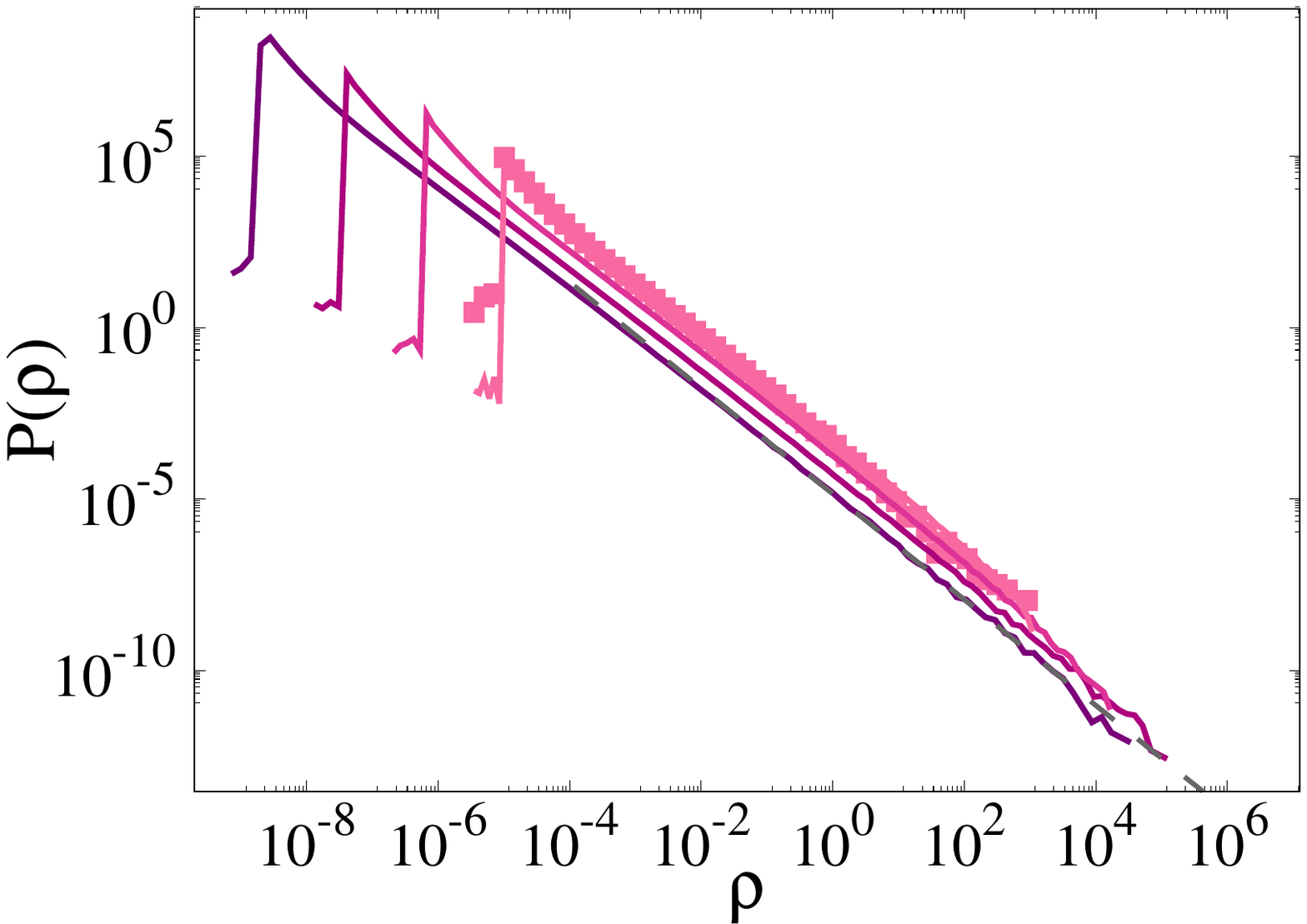} \hspace{-0.36cm} 
\put(-50,60){$\sim \sqrt{\eta} \rho^{-3/2}$} 
\put(-130,22){(a)}
\includegraphics[width=0.338\textwidth]{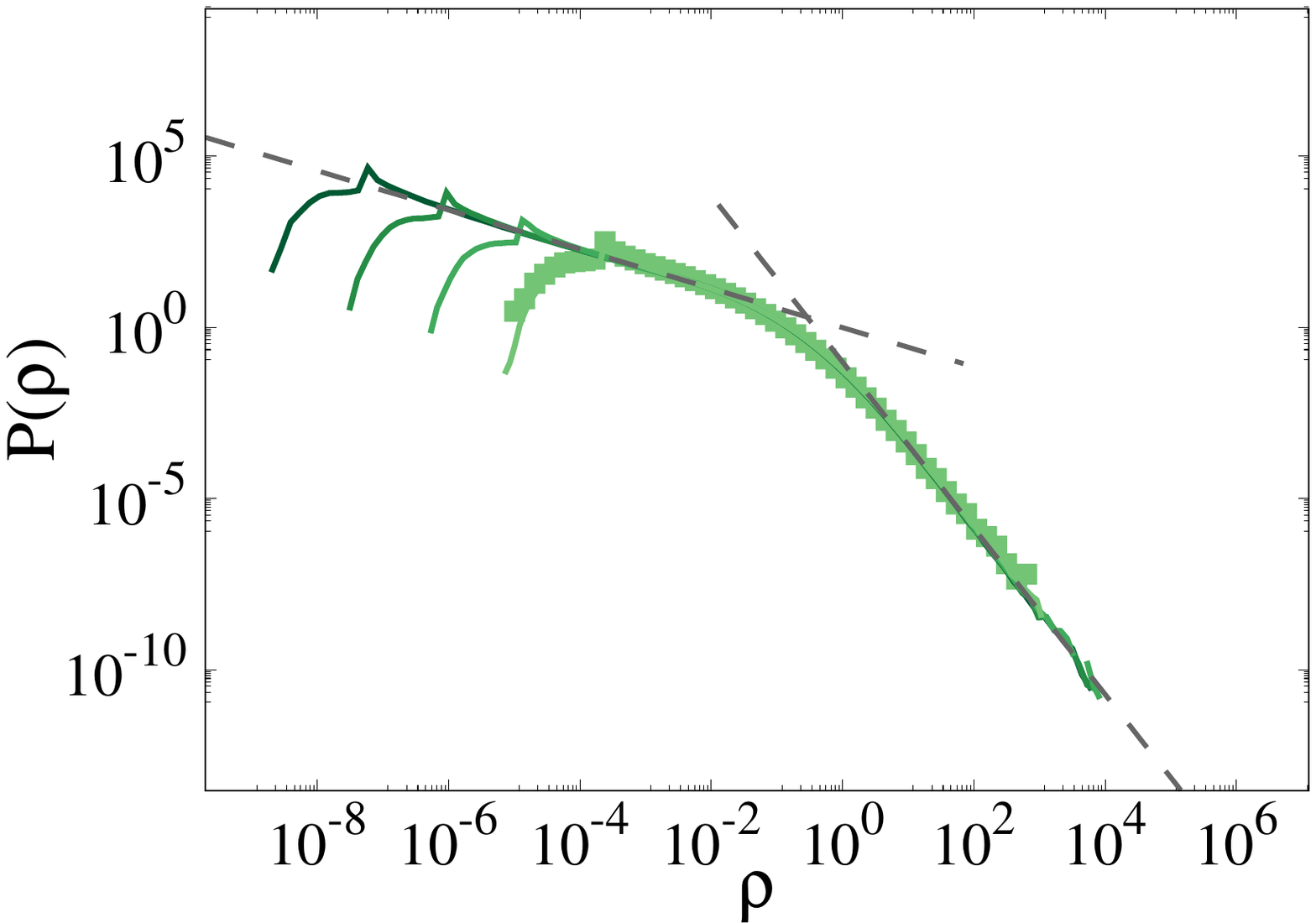} \hspace{-0.36cm} 
\put(-41,58){$\sim \rho^{-(1+\beta)}$}  \put(-103,91){$\sim \rho^{\beta-2}$}
\put(-130,22){(b)}
\includegraphics[width=0.338\textwidth]{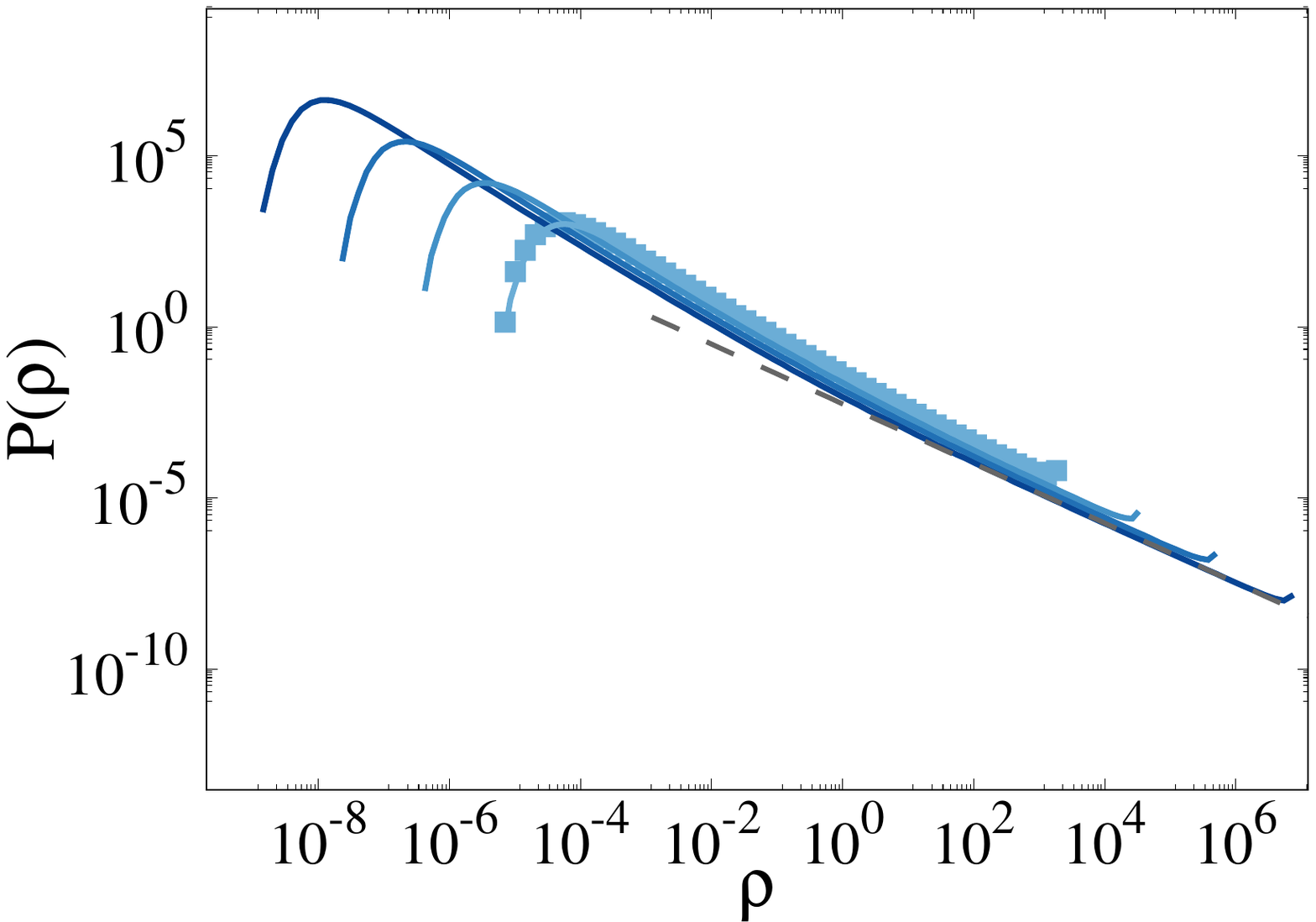} 
\put(-138,22){(c)}
\caption{Probability distributions $P(\rho)$ of the LDoS, for $p=2.4$ and 
(a) $E=1.9$, (b) $E=0.4$, and (c) $E=0$. 
Solid lines
are the distributions obtained by solving Eqs.~\eqref{eq:cavity} and~\eqref{eq:green} for many instances of the WER ensemble with $N \sim 2^{22}$ nodes (we have checked that the system size dependence is negligible within our numerical accuracy). Different colors correspond to different values of the imaginary regulator $\eta$, ranging from $2^{-25}$ (dark colors) to $2^{-13}$ (light colors). 
The cavity approximation results are compared to those obtained by exact diagonalization
(symbols) for the same values of $p$, $E$, and for $N=2^{14}$ and $\eta=2^{-13}$
(similar results were found for other values of $\eta$). 
The dashed lines in the three panels are power-law fits of the tails of 
$P(\rho)$
with exponents $-3/2$ (a), $\beta-2 \approx - 0.56$ and  $-(1+\beta) \approx - 2.44$ (b), and $- 0.87$ (c),  respectively. The origin of the peaks at $\rho \sim \eta$ in panels (a) and (b) and at $\rho \sim \eta^{-1}$ in panel (c) is explained in Sec.~\ref{sec:mechanism}.}
\label{fig:Prho}
\end{figure*}

\subsection{Average spectral density}
\label{sec:spectral_density}
We start by presenting the results for $p=2.4$, focusing first on the average DoS $\rho(E)$, plotted in Fig.~\ref{fig:DoS}. As anticipated, the average DoS 
cannot be used
to distinguish between localized, fully-extended, and delocalized but multifractal states. The plot shows that at small $p$ the average DoS is quite different from the Wigner semicircle (which is recovered in the large-connectivity limit). Most importantly, it features a $\delta$-peak at $E=0$:
its origin, related to degenerate eigenstates localized on small structures, has been
extensively discussed in the literature~\cite{Kirkpatrick_1972,Evangelou_1992,Bauer2001,golinelli2003statistics,Kuhn_2008,Rogers_2008,Tapias2023}.
Note that several other peaks can emerge in the bulk of the spectrum in the case of constant or bimodal hopping amplitudes $h_{ij} = \pm t$, 
as we discuss in App.~\ref{app:delta_peaks}. However, the random nature of the hopping amplitudes smears these peaks out for non-trivial choices of $\pi(h)$. 
In contrast, the peak at $E=0$ persists also in the presence of disorder in the hoppings. 
This peak corresponds to an extensive number (\ie~scaling with the system size) of degenerate eigenstates of ${\cal H}$ of zero energy, exponentially localized on the leaves of the graph (\ie~nodes that have a single connection). 
Indeed,
starting from a leaf, one can explicitly construct an eigenfunction of ${\cal H}$ of zero energy with suitably-designed exponentially-decaying coefficients. A direct inspection of the spatial structure of the eigenvectors at $E=0$ by exact diagonalization confirms that they are linear combinations of these eigenfunctions. Since the  average number of leaves is $N_L = N p /e^p$, these states disappear in the large-connectivity limit, along with the $\delta$-peak at zero energy. 
The degeneracy of these states is due to the fact that (almost all) the diagonal elements $H_{ii}$ are equal to zero in our model. 
The addition of a random on-site potential should remove the degeneracy and broaden the $\delta$-peak throughout the energy band.


As noted in Ref.~\cite{Kuhn_2008}, the $\delta$-peak at $E=0$ is accompanied by a power-law singularity. 
The statistics of the eigenvectors contributing to this divergence has been analyzed
in Ref.~\cite{Tapias2023}, 
revealing that these eigenvectors are typically localized via a mechanism distinct from standard AL.
While this observation is intriguing, our focus in the following is not on the vicinity of $E=0$. We defer the investigation of this region and its peculiar properties, and their potential relationship with our findings, to future studies.

In Fig.~\ref{fig:DoS} we  compare the results obtained by solving the cavity equations (orange curve) with exact diagonalization (ED, green circles). Although exact diagonalization is
limited to much smaller sizes, we already find excellent agreement with the cavity calculation, which implies that the average DoS is not affected by strong finite-size effects, at least in the bulk of the spectrum.


\subsection{Analysis of the distribution of the LDoS}
\label{sec:LDoS}

Next, still at fixed average 
degree 
$p=2.4$, we study how the properties of  the probability distributions of the LDoS vary upon changing $E$ across the energy band. We find three distinct behaviors, illustrated in
Fig.~\ref{fig:Prho}
and described in the following. 

\begin{itemize}
    \item {\bf Anderson-localized states} (Fig.~\ref{fig:Prho}(a)): at large enough energy, $P(\rho)$ exhibits all the features that correspond to AL, as explained in Sec.~\ref{sec:detecting}. Specifically, $P(\rho)$ has a singular behavior in the $\eta \to 0$ limit (\ie~it does not converge to an $\eta$-independent distribution), with power-law tails $\rho^{-3/2}$ at large $\rho$,
    and a cut-off at $\rho \sim \eta^{-1}$. The prefactor of the tails scales as $\sqrt{\eta}$ to ensure that the average DoS is of $\mathcal O(1)$. Small moments of the LDoS (as well as the normalization integral) are dominated by the region where $\rho$ is of 
    $\mathcal{O}(\eta)$,
    while large moments (for $q\ge 1/2$) are dominated by the tails. Anderson-localized eigenvectors at large energy in the WER ensemble are exponentially localized around pairs of neighboring vertices that are connected by anomalously strong matrix elements,
    originating from 
    the tails of the distribution of $\pi(h)$, with $h_{ij}/\sqrt{p} \sim \pm E$.
    \item {\bf Delocalized 
    multifractal states} (Fig.~\ref{fig:Prho}(b)): 
    at intermediate energies, $P(\rho)$ exhibits the peculiar features corresponding to extended but multifractal eigenstates described above in Sec.~\ref{sec:detecting}. The bulk of $P(\rho)$ converges to an $\eta$-independent function for $\eta \to 0$. This function is characterized by an algebraic dependence 
    as $\rho^{\beta-2}$
    at small $\rho$,
    which crosses over for $\rho \sim {\cal O}(1)$ to another power-law decay of the form $\rho^{-(1+\beta)}$ at large $\rho$, with an exponent $\beta(E,p)>1$ (and an $\eta$-independent prefactor).
    These power-law tails are cut-off at $\rho = {\cal O}(\eta^{-1})$. As explained 
    in Sec.~\ref{sec:cavity}, larger and larger values of $\rho_i$ found upon decreasing $\eta$ produce smaller and smaller values of the LDoS on the neighboring nodes, yielding a residual dependence on the regulator in the leftmost part of the distribution. 
    In particular, this is related to the symmetry~\cite{Mirlin_2006,Mirlin_2000}
    \begin{eqnarray} \label{eq:symmetry}
        P(\rho) = \rho^{-3} P(\rho^{-1}),
    \end{eqnarray}
    which is nicely satisfied by our distribution function in the whole interval $\rho \in [\eta,\eta^{-1}]$.
    Small enough moments $\avg{\rho^q}$ (for 
    $q<\beta$, 
    including the average DoS) are dominated by the bulk of the distribution and are $\eta$-independent,
    while large moments are dominated by the tails (see Eq.~\eqref{eq:rhoqI}). The exponent $\beta$ increases with the average degree, implying that the system 
    progressively recovers a standard fully-\rev{extended} 
    behavior 
    as $p$ is increased. 
    \item {\bf Degenerate localized eigenstates at $E=0$} (Fig.~\ref{fig:Prho}(c)): the distributions of the LDoS at $E=0$ resemble those of Anderson-localized states shown in 
    Fig.~\ref{fig:Prho}(a),
    with a strong singular behavior in the $\eta \to 0$ limit. Yet, the power-law tails are characterized by a much smaller exponent (slightly less than $1$), and a prefactor scaling differently with $\eta$. This corresponds to the fact that the average DoS does not have a finite limit for $\eta=0$, 
    \revv{i.e.~(see Eq.~\eqref{eq:Prho})}
    \begin{equation}
        \rho_{\rm av} (0)=\int^{\eta^{-1}} \de \rho\, \rho \,P(\rho; E=0) \xrightarrow[]{\eta \to 0} \infty,
    \end{equation}
    which is a manifestation of the $\delta$-peak 
    in Fig.~\ref{fig:DoS}.
\end{itemize}

Figure~\ref{fig:Prho} also shows that the probability distributions found by solving the self-consistent cavity equations provide an excellent approximation of those obtained from exact diagonalization, already at moderately large sizes. (We observe that the finite-size corrections to the cavity results are slightly stronger in the delocalized multifractal phase, but they are already quite small for $N = 2^{14}$, at least at $E=0.4$, far from the localization transition.)

\begin{figure*}
\includegraphics[width=0.336\textwidth]{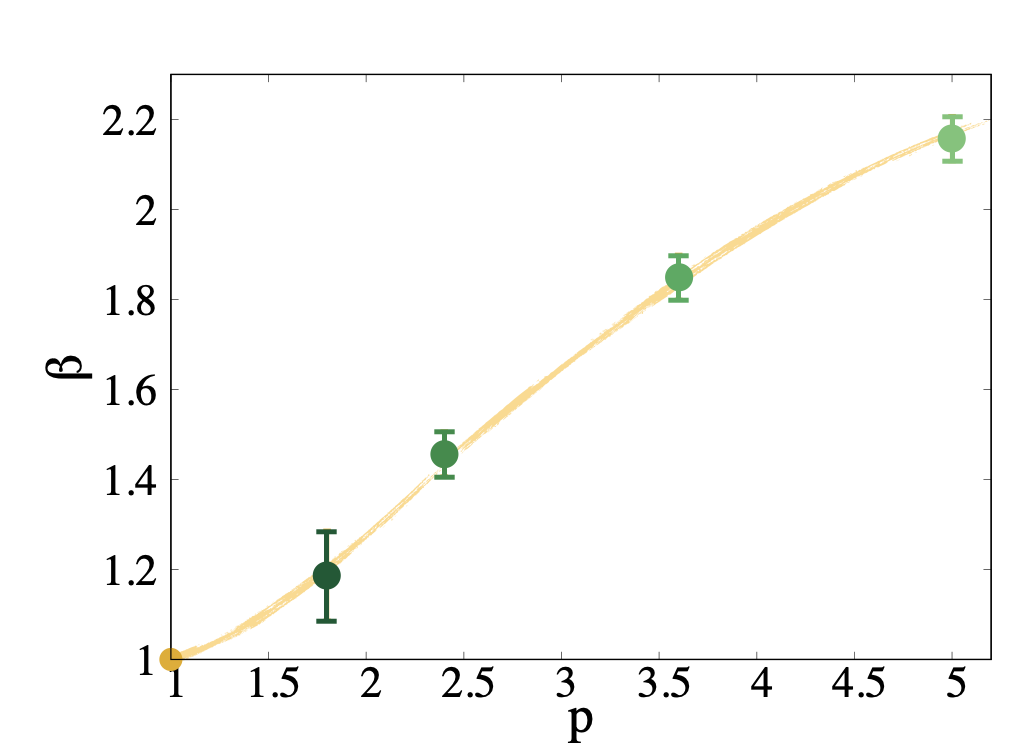} 
\put(-20,20){(a)}
\hspace{-0.31cm} \includegraphics[width=0.338\textwidth]{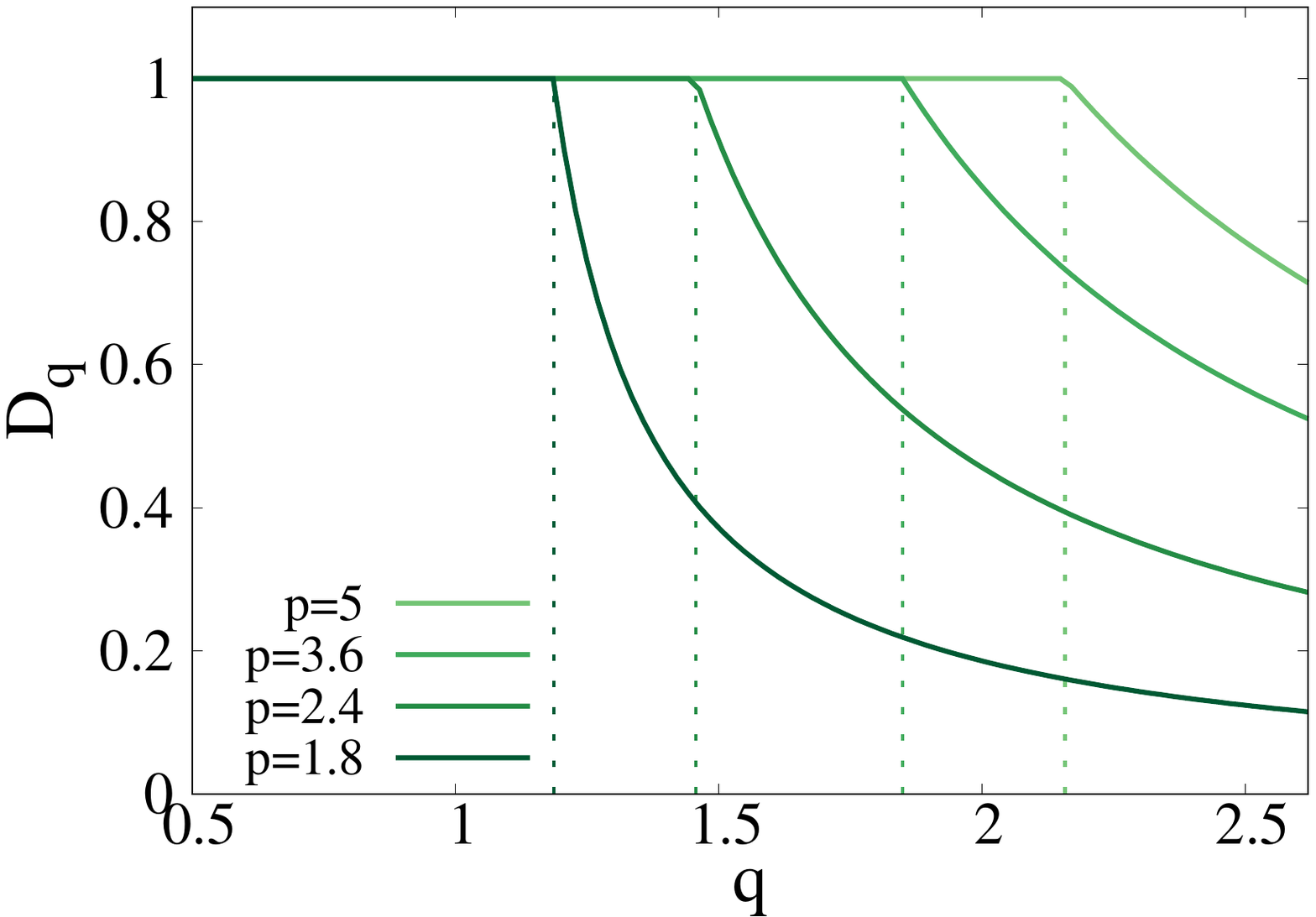} 
\put(-21,102){(b)}
\hspace{-0.36cm} 
\includegraphics[width=0.338\textwidth]{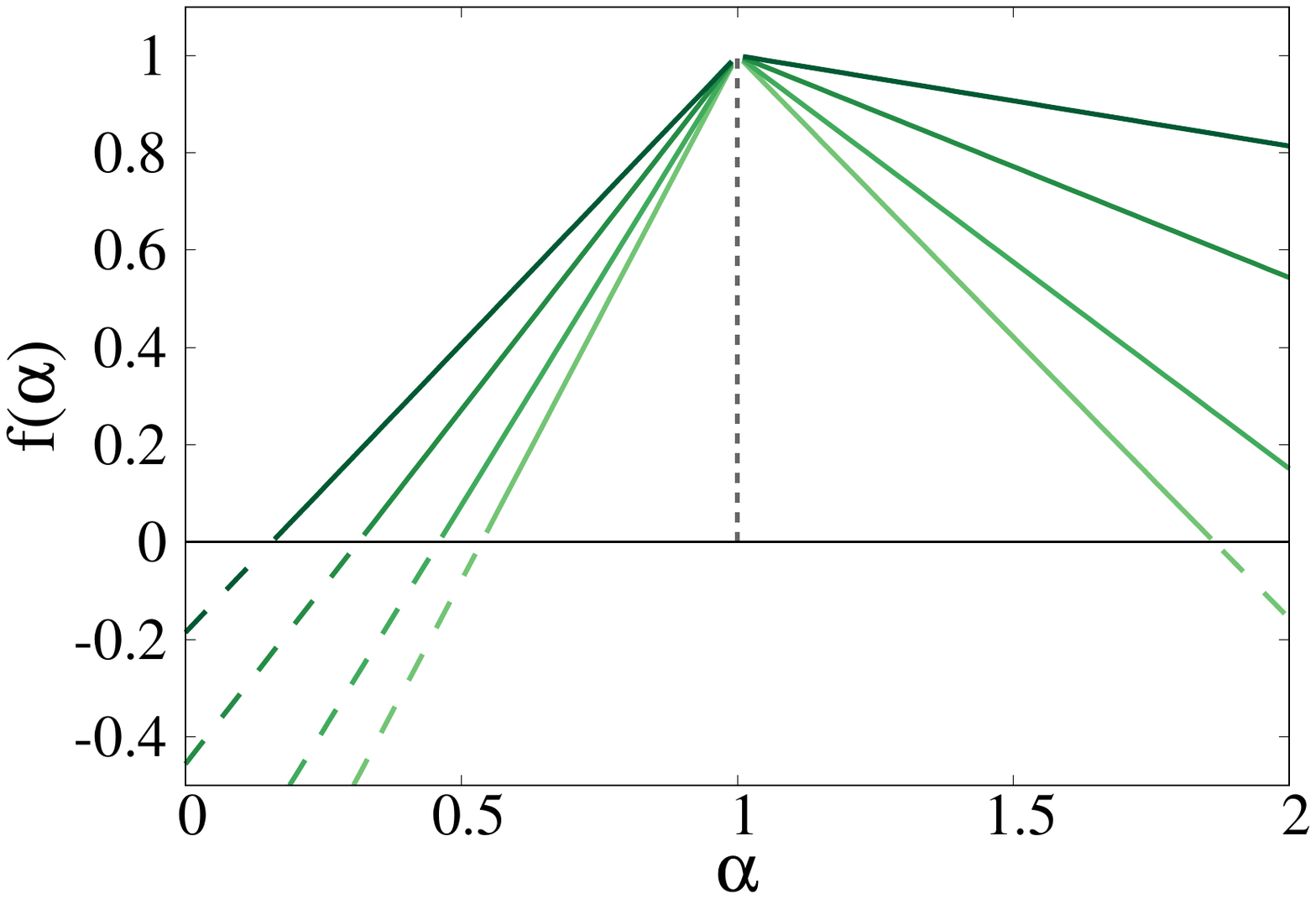}
\put(-20,20){(c)}
\caption{(a) Exponent $\beta$ obtained by fitting the tails of the probability distributions $P(\rho)$ with a power-law for $E=0.4$, and several values of the average degree $p$. The yellow line is a guide to the eye.
(b) Fractal dimensions $D_q$ obtained from Eq.~\eqref{eq:Dqbeta} for several values of $p$ (and $E=0.4$). The vertical dotted lines represent the breaking point $q=\beta$. (c) Spectrum of fractal dimensions $f(\alpha)$, see Eq.~\eqref{eq:falpha}, for the same values of $E$ and $p$ as in panels (a)-(b). 
\label{fig:Dq}}
\end{figure*}

Performing this analysis for several values of 
$p$ and $E$
we obtain the phase diagram reported in Fig.~\ref{fig:PD}(a). The yellow curve corresponds to the so-called {\it mobility edge} $E_{\rm loc} (p)$~\cite{abou1974self,biroli2010anderson}, separating the region of extended (although \rev{multifractal}) 
states in the middle of the spectrum from the Anderson-localized states closer to the band edges. The transition points are obtained by direct inspection of $P(\rho)$ upon varying the regulator $\eta$, and varying the energy $E$ on a small grid at fixed $p$. The mobility edge tends
to $2$ for $p \to \infty$. Concomitantly, in this limit the DoS approaches the Wigner semicircle for $E \in [-2,2]$, implying that the Anderson-localized states gradually disappear for large $p$. (However, as discussed in Refs.~\cite{Semerjian_2002,Alt2021,Alt2021_2,Alt2022,Alt2023,tarzia2022fully}, a subextensive number of AL states persist in the Lifshitz tails of the spectrum, associated with rare nodes with anomalously large 
\rev{degree}.)
The thick vertical line at $E=0$ signals the presence of degenerate eigenstates localized on the leaves. The horizontal dashed line corresponds to the percolation transition, such that for $p<1$ the graph breaks 
into many finite clusters, while for $p>1$ a single giant connected component exists with probability approaching one at large $N$. For $p \to 1^+$ the connected component is a one-dimensional chain with Gaussian random hoppings, and all eigenstates are Anderson localized~\cite{mott1961theory}.

As explained below, the multifractality of the eigenstates in the intermediate region progressively 
weakens
upon increasing $p$ 
(see Fig.~\ref{fig:Dq} below), and the standard fully-\rev{extended} 
behavior described by the GOE universality class 
is recovered in the large-connectivity limit, in agreement with the conjecture of Ref.~\cite{mirlin1991localization}.

\subsection{Fractal dimensions of the delocalized multifractal states}
As explained in Sec.~\ref{sec:detecting}, the statistics of the wavefunctions' coefficients is 
encoded in the distributions of the LDoS. In particular, the fractal dimensions $D_q$ are directly related, through Eq.~\eqref{eq:Dqbeta}, to the exponent $\beta$ describing the power-law decay of the tails of $P(\rho)$. We find that, while $\beta$ does not vary significantly with $E$ within our numerical accuracy, it strongly depends on the average degree $p$. The values of $\beta$ extracted from the fits of the tails of the distributions of the LDoS for fixed energy $E=0.4$ and for several values of $p$ are shown in Fig.~\ref{fig:Dq}(a), indicating that $\beta$ grows upon increasing $p$. In Fig.~\ref{fig:Dq}(b)
we show our prediction for the fractal dimensions $D_q$, obtained using Eq.~\eqref{eq:Dqbeta}. The plot clearly illustrates the mechanism through which for $p \to \infty$ the system reverts to standard fully-delocalized behavior, with $D_q$ equaling 1 for all values of $q$. This behavior corresponds to a specific kind of multifractality, in which small enough moments of the wavefunction's coefficients behave as in the standard metallic 
phase, while a non-trivial multifractal behavior emerges at large $q$ due to the presence of a sub-extensive fraction of nodes on which the amplitudes are anomalously large. This means, in particular, that the support set of the eigenvectors contains a finite fraction of the total number of sites, at odds with other forms of multifractality (see \eg~Refs.~\cite{rodriguez2011multifractal,Kravtsov_2015,vonSoosten_2019,Facoetti_2016,Truong_2016,Bogomolny_2018,DeTomasi_2019,amini2017spread,pino2019ergodic,berkovits2020super,us_2023,kravtsov2020localization,khaymovich2020fragile,monthus2017multifractality,biroli2021levy,buijsman2022circular,khaymovich2021dynamical}).

An alternative characterization of the multifractality of the wavefunctions' amplitudes, which is
commonly used in the literature, is provided by the multifractal spectrum $f(\alpha)$ (see \eg~Refs.~\cite{Kutlin_2024,Mirlin_2006}).
It is defined so that 
the number of nodes having wavefunctions' amplitudes scaling as $N^{-\alpha}$ behaves as $N^{f(\alpha)}$. 
In the large-$N$ limit, the anomalous dimensions $D_q$ are related to $f(\alpha)$ via a Legendre transformation, as detailed in App.~\ref{app:multifractal}. Since inverting such transformation is delicate in our case, in App.~\ref{app:multifractal} we rather compute $f(\alpha)$ directly, starting from its definition. The result reads
\begin{equation} \label{eq:falpha}
    f(\alpha) = \begin{dcases}
        \alpha(1-\beta)+\beta, & \textrm{for} \; \; 
        \alpha > 1
        ,\\
        1  - \beta (1 - \alpha) \, , & \textrm{for} \; \; 
        0\leq \alpha \leq 1,
    \end{dcases} 
\end{equation}
which is plotted in Fig.~\ref{fig:Dq}(c) for several values of $p$.
This provides another illustration of how the system recovers a fully-\rev{extended} 
behavior (which corresponds to
$\beta \to \infty$)
upon increasing $p$.
Moreover, it is simple to show that $f(\alpha)$ given in Eq.~\eqref{eq:falpha} verifies the exact symmetry relation~\cite{Mirlin_2000,Fyodorov2004,Mirlin_2006}
\begin{equation}
    f(\alpha+1) = f(1-\alpha)+\alpha.
\end{equation}

\begin{figure*}
\includegraphics[width=0.42\textwidth]{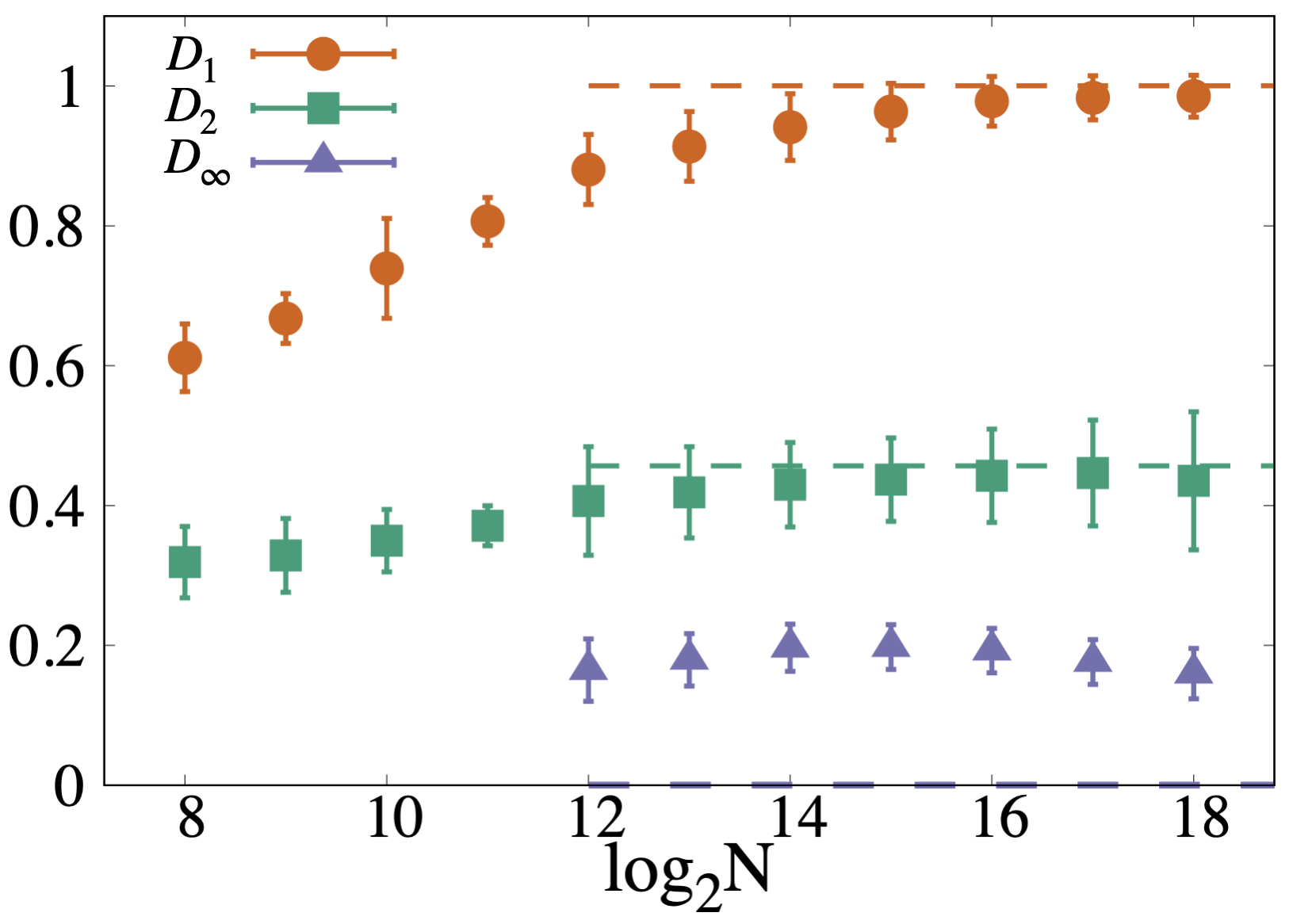} 
\put(-188,29){(a)}
\hspace{0.5cm} 
\includegraphics[width=0.43\textwidth]{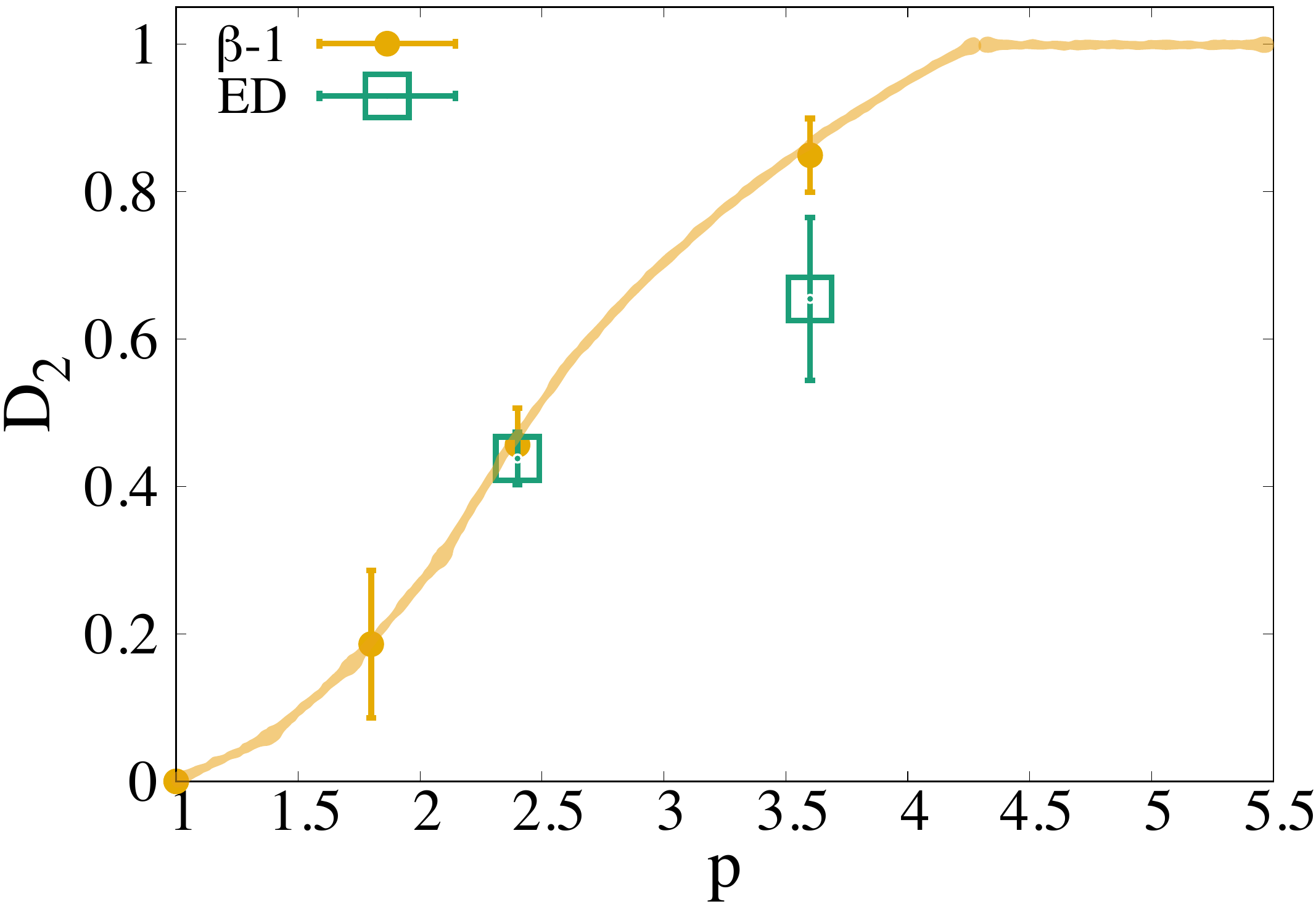} 
\put(-23,24){(b)}
\caption{(a) Comparison between the theoretical predictions obtained within the cavity approximation at large $N$ (dashed lines), and the exact diagonalization results for 
$D_1$, $D_2$, and $D_\infty$,
upon increasing the system size $N=2^n$ with $E=0.4$ and $p=2.4$, showing a good agreement. (b) Comparison between the theoretical prediction in Eq.~\eqref{eq:Dqbeta} for the fractal dimension $D_2$
 (filled circles, corresponding to $D_2 = 1 - \beta$ if $\beta<2$, and $D_2=1$ if $\beta>2$, while the yellow line is a guide to the eye)
 and the exact diagonalization results for large sizes (empty squares), for $E=0.4$. The values of $\beta$ are obtained from the power-law fits of the tails of $P(\rho)$, and are reported in Fig.~\ref{fig:Dq}(a). The exact diagonalization values are obtained by averaging the estimates of $D_2$ for the three largest available sizes.
\label{fig:Dq_ED}}
\end{figure*}

We have compared the theoretical predictions obtained within the cavity approximation for large $N$ with the results of exact diagonalization of several instances of the WER, for $p=2.4$ and $E=0.4$. We have measured the flowing fractal exponents $D_1$, $D_2$, and $D_\infty$ by performing (discrete) logarithmic derivatives of the corresponding generalized IPR (see Eqs.~\eqref{eq:Dq}--\eqref{eq:Dinf}). The results of this procedure are plotted in
Fig.~\ref{fig:Dq_ED}(a). According to the cavity calculations, $D_1$ should tend to $1$ at large $N$, $D_2$ should approach $\beta-1 \approx 0.44$, and $D_\infty$ should tend to zero. The plot shows that our ED data converge very well towards these predictions for $D_1$ and $D_2$. The numerical value of $D_\infty$ for the accessible sizes is still quite above zero, yet a slower convergence for $D_\infty$ is somewhat expected (as also empirically observed in other similar models~\cite{biroli2021levy}), since it is related to the scaling with
the system size of a \textit{single} coefficient of each eigenstate.

In Fig.~\ref{fig:Dq_ED}(b) we compare the value of $D_2$ obtained from exact diagonalizations 
(for the largest available sizes and for $E=0.4$, with $p=2.4$ or $p=3.6$)
with the theoretical prediction given in Eq.~\eqref{eq:Dqbeta}, \ie~$D_2 = \beta-1$ if $\beta<2$, and $D_2=1$ if $\beta>2$. The plot demonstrates a reasonably good agreement, within our numerical accuracy. However, it also distinctly shows that for excessively large values of the average degree, \ie~$p \gtrsim 4.2$, the fractal dimension $D_2$ 
converges
to one. This implies that the singularity of the spectrum of the wavefunction's amplitudes only manifests in even larger moments, as the system progressively becomes
\rev{less multifractal} 
with increasing $p$.
Note that the typical value of $D_2$ given by the geometric mean of the IPR is larger than the one obtained from the arithmetic average (as defined in Eq.~\eqref{eq:Iq}, and shown in Fig.~\ref{fig:Dq_ED}). This implies that the IPR is broadly distributed, and that the typical wavefunctions are more delocalized than the typical ones that dominate the average of $D_2$.

We have also studied the spatial decay of the correlations 
\begin{eqnarray}
    C(r) =  N \avg{\abs{ \psi_\alpha (i) }^2 \abs{\psi_\alpha (i+r) }^2} - N^{-1}
\end{eqnarray}
between two points at distance $r$ on the graph, which are related to the two-point connected correlation functions of the LDoS $\avg{\rho_i \rho_{i+r}}_c$. Our analysis, reported in detail in App.~\ref{app:two-point}, reveals in particular that $C(r)$ multiplied by the number of nodes $\Omega(r)$ at distance $r$ from a given node decays algebraically in the multifractal extended phase (see Fig.~\ref{fig:2P}(a)), as expected for delocalized eigenfunctions, but with a distinct $N$ dependence in the prefactor with respect to the standard metallic regime~\cite{tikhonov2019statistics}. Conversely, in the AL phase $C(r) \Omega(r)$ decays exponentially as $e^{-r / \xi_{\rm loc}}$ (see Fig.~\ref{fig:2P}(b)), $\xi_{\rm loc}$ being the localization length.

\section{Understanding the origin of multifractal extended states} \label{sec:mechanism}

As explained in Sec.~\ref{sec:detecting}, the presence of multifractal extended states manifests itself in the fact that at least some part of the probability distribution $P(\rho)$ of the LDoS has a singular behavior in the $\eta \to 0$ limit, and conserves a dependence on the imaginary regulator. 
Inspecting
the structure of the cavity self-consistent equations~\eqref{eq:cavity} one immediately realizes that, for the leaves of the graph (\ie~nodes that are connected to the bulk via a single edge), the cavity Green's function
(computed by removing their unique neighbour)
is simply given by
\begin{equation}
    G_{i \to j} (z) = 
    -(E+ {\rm i} \eta)^{-1} \, .
\end{equation}
For $E \approx 0$ one has $\im G_{i \to j} (z) \approx 1/\eta$, while $\im G_{i \to j} (z) \approx \eta/E^2$ for $E>0$. 
Such $G_{i \to j} (z)$ enters the recursion relation for $G_{j \to k} (z)$ on the neighboring node $j$ (with one of its other neighbors $k\neq i$ removed), thus producing an $\eta$-dependence of the cavity Green's functions that propagates through the graph. 
In particular, values of $\im G_{i \to j}$ of $\mathcal O(\eta)$ can generate very large $\mathcal O(1/\eta)$ values of the imaginary part of the cavity Green's functions on the neighboring nodes --- provided that the real part in the denominator of Eq.~\eqref{eq:ImGii} is also very small (\textit{resonance condition}~\cite{abou1974self}).
This results
in a peak of $P(\rho)$ for $\rho \sim \eta^{-1}$ at $E=0$, as in 
Fig.~\ref{fig:Prho}(c). Conversely, values of $\im G_{i \to j}$ of $\mathcal O(1/\eta)$ produce very small imaginary parts of the cavity Green's functions on the neighboring nodes --- resulting in a peak of $P(\rho)$ for $\rho \sim \eta$ at $E>0$, as in 
Fig.~\ref{fig:Prho}(a--b). These features are related to the symmetry~\eqref{eq:symmetry} satisfied by the probability distribution of the LDoS. 

The behavior of the Green's function on the leaves of the graph is in fact intimately related to the existence of  eigenstates of ${\cal H}$ exponentially localized around the leaves, associated to the $\delta$-peak at $E=0$ of the average spectral density. 
Since the number of leaves $N_L = N p /e^p$ is extensive, 
one necessarily finds
a residual $\eta$-dependence on a finite fraction of the nodes that are sufficiently close to a leave, 
which causes
a singular behavior of $P(\rho)$. (This phenomenon is well known in the case of the Anderson model on loop-less Cayley trees, and produces a genuinely multifractal phase in the whole delocalized regime~\cite{sonner2017multifractality,monthus2011anderson,biroli2020anomalous,nosov2022statistics}.)

Therefore, one could (naively) 
expect
that removing all  leaves from the graph should restore a standard, fully-\rev{delocalized} 
behavior. Yet, this is not the case. In particular, we proceeded to eliminate all the leaves by performing a rewiring of the connections of the giant connected component of the graph
as follows: (i)~for each leaf of the graph, labeled $i$, we pick at random another node, 
labeled $j$, with 
\rev{degree}
strictly larger than 1; (ii)~we pick at random one of the neighbors of $j$, labeled  $k$, and check that it has a degree strictly larger than $2$; (iii)~we add a new edge between $i$ and $j$ with hopping amplitude equal to $h_{jk}/\sqrt{p}$, and we eliminate the edge between $j$ and $k$. 
By repeating this 
procedure
for all the leaves, 
we obtain
a graph in which every vertex has 
\rev{a degree}
strictly larger than one, while conserving the total number of edges (and hence the average effective degree $\tilde{p}$). Of course, this algorithm necessarily produces a small modification of the degree distribution. 
We then solved again
the self-consistent recursion relations for the cavity Green's functions, Eqs.~\eqref{eq:cavity} and~\eqref{eq:green}, 
on the new \textit{rewired} graph,
and computed the corresponding distribution of the LDoS. 
Surprisingly,
for values of $p$ and $E$ within the multifractal extended phase, we
found that $P(\rho)$ is only weakly modified compared to the distribution obtained for the original graph (containing an extensive number of leaves), and conserves all the singular features corresponding to delocalized but \rev{multifractal} 
eigenstates. Concomitantly, one still finds a singular behavior of $P(\rho)$ at $E=0$, accompanied by a $\delta$-peak of the average DoS, 
which corresponds to an extensive number of degenerate eigenstates of zero energy.

\begin{figure}
\includegraphics[width=0.482\textwidth]{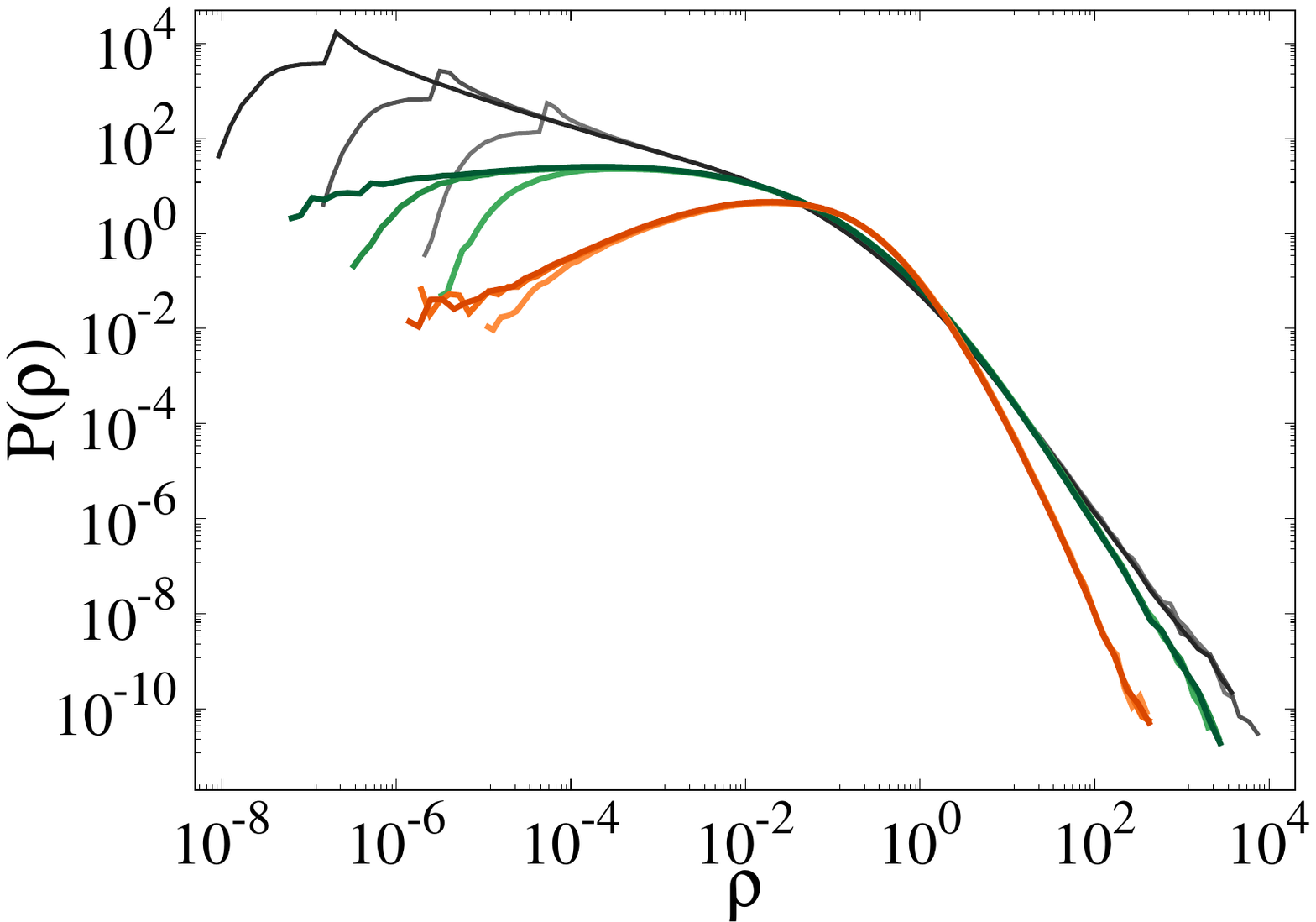}
\put(-195,28){\includegraphics[scale=0.11]{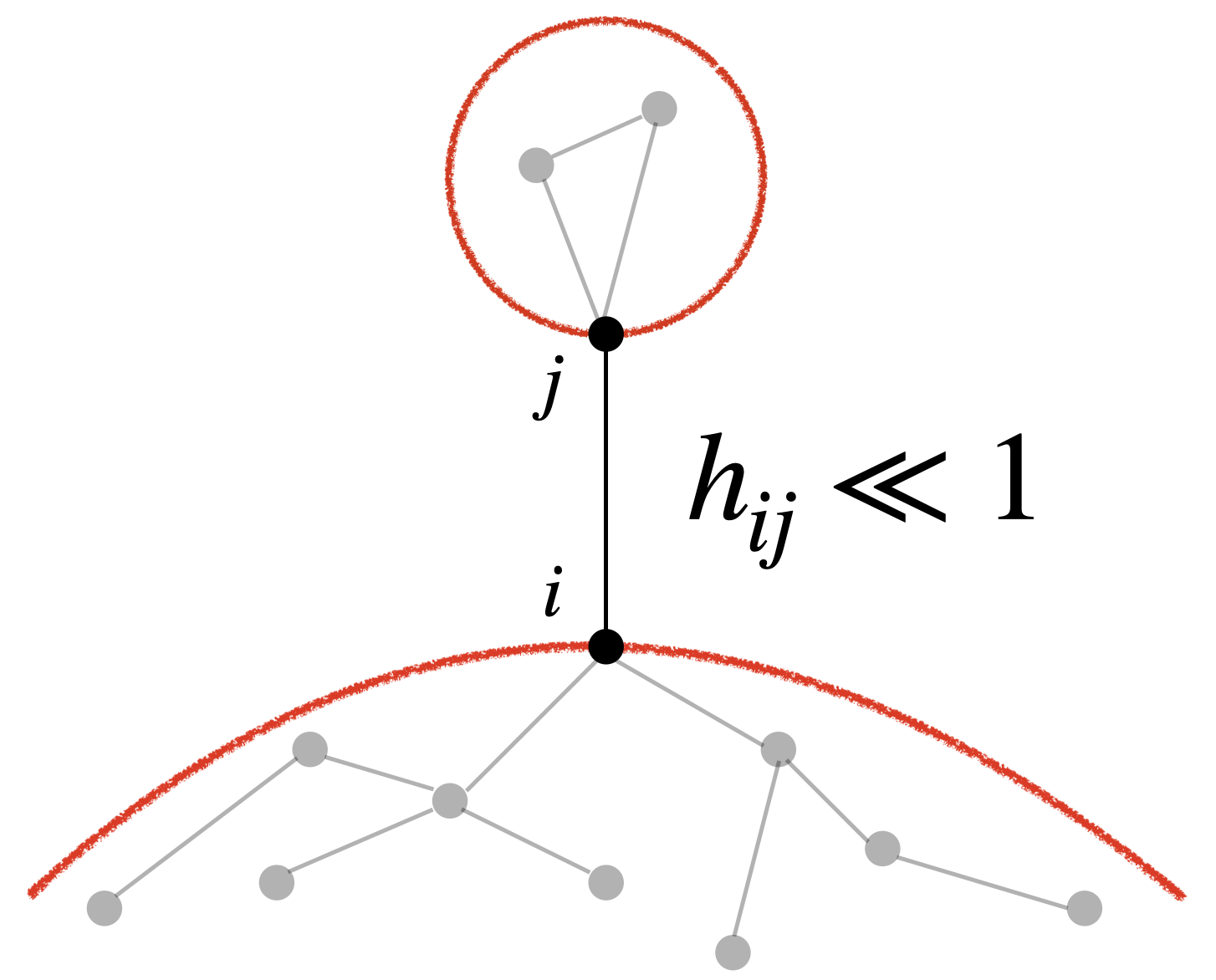}}
 \caption{Probability distribution of the LDoS (for $p=2.4$, $E=
0.4$, and several values of the regulator $\eta$) of the modified model in which all the nodes with degree one have been rewired, and small hopping amplitudes below a threshold $\nu$ have been suppressed (see Eq.~\eqref{eq:Phijnu}). Different colors correspond to different values of the threshold $\nu$: 
$\nu=0$ (gray, \ie~the original model),
$\nu=0.1$ (green), and $\nu=0.3$ (orange). 
The curves are obtained by solving the self-consistent cavity equations for systems of size $N=2^{20}$. The regulator varies from $\eta=2^{-23}$ to $\eta=2^{-15}$ (for each $\nu$, darker curves correspond to smaller $\eta$). 
Notably, the singular 
features of $P(\rho)$, namely the $\eta$-dependence and the power-law tails, disappear for $\nu \approx 0.3$, for which a 
standard metallic behavior (similar to the one of the Anderson model on the Bethe lattice below the critical disorder) is recovered.
Inset: schematic representation of a \textit{hidden} isolated cluster, showing the effective fragmentation of the graph (see the main text). 
\label{fig:PrhoERGO}
} 
\end{figure}

This implies that rewiring the leaves is not enough to eliminate the exotic features of the spectral statistics of the model, and to restore a standard metallic behavior 
of the eigenstates. 
The reason is that, in fact, small values of the hopping amplitudes $h_{ij}$ generate a greater number of effective or ``hidden'' leaves, as well as ``hidden'' effectively isolated clusters (see Fig.~\ref{fig:PD}(b)), than one might naively estimate by solely examining the degrees of the nodes.
These structures are still present even after all the ``true'' leaves have been removed. 
To see this, let us consider a situation such as the one depicted in 
the inset of Fig.~\ref{fig:PrhoERGO}.
Let us assume that the hopping amplitude $h_{ij}$ 
is very small and can be treated perturbatively. In the absence of hopping, one has a finite number of states localized on the small cluster. In this situation, a too small hopping 
might not 
effectively
hybridize these states with those delocalized on the bulk of the graph. The presence of many structures of this kind may lead to the emergence of \rev{anomalously delocalized} 
eigenstates.

To check these ideas we have implemented another modification of the graph: after rewiring all the nodes with degree $1$ using the algorithm described above, on each edge $(i \leftrightarrow j)$ we extract again the hopping amplitudes $h_{ij}$ from a Gaussian distribution, but
this time requiring
$|h_{ij}| \ge \nu$ (\ie~we repeat the extraction until $|h_{ij}|$ is larger than the threshold $\nu$). 
This procedure essentially consists in modifying the probability distribution of the hoppings as
\begin{equation} \label{eq:Phijnu}
    \pi_\nu (h_{ij}) = {\cal N}_\nu \, e^{- h_{ij}^2} \theta(|h_{ij}| - \nu) \, ,
\end{equation}
with a $\nu$-dependent normalization factor ${\cal N}_\nu$. As shown in Fig.~\ref{fig:PrhoERGO} for $p=2.4$ and $E=0.4$, upon increasing the threshold $\nu$ the probability distribution of the LDoS progressively loses its singular 
features: the power-law tail at large values of $\rho$ and the $\eta$-dependence both disappear, corresponding to a 
situation in which \rev{a standard metallic behavior with fully-extended eigenstates} 
is restored. 
Concomitantly, the singular behavior of $P(\rho)$ at zero energy and the $\delta$-peak in the average DoS disappear 
(the peak
is in fact replaced by a cusp), signaling the disappearance of the degenerate localized states at $E=0$. We reasonably expect that the larger the average degree $p$, the smaller the threshold $\nu$ needed to recover a standard 
delocalized
phase. 

Finally, we have also checked that upon removing the small hopping amplitudes, but not rewiring the leaves, one still finds a residual singular behavior in $P(\rho)$, 
albeit with strongly reduced \rev{multifractal} 
features.

We thus conclude that the physical origin of the extended multifractal states in this model 
stems from
the strong spatial heterogeneities of the graph, 
and from
the interplay between the fluctuations of the degree and of the hopping amplitudes. Particularly for small $p$, the fluctuations of the degree generate many nodes or groups of nodes connected to the giant component by a single edge (\textit{grafted trees}~\cite{Bauer2001,golinelli2003statistics}), while the fluctuations of the hopping 
lead to
an effective fragmentation of the graph, as illustrated in Fig.~\ref{fig:PD}(b). Such heterogeneity is a generic feature not only of the WER model, but also of other related sparse random matrix ensembles with fluctuating connectivity. In particular, we expect that choosing a different weight distribution $\pi(h)$ will not change the qualitative features of the phase diagram presented in Fig.~\ref{fig:PD}(a), but only possibly impact the strength of the \rev{multifractal} 
properties delineated here. Furthermore, choosing a bimodal $\pi(h)$ or constant weights $h$ (as we do in App.~\ref{app:delta_peaks}) prevents the fragmentation of the graph, but even in that case ordinary leaves remain statistically relevant. 


The simultaneous disappearance of multifractal states within the bulk of the spectrum and of the $\delta$-peak at $E=0$ in the average DoS (associated with an extensive number of localized eigenstates at zero energy) offers an additional perspective on the origin of the multifractal behavior. Indeed, this phenomenon can be elucidated as the consequence of the \textit{orthogonalization} between fully-\rev{extended} 
states, with support set on the bulk of the graph, and the finite fraction of states with $E=0$ localized on the graph's leaves. Such mechanism has been elaborated upon in Ref.~\cite{haque2022entanglement} (see also  Refs.~\cite{kutlin2021emergent,nosov2019correlation}). The relevance of this orthogonalization-mediated multifractality is further underscored by the observation that the exponent $\beta$, which describes the tails of $P(\rho)$ (see Fig.~\ref{fig:Prho}(b)) and hence the fractal dimensions, exhibits minimal dependence on the energy $E$. In the WER ensemble, the origin of degenerate localized states at $E=0$ also stems from the pronounced heterogeneity in the graph's topology. Consequently, the ``geometric'' mechanism associated with the effective fragmentation of the graph discussed above and the orthogonalization-mediated mechanism mentioned here are intricately linked, and probably represent two different facets of the same underlying phenomenon.

\section{Level statistics in the multifractal 
extended phase} \label{sec:level}

\begin{figure*}
\includegraphics[width=0.26\textwidth]{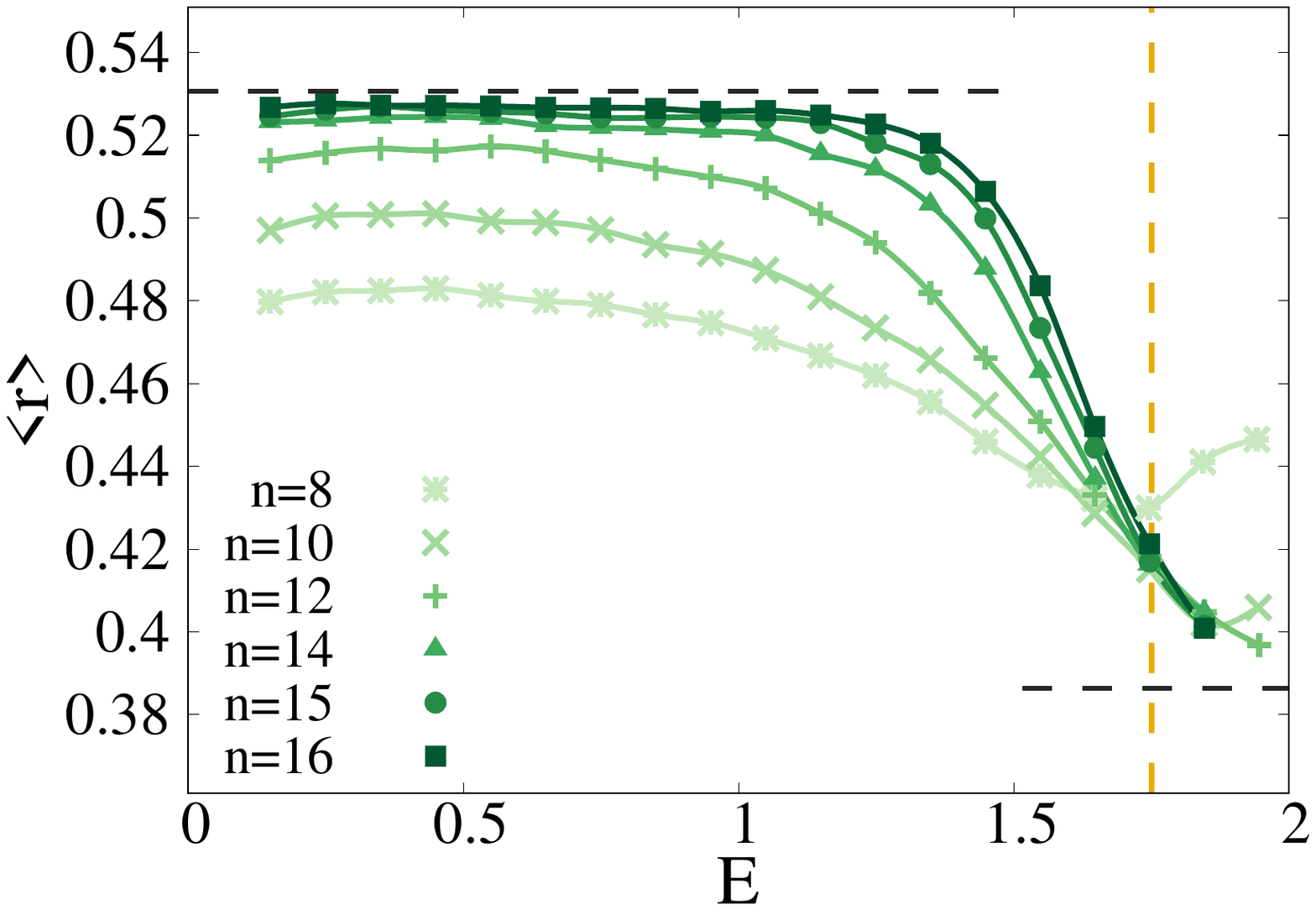} \put(-20,75){(a)} \hspace{-0.36cm} \includegraphics[width=0.26\textwidth]{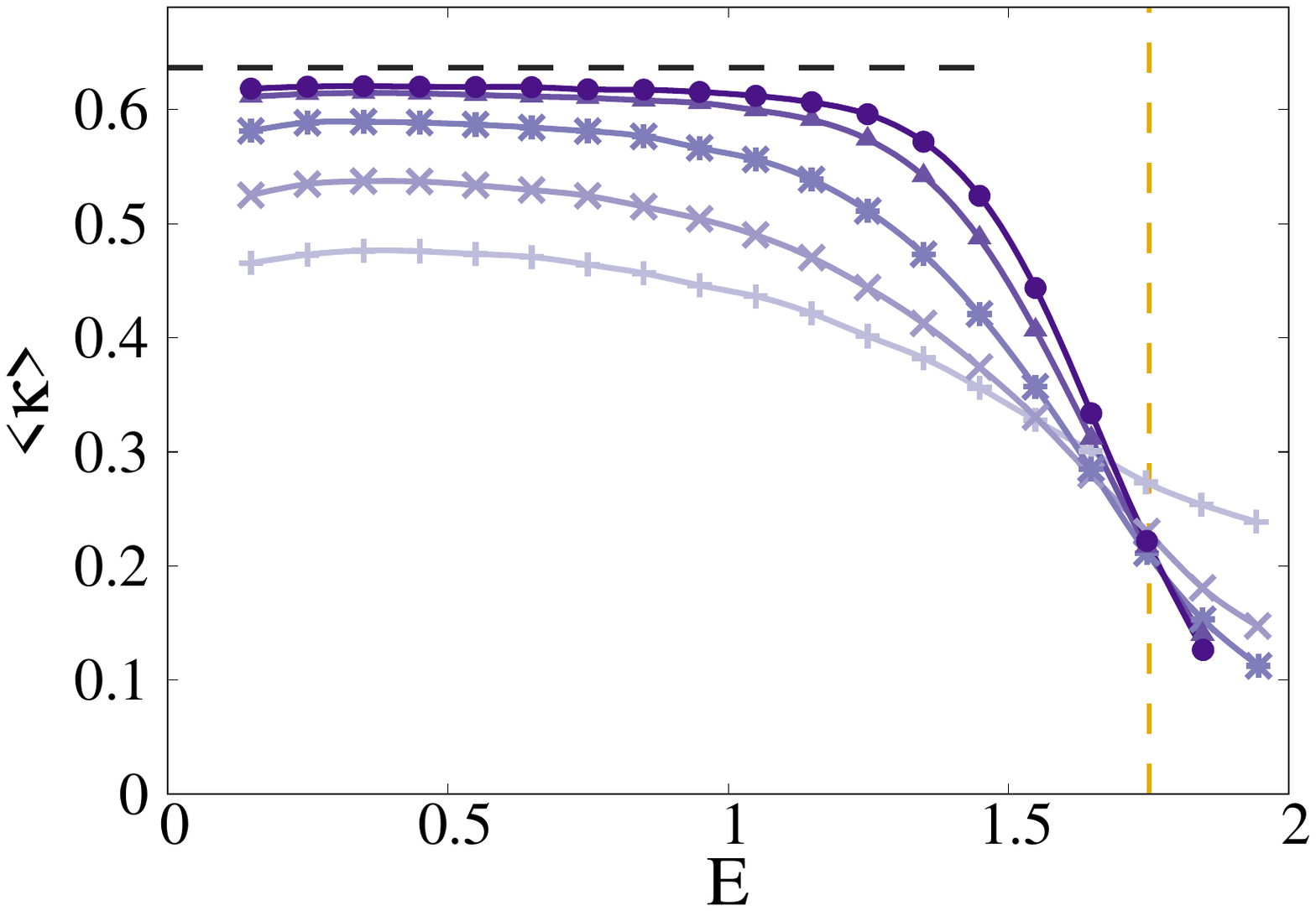} \put(-20,75){(b)} \hspace{-0.31cm} \includegraphics[width=0.26\textwidth]{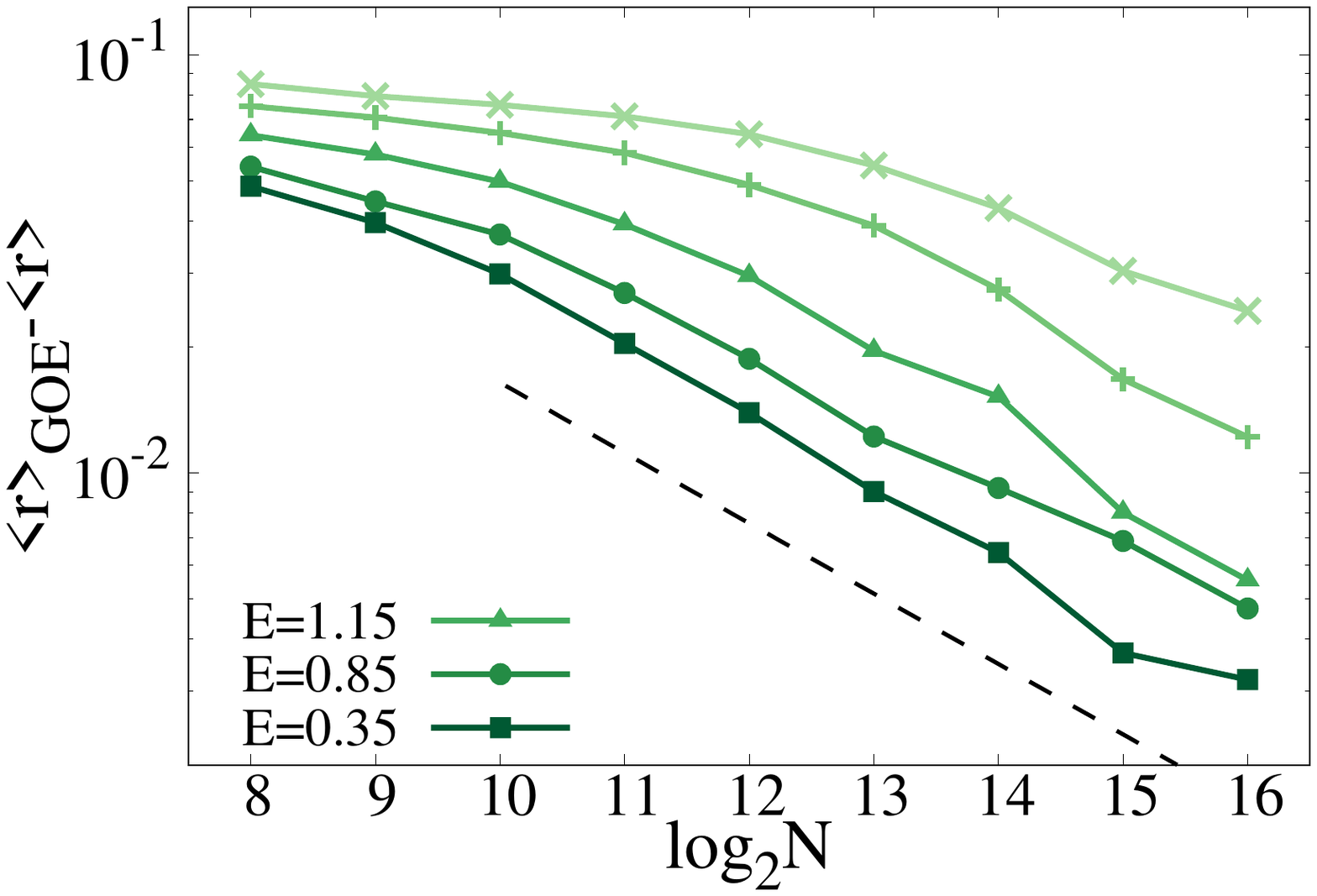} \put(-20,75){(c)} \hspace{-0.31cm} \includegraphics[width=0.26\textwidth]{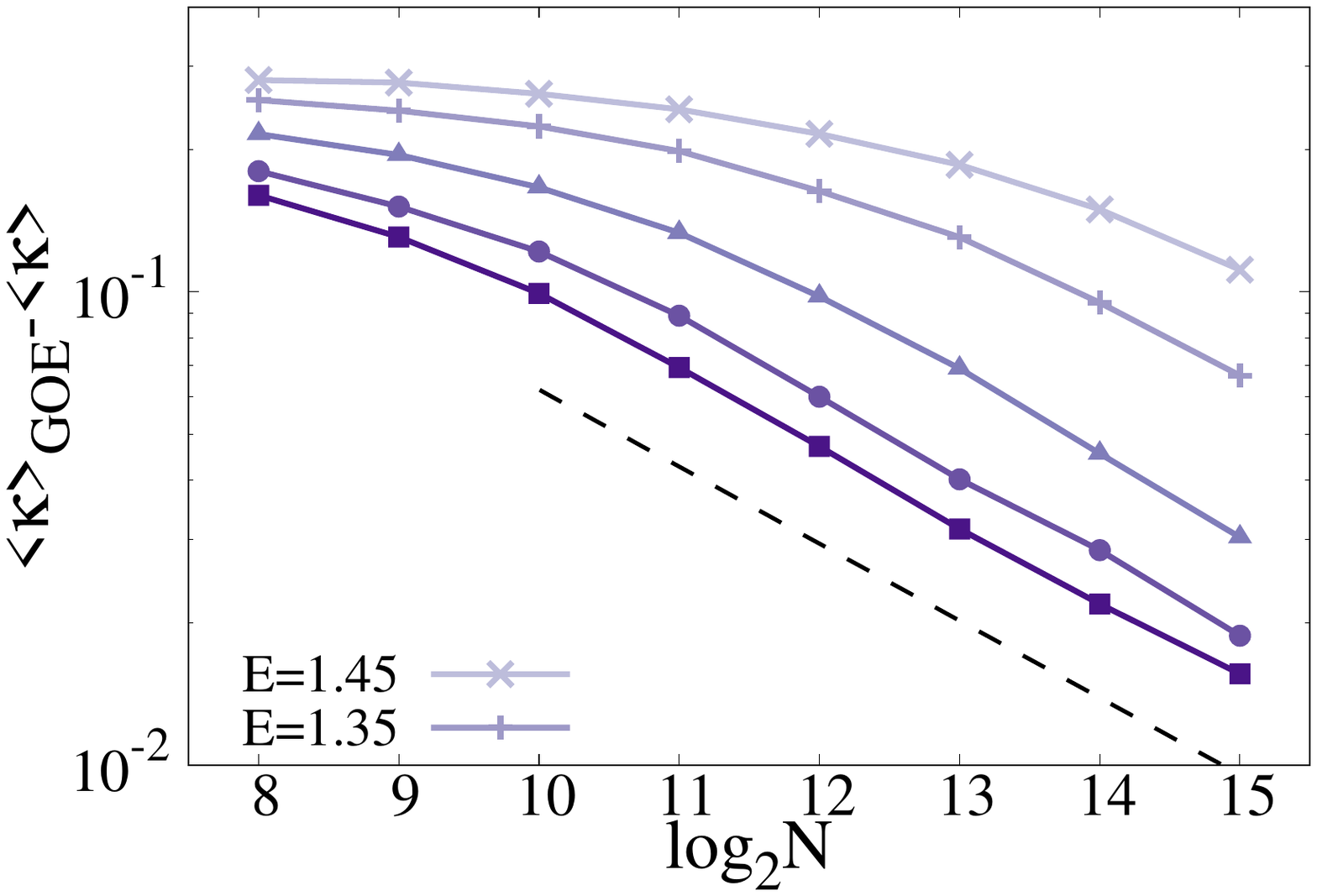} \put(-20,75){(d)}
\caption{(a) Average gap ratio and (b) average overlap between subsequent eigenvectors, see Eqs.~\eqref{eq:erre}~and~\eqref{eq:kappa}, for $p=2.4$, varying $E$ across the energy band for several system sizes $N=2^n$. The dashed horizontal gray lines show the GOE and Poisson asymptotic values, 
which are respectively roughly equal to $0.53$ and $0.39$ in (a), or to $0.64$ and $0$ in (b).
The vertical dashed yellow line corresponds to the position of the mobility edge $E_{\rm loc}$ found within the cavity calculation. Panels (c) and (d) show the convergence of $\avg{r}$ and $\avg{\kappa}$ to their GOE asymptotic limits in the delocalized
 multifractal phase, $E<E_{\rm loc}$. The values of the energies used in (c) and (d) are shared between the keys of the two panels.
\label{fig:gap_statistics}}
\end{figure*}

The exotic nature of the eigenstates also manifests itself in the statistics of the energy levels. In fact, fully-extended eigenstates exhibit level repulsion: the level statistics is universal and is described by the GOE. Conversely, two Anderson-localized eigenstates close in energy typically occupy distinct support sets, and do not overlap: hence, the energy levels are thrown as random points on a line, and obey the Poisson statistics. In the case of fractal eigenstates several intermediate situations can arise, depending on their specific spectral properties~\cite{Mirlin_2000}. 
In 
this section
we inspect the correlations of the energy levels by measuring 
the statistics of the gaps,  
the overlap correlation function, 
and the level compressibility.

\subsection{Gap statistics}
\label{sec:gap}
We start by focusing on the statistics of the gaps between subsequent levels, \ie~on the scale of the mean level spacing
\begin{equation}
    \delta =[\tilde{N} \rho_{\rm av} (E)]^{-1} \, .
    \label{eq:delta}
\end{equation}
Two commonly used diagnostics for the statistics of neighboring gaps are provided by the average gap ratio $\avg{r}$~\cite{Luitz_2015,oganesyan2007localization} and the average overlap $\avg{\kappa}$ between subsequent eigenvectors~\cite{Atas_2013,biroli2018delocalization}.
Indeed, these converge to distinct universal values depending on whether the level statistics is described by the GOE, or instead is of the Poisson type.
The average gap ratio is defined as
\begin{equation} \label{eq:erre}
\avg{r} = \avg{ \frac{\min \{ \lambda_{\alpha + 1} - \lambda_\alpha, \lambda_{\alpha+2} - \lambda_{\alpha + 1} \}}{\max \{ \lambda_{\alpha + 1} - \lambda_\alpha, \lambda_{\alpha+2} - \lambda_{\alpha + 1} \}} }\, ,
\end{equation}
where $\lambda_\alpha$ are the eigenvalues of ${\cal H}$ labeled in ascending order ($\lambda_{\alpha + 1} > \lambda_\alpha$). The average is performed over several realizations of the disorder, and within a small energy window around a certain value of the energy $E$. If the level statistics is described by the GOE (at least on the scale of the mean level spacing), then $\langle r \rangle$ converges to $\langle r \rangle_\text{GOE} \simeq 0.53$, while for independent energy levels described by Poisson statistics one has $\langle r \rangle_\text{P} \simeq 2 \ln 2 - 1 \simeq 0.39$. 

Similarly, the overlap between subsequent eigenvectors is defined as
\begin{equation} \label{eq:kappa}
\avg{\kappa} = \avg{ \sum_{i=1}^{\tilde{N}} | \psi_\alpha (i)  | | \psi_{\alpha+1} (i) | } \, ,
\end{equation}
where the meaning of the average is the same as in Eq.~\eqref{eq:erre}.
In the GOE regime the wavefunctions' amplitudes are i.i.d. Gaussian random variables of zero mean and variance $1/\tilde{N}$, hence $\langle \kappa \rangle$ converges to $\langle \kappa \rangle_\text{GOE} = 2/\pi \simeq 0.64$. Conversely, in the localized phase two successive eigenvectors are typically peaked around very distant sites and do not overlap, and therefore $\langle \kappa \rangle \to 0$. 

We performed exact diagonalizations of the giant connected component of 
samples of total size $N \in [2^8,2^{16}]$ --- the corresponding numerical results are shown in the plots of Fig.~\ref{fig:gap_statistics} for $p=2.4$, varying $E$ across the energy band. In agreement with the analysis of the distributions of the LDoS, which allowed us to locate the mobility edge at $E_{\rm loc} \simeq 1.75 \pm 0.05$ for this value of 
$p$,
panels (a) and (b) show that, at large energy, $\avg{r}$ and $\avg{\kappa}$ seem to approach their Poisson limiting value upon increasing the system size $N$; conversely, for $E < E_{\rm loc}$ 
they
seem to tend to their GOE asymptotic values upon increasing $N$.  
Yet, the convergence to the GOE values is quite slow even very far from the transition point. This is illustrated in panels (c) and (d), which indicate that, even for energies far below $E_{\rm loc}$, one has 
\begin{align}
\avg{r}_{\rm GOE} - \avg{r} &\simeq c N^{-a} \, , \\
\avg{\kappa}_{\rm GOE} - \avg{\kappa} &\simeq c^\prime N^{-a^\prime} \, ,
\end{align}
with $a,a^\prime \approx 0.54$, and prefactors $c,c^\prime$ that increase (exponentially) upon approaching the transition. 

These findings suggest that in the delocalized \rev{multifractal} 
phase, although the spacing between neighboring levels is described by universal ensembles in the large-$N$ limit, the level statistics might be somewhat non-standard. This is confirmed by a close inspection of the correlations beyond the scale of the average gap, as described in the next subsection.

\subsection{Overlap correlation function}
\label{sec:overlap}
Next, we study the overlap correlation function
\begin{widetext}
\begin{equation} \label{eq:K2}
\begin{aligned}
K_2 (\omega;E) &= \frac{N \avg{\sum\limits_{\alpha,\beta} \sum\limits_i \abs{\psi_\alpha(i)}^2 \abs{\psi_\beta(i)}^2 \delta \left(E + \frac{\omega}{2} - \lambda_\alpha \right) \delta \left(E - \frac{\omega}{2} - \lambda_\beta \right)}}{\avg{\sum\limits_{\alpha,\beta} \delta \left(E + \frac{\omega}{2} - \lambda_\alpha \right) \, \delta \left(E - \frac{\omega}{2} - \lambda_\beta \right)}} 
= \lim_{\eta \to 0} \frac{N \avg{\sum\limits_i \rho_i \left(E + \frac{\omega}{2} \right) \rho_i \left(E - \frac{\omega}{2}  \right) }}{\avg{\sum\limits_i \rho_i \left(E + \frac{\omega}{2}  \right) \sum\limits_j \rho_j \left(E - \frac{\omega}{2} \right) }} \, ,
\end{aligned}
\end{equation}
\end{widetext}
where the average is performed over several realizations of ${\cal H}$. Note that for $\omega \to 0$ the overlap correlation function is related to the IPR in Eq.~\eqref{eq:Iq} via $\lim_{\omega \to 0} K_2(\omega) = N I_2/3$~\cite{fyodorov1997strong}. 

The properties of $K_2(\omega)$ and its relationship with other spectral probes have been discussed extensively in the literature (see \eg~Ref.~\cite{Mirlin_2000} for a review).
For GOE matrices the eigenvectors are random vectors on the $N$-dimensional unit sphere, and thus $K_2(\omega) = 1$ identically, independently of $\omega$, over the entire spectral bandwidth~\cite{fyodorov1997strong}. In the standard fully-delocalized metallic phase $K_2(\omega)$ has a plateau at small energies, for $\omega < E_{\rm Th}$, followed by a fast-decay, often described by a power-law with a system-dependent exponent~\cite{chalker1990scaling,chalker1988scaling}. The height of the plateau is larger than one, which implies an enhancement of correlations compared with the case of independently fluctuating Gaussian wavefunctions. The threshold $E_{\rm Th}$ is known as the Thouless energy, which separates the plateau from the fast decay. 
The Thouless energy thus corresponds to the energy band over which GOE-like correlations establish~\cite{altshuler1986repulsion}. 

\begin{figure*}
\includegraphics[width=0.259\textwidth]{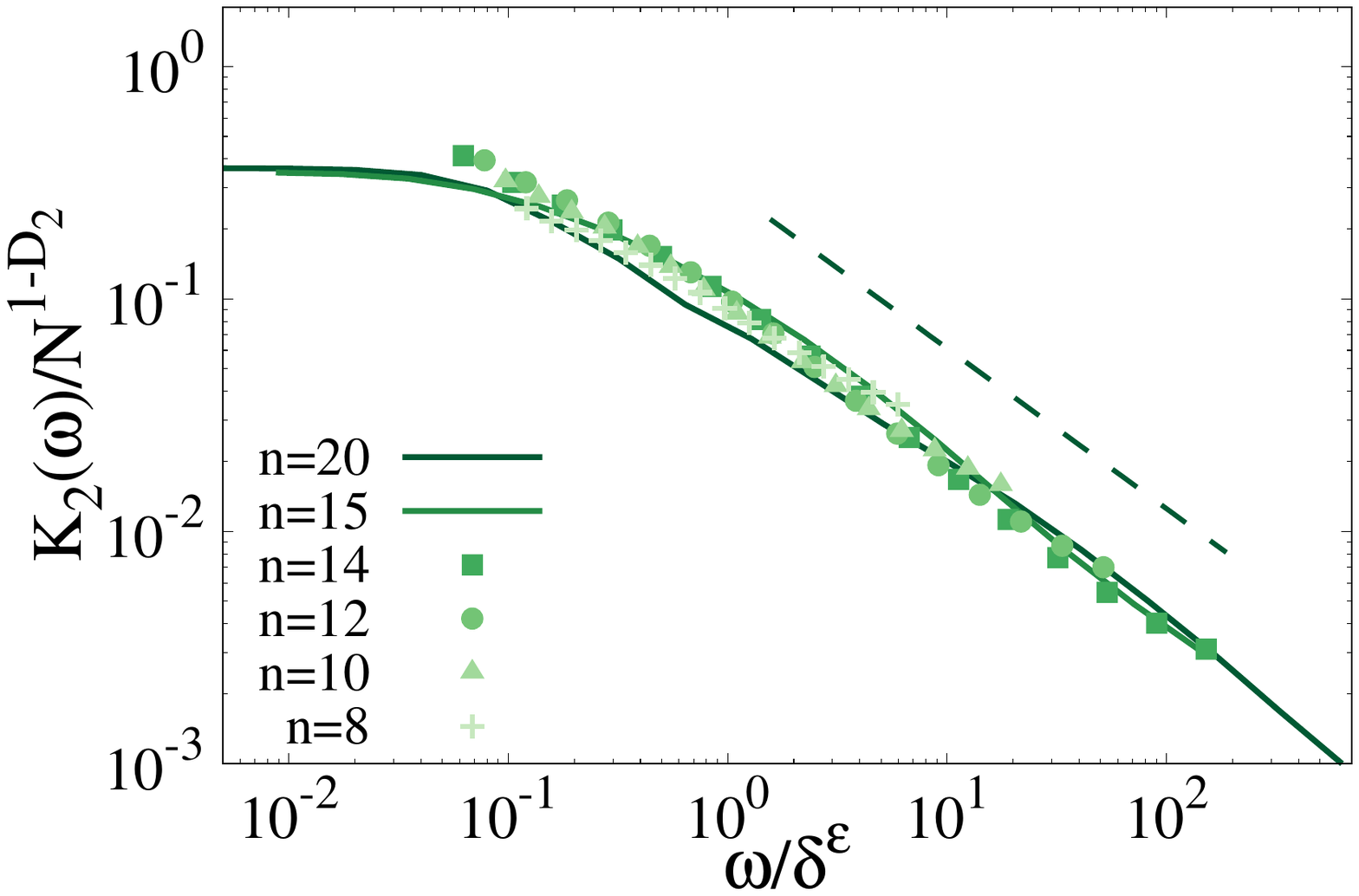} 
\put(-20,75){(a)}
\hspace{-0.23cm} 
\includegraphics[width=0.259\textwidth]{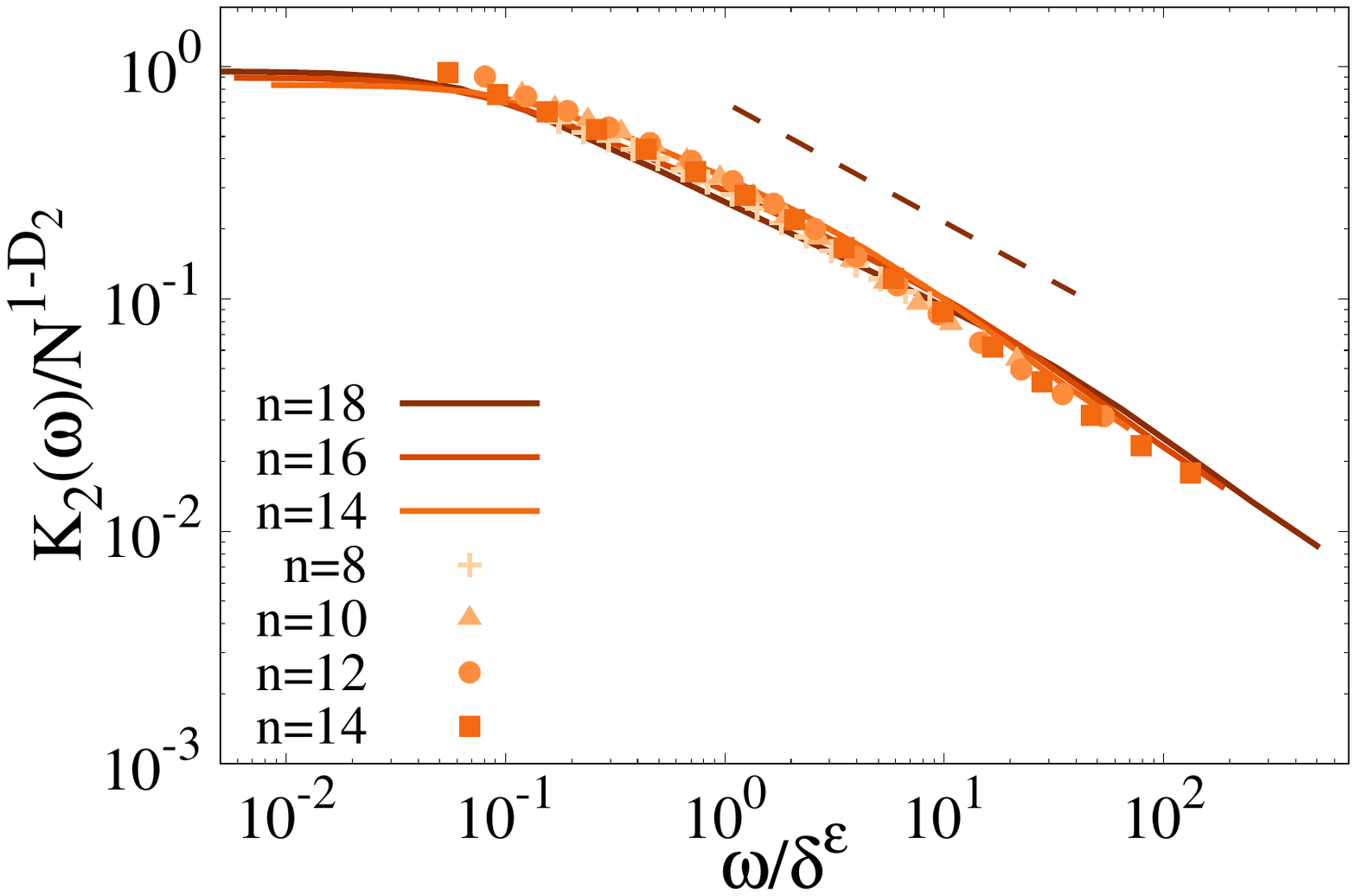} 
\put(-20,75){(b)}\hspace{-0.30cm} 
\includegraphics[width=0.259\textwidth]{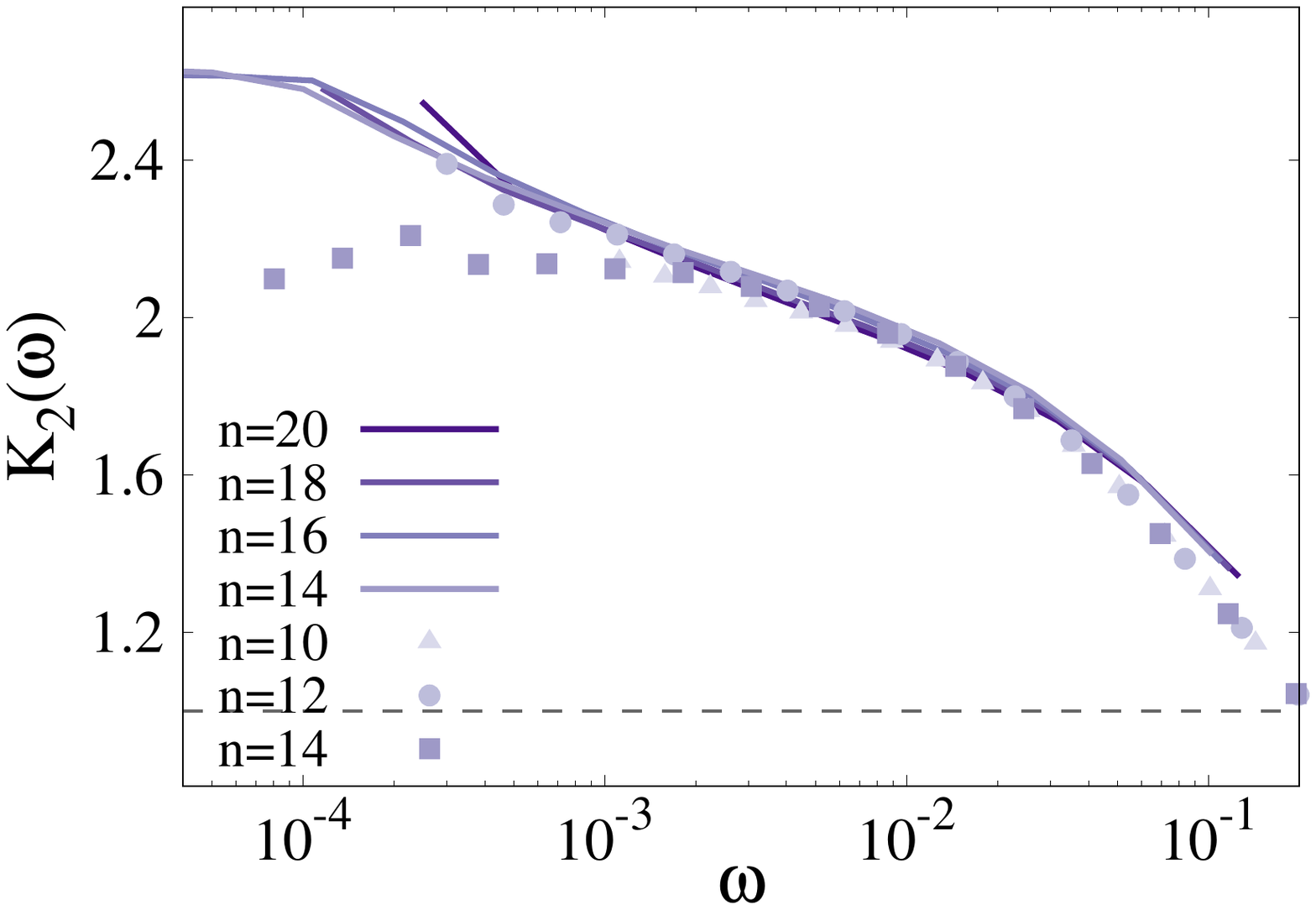} 
\put(-20,75){(c)}
\hspace{-0.36cm} 
\includegraphics[width=0.259\textwidth]{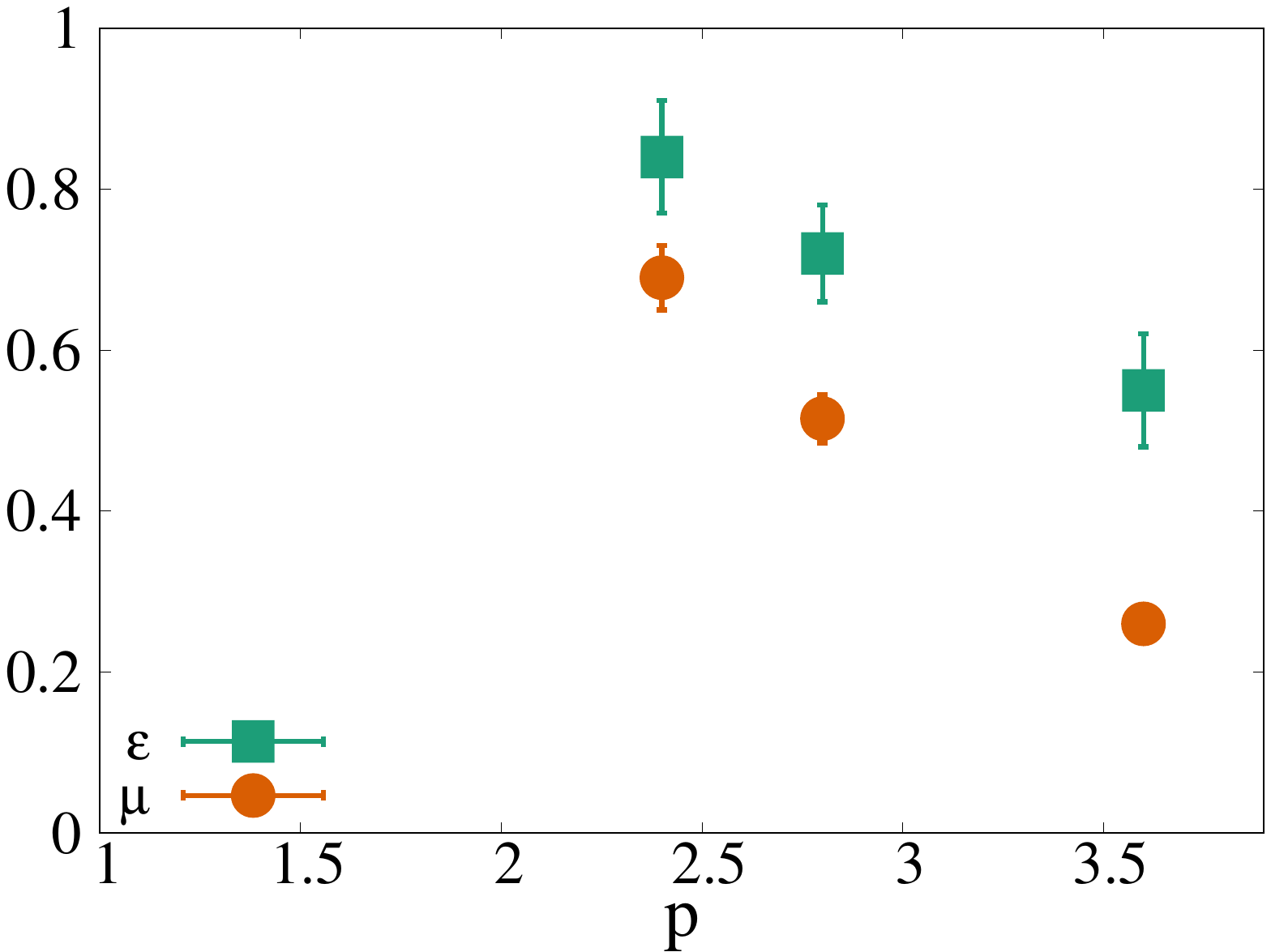}
\put(-20,75){(d)}
\caption{Overlap correlation function $K_2(\omega;E)$, see Eq.~\eqref{eq:K2}, for $E=0.4$ and $p=2.4$ (a), $p=2.8$ (b), and $p=5$ (c). In panels (a) and (b) the energy separation $\omega$ is rescaled by the Thouless energy $\delta^{\epsilon}$ --- with a $p$-dependent exponent $\epsilon \in [0,1]$ --- and the $y$-axis is rescaled by $N^{1-D_2}$ --- where $D_2 = \beta -1$, with $\beta$ given in Fig.~\ref{fig:Dq}(a). Continuous lines correspond to the results obtained by solving the self-consistent cavity equations, while symbols  show the exact diagonalization results. The dashed lines in panels (a) and (b) represent the power-law decay of correlations for $\omega \gg \delta^\epsilon$ as $K_2 (\omega) \propto (\omega/\delta^\epsilon)^{-\mu}$, with a $p$-dependent exponent $\mu$. The horizontal gray line in panel (c) corresponds to the Wigner-Dyson behavior for GOE matrices, $K_2(\omega)=1$. In panel (d) we plot our numerical estimates for the exponents $\epsilon$ (squares) and $\mu$ (circles) varying the average degree $p$, showing that a standard fully-delocalized behavior (with $\epsilon,\mu\to 0$) is progressively reached
upon increasing $p$.
\label{fig:K2}}
\end{figure*}

The benchmark example for fractal extended eigenstates is provided by the Gaussian RP ensemble and its generalizations~\cite{Kravtsov_2015,vonSoosten_2019,Facoetti_2016,Truong_2016,Bogomolny_2018,DeTomasi_2019,amini2017spread,pino2019ergodic,berkovits2020super,us_2023,kravtsov2020localization,khaymovich2020fragile,monthus2017multifractality,biroli2021levy}. In this case the plateau develops only in a narrow energy interval which shrinks with the system size, yet remaining much larger than the mean level spacing $\delta$~\cite{Kravtsov_2015,DeTomasi_2019,khaymovich2020fragile,biroli2021levy}. This implies that the Thouless energy shrinks to zero in the thermodynamic limit as $E_{\rm Th} \propto \delta^{\epsilon}$, with $0 < \epsilon< 1$. This corresponds to the formation of {\it mini-bands} containing a sub-extensive fraction of eigenstates characterized by GOE-like correlations. 
Beyond $E_{\rm Th}$ the eigenfunctions poorly overlap with each other, and the statistics is no longer GOE: $K_2(\omega)$ thus decays as a power-law, $K_2(\omega) \propto (\omega/E_{\rm Th})^{-\mu}$, with a system-dependent exponent $\mu$.
Since in the multifractal extended phase the IPR behaves as $I_2 \propto N^{-D_2}$, the height of the plateau at small energies grows as $K_2(\omega \to 0) \propto N^{1-D_2}$.  This suggests that, in the \rev{multifracrtal} 
extended regime, $K_2(\omega)$ for different $N$ should collapse onto the same curve if the energy axis is rescaled by 
$E_{\rm Th} \propto \delta^{\epsilon}$, and the vertical axis is rescaled by $N^{1-D_2}$. This is indeed what we observe in 
Fig.~\ref{fig:K2}(a--b), for $p=2.4$ and $p=3.6$ (and for $E=0.4$). 
Yet, differently from the RP ensemble and its extensions, in the multifractal extended phase of the WER ensemble a genuine plateau of $K_2(\omega)$ only establishes at very small energy separations, even smaller than
$\delta$ for the accessible sizes. (In other words, the Thouless energy behaves as $b \delta^\epsilon$ with a prefactor $b \ll 1$.)
This explains the strong finite-size corrections to the GOE asymptotic behavior observed even at the scale of the average gap.

In Fig.~\ref{fig:K2}(d) we plot the values of the exponent $\epsilon$ describing the scaling of 
$E_{\rm Th}$ with the system size, and that of the exponent $\mu$, which describes the power-law decay of the overlap correlations for an energy separation larger than $E_{\rm Th}$. 
Note that, since $K_2(\omega) \approx N^{1-D_2}$ for $\omega \ll E_{\rm Th}$ and $K_2(\omega) \sim {\cal O}(1)$ for $\omega \sim {\cal O} (1)$ (owing to the lack of correlation between wavefunctions at large energy separations), one finds $\delta^{-\epsilon \mu} \simeq N^{1 - D_2}$ on the scale of the Thouless energy, implying that $\epsilon \mu = 1 - D_2 = 2 - \beta$. We have verified that this consistency relation holds very well within our numerical accuracy.

Both $\epsilon$ and $\mu$ decrease with the average degree $p$, 
consistently with the system 
gradually becoming \rev{less multifractal}
as $p$ is increased. 
This is confirmed by Fig.~\ref{fig:K2}(c), where the overlap correlation function is plotted for $p=5$. In this case one recovers the behavior of the standard metallic phase, in which $K_2(\omega)$ becomes essentially independent of $N$. (Note, however, that for $p=5$ one has $D_2=1$ but $D_q<1$ for sufficiently large $q$, see Fig.~\ref{fig:Dq}(b). Hence we expect that the singular 
features of the spectral statistics should show up in generalized overlap correlations involving higher powers of the wavefunctions' coefficients.)

\subsection{Level compressibility}
\label{sec:level_comp}

\begin{figure*}
\includegraphics[width=0.42\textwidth]{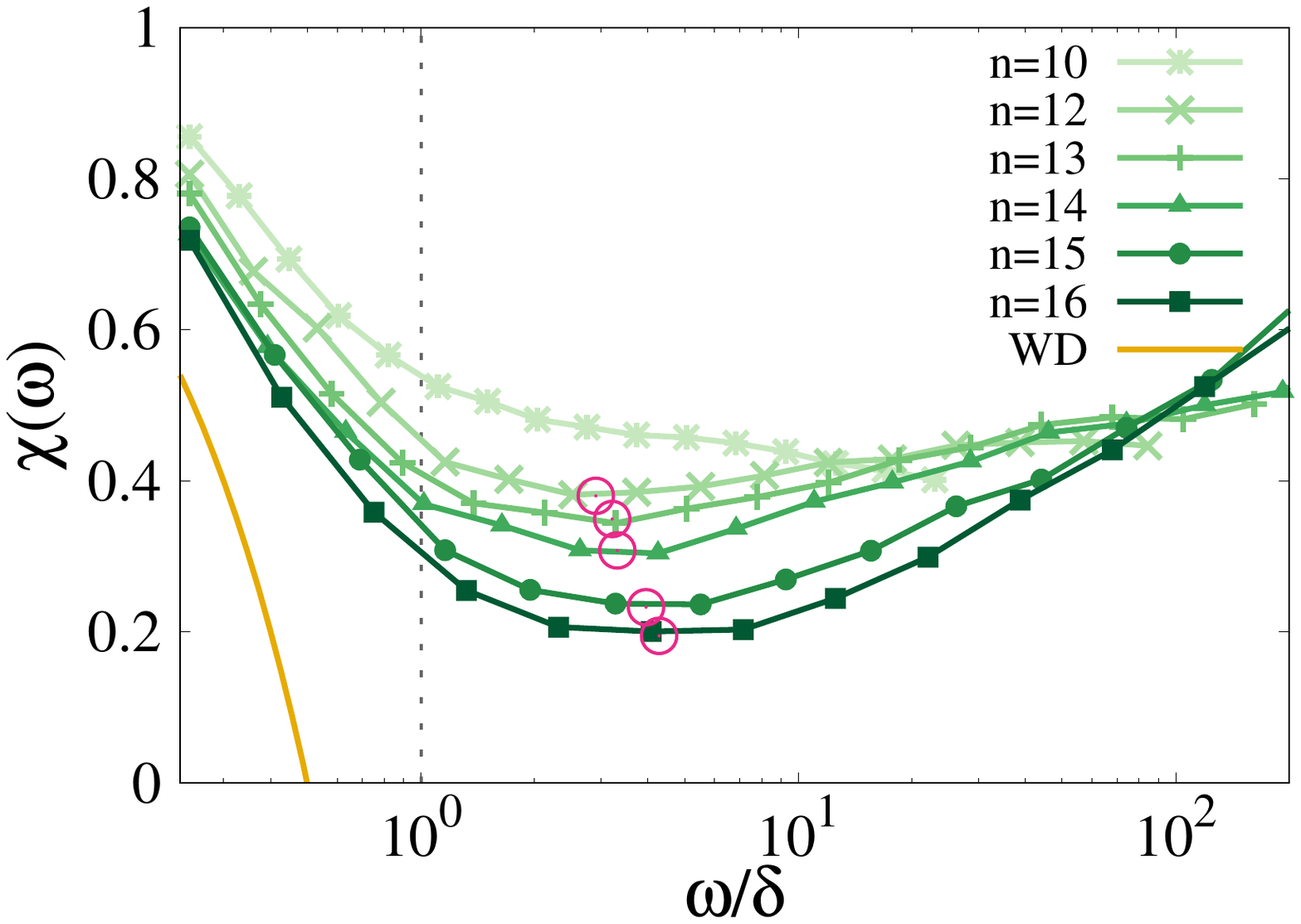}  
\put(-25,30){(a)} 
\hspace{0.5cm}
\includegraphics[width=0.42\textwidth]{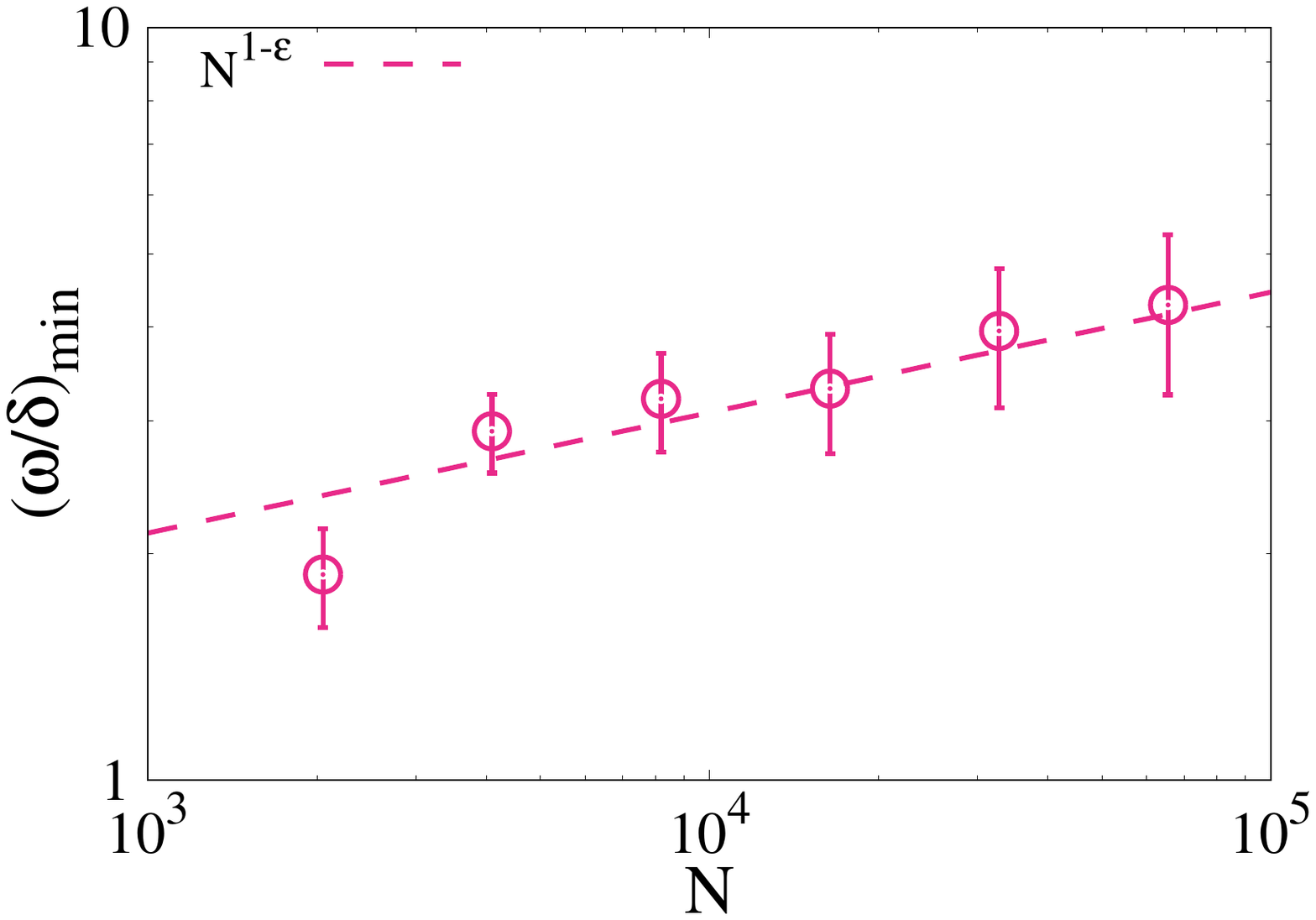}
\put(-25,30){(b)} 
 \caption{(a) Level compressibility $\chi(\omega)$, see Eq.~\eqref{eq:level_compress}, for $p=2.4$ and $E=0.4$, and for several values of $N=2^n$. The width of the energy interval is rescaled by the mean level spacing $\delta$ (see Eq.~\eqref{eq:delta}). The orange solid line shows the limiting behavior $\chi(\omega) \simeq 1 - 2 \omega/\delta$ of the level compressibility in the GOE at small $\omega$. The empty circles indicate the position of the minima of $\chi(\omega)$, signaling the crossover from GOE to Poisson behavior. (b) Position of the minima of $\chi(\omega)$ (in units of the mean level spacing $\delta$) as a function of the system size $N$. The dashed line corresponds to the power-law behavior $(\omega/\delta)_{\rm min} \propto N^{1 - \epsilon}$, with the exponent $\epsilon \approx 0.84$ obtained from the rescaling of the overlap correlation function by the Thouless energy $E_\text{Th}\sim \delta^{\epsilon}$, see Fig.~\ref{fig:K2}(a).
 \label{fig:chi}}
\end{figure*}

The emerging picture on the correlations of the energy levels 
in the delocalized multifractal phase of the WER ensemble are also supported by the inspection of the level compressibility $\chi(\omega)$, which is a simple indicator of the degree of level repulsion~\cite{altshuler1988repulsion,Mirlin_2000,chalker1996spectral,bogomolny2011eigenfunction,metz2017level}, and is defined as follows. Let $I_N [ E-\omega/2,E+\omega/2 ]$ denote the number of eigenvalues lying in the interval 
$[ E- \omega/2,E+\omega/2]$, which is a random variable; then 
\begin{equation}
\begin{aligned}
    \chi(\omega;E) & \equiv  \frac{\avg{I_N^2\left[ E-\frac{\omega}{2},E+\frac{\omega}{2} \right]}-\avg{I_N\left[ E-\frac{\omega}{2},E+\frac{\omega}{2} \right]}^2}{\avg{I_N\left[ E-\frac{\omega}{2},E+\frac{\omega}{2} \right]}}
    \, .
    \end{aligned}
    \label{eq:level_compress}
\end{equation}
For Poisson statistics, one finds $\chi(\omega)\simeq 1$ (for small $E$) since $\avgg{I_N^2 
}_c \simeq \avgg{I_N 
}$. On the contrary, for a rigid spectrum with GOE correlations the mean number of eigenvalues behaves as $\avg{I_N 
} \propto \omega/\delta$, while $\avg{I_N^2 
}_c \propto \ln (\omega/\delta)$ for $\omega \gg \delta$. Hence in the GOE case one finds $\chi(\omega) \to 0$ for $\omega/\delta \to \infty$, while $\chi (\omega)$ is described by a universal function for $\omega/\delta$ of ${\cal O}(1)$ (see {\it e.g.}~\cite{us_2023,Mirlin_2000}).

In Fig.~\ref{fig:chi}(a) we plot $\chi(\omega)$ for $p=2.4$ and $E=0.4$ (similar results are obtained for other values of the parameters across the \rev{multifractal} 
extended phase). The plot shows that at small $\omega$ the level compressibility decreases with $N$ and slowly approaches the GOE limiting curve, until a minimum is reached. The position of the minimum scales as $\delta^\epsilon$ (see Fig.~\ref{fig:chi}(b)), with the value of $\epsilon$ given in Fig.~\ref{fig:K2}(d), confirming that the Thouless energy $E_\text{Th}\propto \delta^\epsilon$ corresponds to the width of the energy (mini)bands within which GOE-like correlations of the eigenvalues are established. Conversely, for much broader energy intervals, $\chi(\omega)$ seems to approach a value of $\mathcal O(1)$, corresponding to uncorrelated levels.

\section{Return probability}
\label{sec:return_probability}

\begin{figure*}
\includegraphics[width=0.257\textwidth]{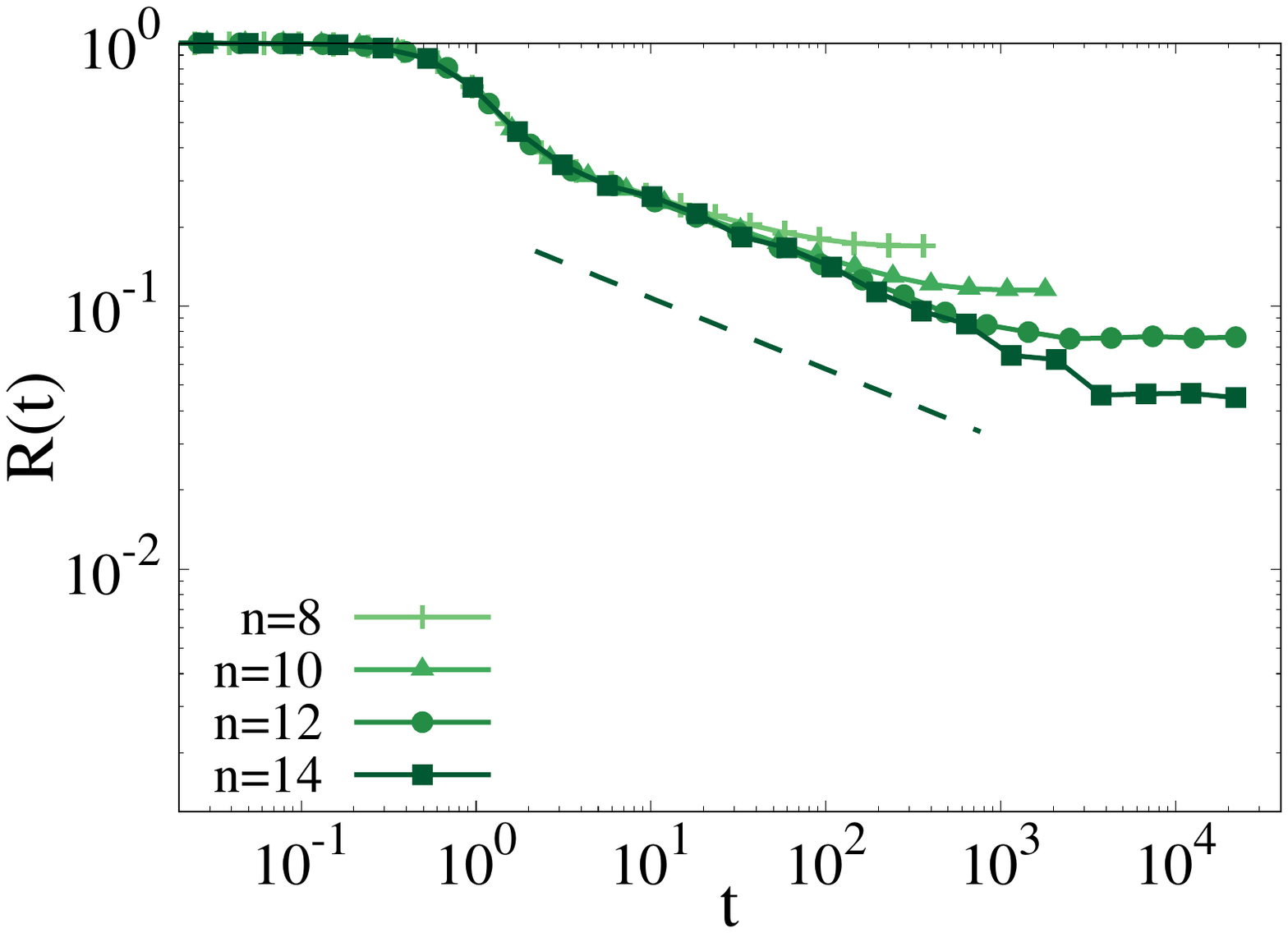} \put(-20,17){(a)} \hspace{-0.31cm} \includegraphics[width=0.257\textwidth]{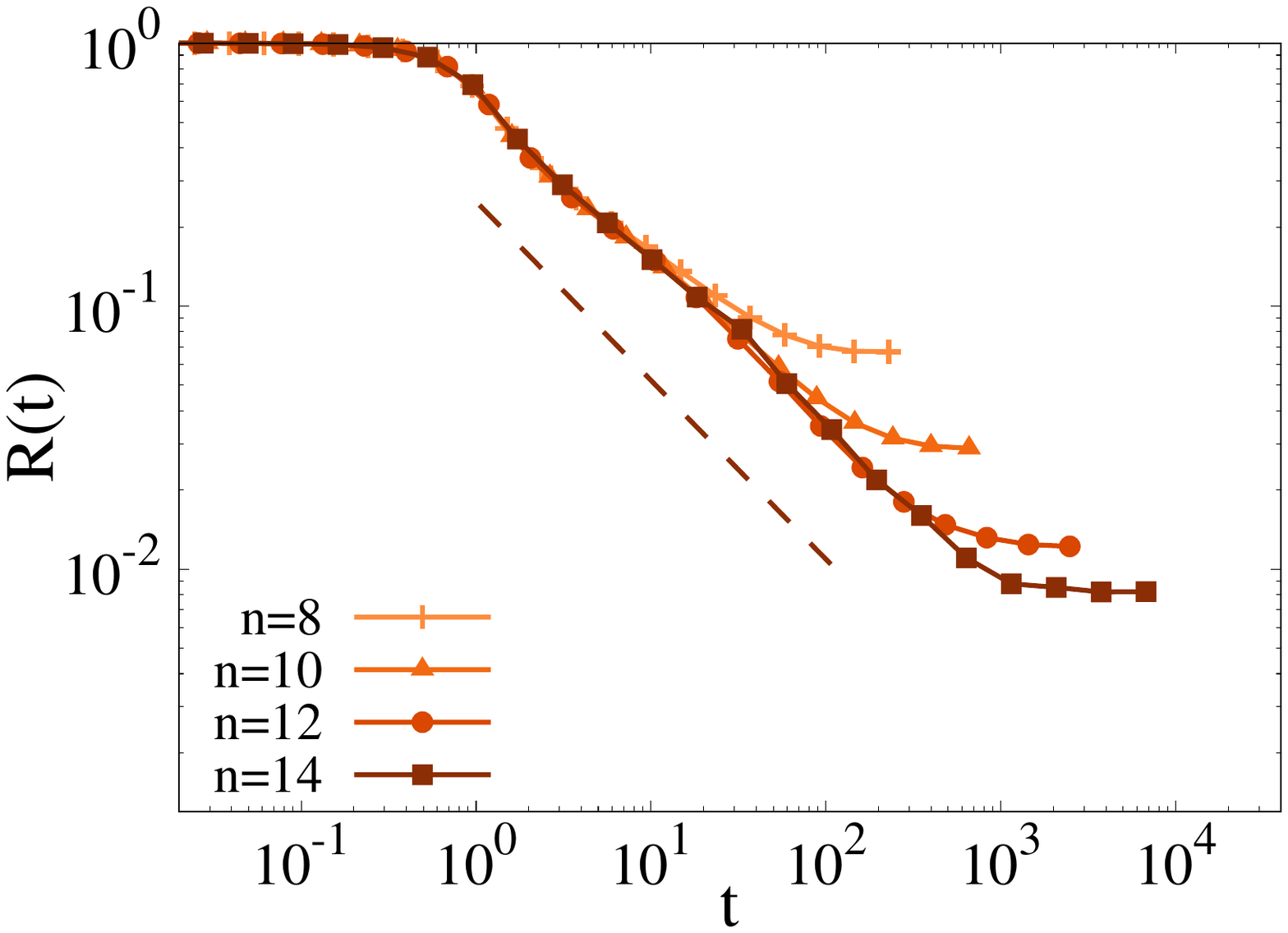} \put(-20,17){(b)} \hspace{-0.31cm} \includegraphics[width=0.257\textwidth]{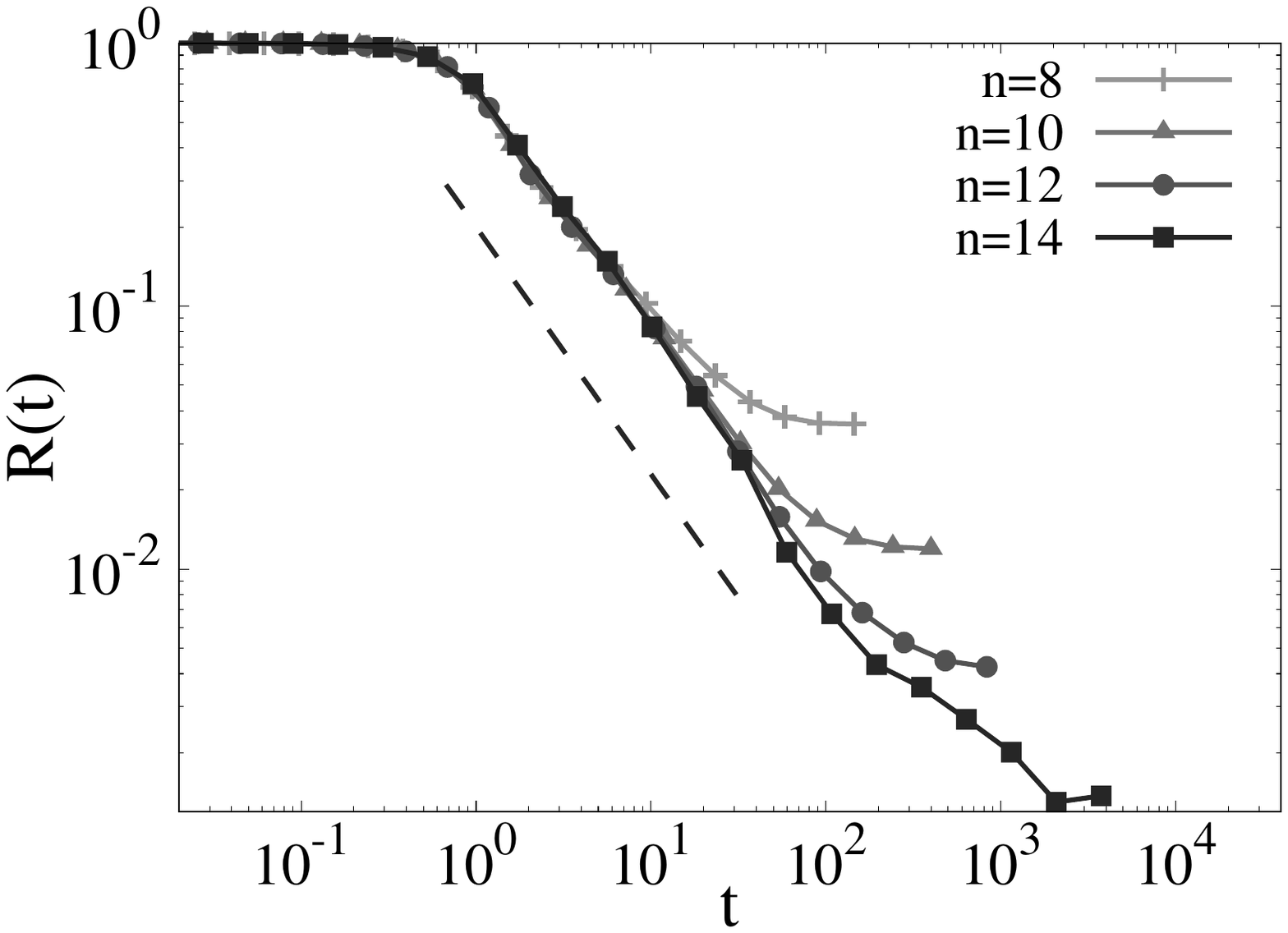} \put(-20,17){(c)} \hspace{-0.31cm} \includegraphics[width=0.257\textwidth]{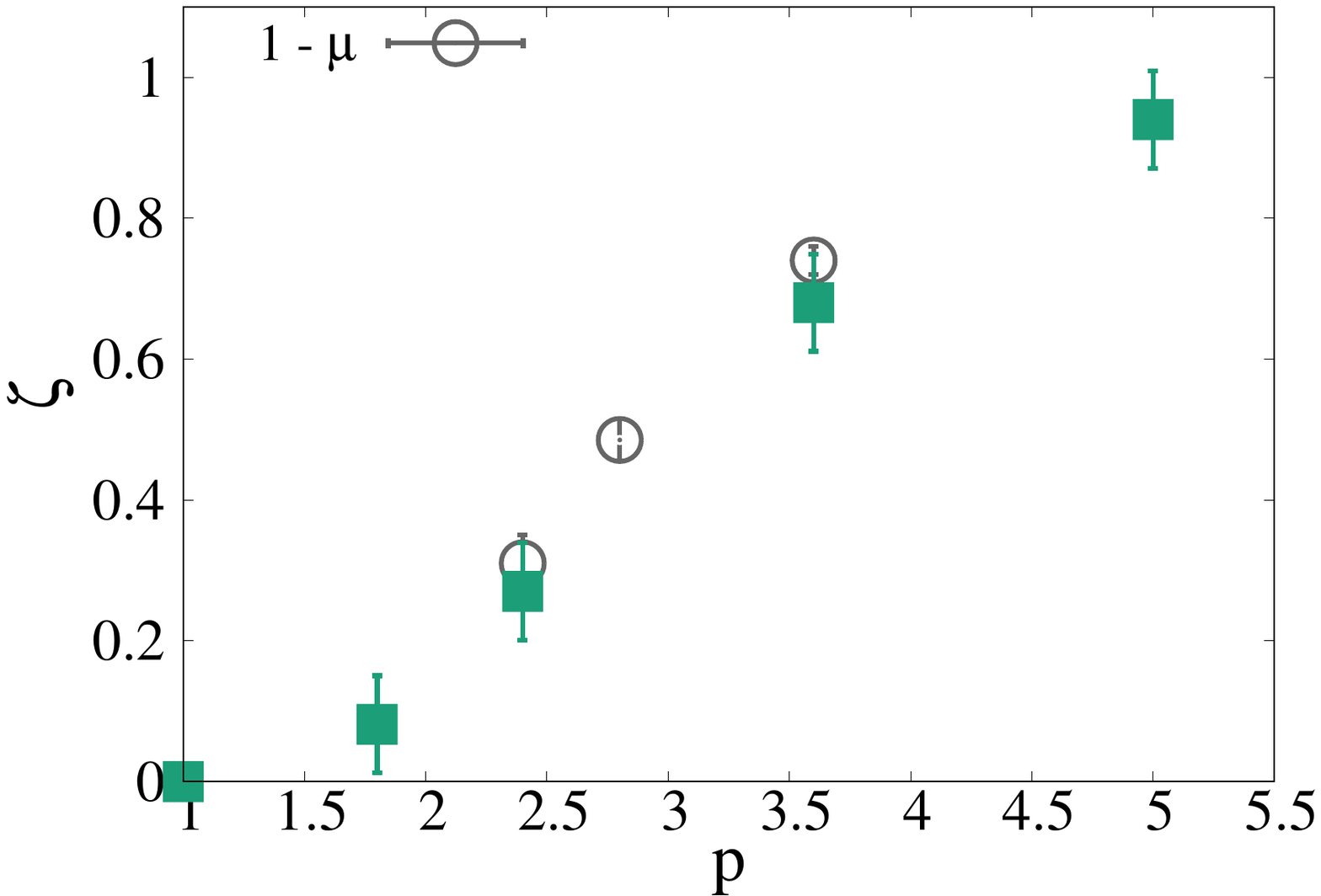} \put(-20,17){(d)}
\caption{Return probability, see Eq.~\eqref{eq:Rt}, as a function of time, for $p=2.4$ (a), $p=3.6$ (b), and $p=5$ (c). The dashed lines show the power-law decay of the return probability as $R(t) \propto t^{- \zeta}$. In panel (d) we plot the exponent $\zeta$ upon varying the average degree $p$, showing that its value is very close to $\zeta \simeq 1-\mu$ within our numerical accuracy, $\mu$ being the exponent that describes the power-law decay of the overlap correlation function $K_2(\omega)$ --- see Figs.~\ref{fig:K2}(a--b).} 
\label{fig:R}
\end{figure*}

The existence of multifractal extended eigenstates in a broad region of the phase diagram is expected to 
have significant consequences on the quantum dynamics. In particular, 
in the \rev{multifractal} 
extended regime it is natural to expect both transport and relaxation 
to be anomalously slow~\cite{bera2018return,de2020subdiffusion,biroli2017delocalized,jagannathan2022electronic,biroli2020anomalous}. To check these ideas, we have run a few dynamical experiments in which a wave packet $\vert \varphi_0 \rangle$,
initially
localized in a
small spatial region, spreads under the unitary evolution
induced by $\mathcal H$.
Note that choosing
the initial amplitude to be localized on a single
node would correspond
to a wave packet of energy $E_0 = \langle  \varphi_0 \vert {\cal H} \vert \varphi_0 \rangle = 0$: in this case, however, the presence of an extensive number of degenerate eigenstates of zero energy exponentially localized on the leaves of the graph could affect the dynamical properties. 
To avoid a strong projection of our initial state onto
such localized eigenvectors, in the following we choose initial conditions such that the wave packet is localized on two randomly selected neighboring sites $i$ and $j$, 
both having degree strictly larger than one. Hence we take
\vspace{-10pt}
\begin{equation} \label{eq:phi0}
    \vert \varphi_0 \rangle = \sin (\theta)\, \vert i \rangle \pm \cos (\theta)\, \vert j \rangle \, ,
\end{equation}
where the angle $\theta$ is extracted uniformly within the interval $\theta \in (\pi/4 - \pi/8,\pi/4+\pi/8)$; with this choice, the energy of the wave packet fluctuates as 
\begin{eqnarray}
    E_0 = \langle  \varphi_0 \vert {\cal H} \vert \varphi_0 \rangle = \pm h_{ij} \sin (2 \theta) / \sqrt{p},
\end{eqnarray}
where $h_{ij}/\sqrt{p}$ is the hopping amplitude on the edge $\langle i, j \rangle$.
The time evolution of the wavefunction $\vert \varphi_0 \rangle$ at time $t$ can then be written in terms of the eigenvalues $\lambda_\alpha$ and the eigenfunctions $\vert \psi_\alpha \rangle$ of the Hamiltonian as
\begin{equation}
    \vert \varphi(t) \rangle = \sum_\alpha e^{-{\rm i} \lambda_\alpha t} \left[ \sin (\theta)\, \psi_\alpha(i) \pm \cos (\theta)\, \psi_\alpha (j) \right] \vert \psi_\alpha \rangle \, .
\end{equation}
To simplify the notations, we have set $\hbar=1$.
Below we focus on the so-called return probability $R(t)$, defined as the probability that a particle starting from the state~\eqref{eq:phi0} at $t = 0$ is found either on node $i$ or $j$ after a time~$t$:
\begin{widetext}
\begin{eqnarray}
 \label{eq:Rt}
    R(t) &=& \avg{\left \vert \langle i \vert \varphi(t) \rangle \right \vert^2 + \left \vert \langle j \vert \varphi(t) \rangle \right \vert^2} \\
    &=& \avg{\sum_{\alpha,\beta} \cos \left[ (\lambda_\alpha - \lambda_\beta) t \right]  \left[ \psi_\alpha(i) \psi_\beta(i) + \psi_\alpha(j) \psi_\beta(j) \right]  \left[ \sin (\theta)\, \psi_\alpha(i) \pm \cos (\theta)\, \psi_\alpha (j) \right] \left[ \sin (\theta)\, \psi_\beta(i) \pm \cos (\theta)\, \psi_\beta (j) \right]} \, .\nonumber
\end{eqnarray}
\end{widetext}
This represents a measure of how quickly the wave packet spreads over the lattice and decorrelates from the initial condition.
The averages in Eq.~\eqref{eq:Rt} are performed over several realizations of the disorder and several initial conditions~\eqref{eq:phi0}. It is easy to check that in the limit $t = 0$ one has $R(t=0)=1$, while in the $t \to \infty$ limit $R(t)$ approaches a plateau value closely related to the average IPR.
The corresponding numerical results are plotted in Fig.~\ref{fig:R}. Their most striking feature is the emergence of a power-law decay $R(t) \propto t^{- \zeta}$ of the return probability,
which extends to larger and larger times as the system size is increased. By contrast, the standard diffusive spreading of the wave-packet in a standard fully-delocalized
phase is expected to be exponential at long times on tree-like graphs. We recall that the return probability is tightly related to the Fourier transform of the overlap correlation function $K_2(\omega)$ (which we studied in Sec.~\ref{sec:level})~\cite{tikhonov2019statistics}, hence the power-law decay of $R(t)$ at large times is directly connected  to the power-law decay of the spectral correlations at small energy separation. 
A direct numerical check is 
reported in  Fig.~\ref{fig:R}(d), showing that within our numerical accuracy we have $\zeta \approx 1 - \mu$.
Concomitantly, we also expect anomalous sub-diffusion, meaning that the mean-square displacement grows with an exponent smaller than one.

Anomalous power-law relaxation in disordered quantum systems has been intensively investigated in the past years~\cite{bera2018return,de2020subdiffusion,biroli2017delocalized,jagannathan2022electronic,biroli2020anomalous}, especially in relation to the anomalous transport and thermalization properties on the metallic side of the MBL transition~\cite{luitz2017ergodic,agarwal2017rare}. A similar 
slow
decay of the return probability has 
also
been reported in the delocalized phase of the Anderson model on the RRG of finite size at short time scales~\cite{bera2018return,de2020subdiffusion,biroli2017delocalized}, although a crossover to 
a stretched-exponential
decay is found in that case in the infinite-size and infinite-time limits~\cite{tikhonov2019statistics}. Similarly we expect that
the power-law decay of $R(t)$ should eventually cross over to a standard exponential relaxation at large 
\rev{average degree}
$p$, while it should persist up to arbitrarily long times at small enough $p$.

Interestingly, even at the classical level, the fluctuating connectivity of vertices in \ER~graphs is recognized for inducing slow dynamics and anomalous diffusion~\cite{metzler2014anomalous}.
These models essentially behave like spatially extended ``trap models''~\cite{bouchaud1990anomalous}, capturing various aspects of the slow glassy dynamics~\cite{margiotta2019glassy}.

\begin{figure*}
    \includegraphics[width=0.7\textwidth]{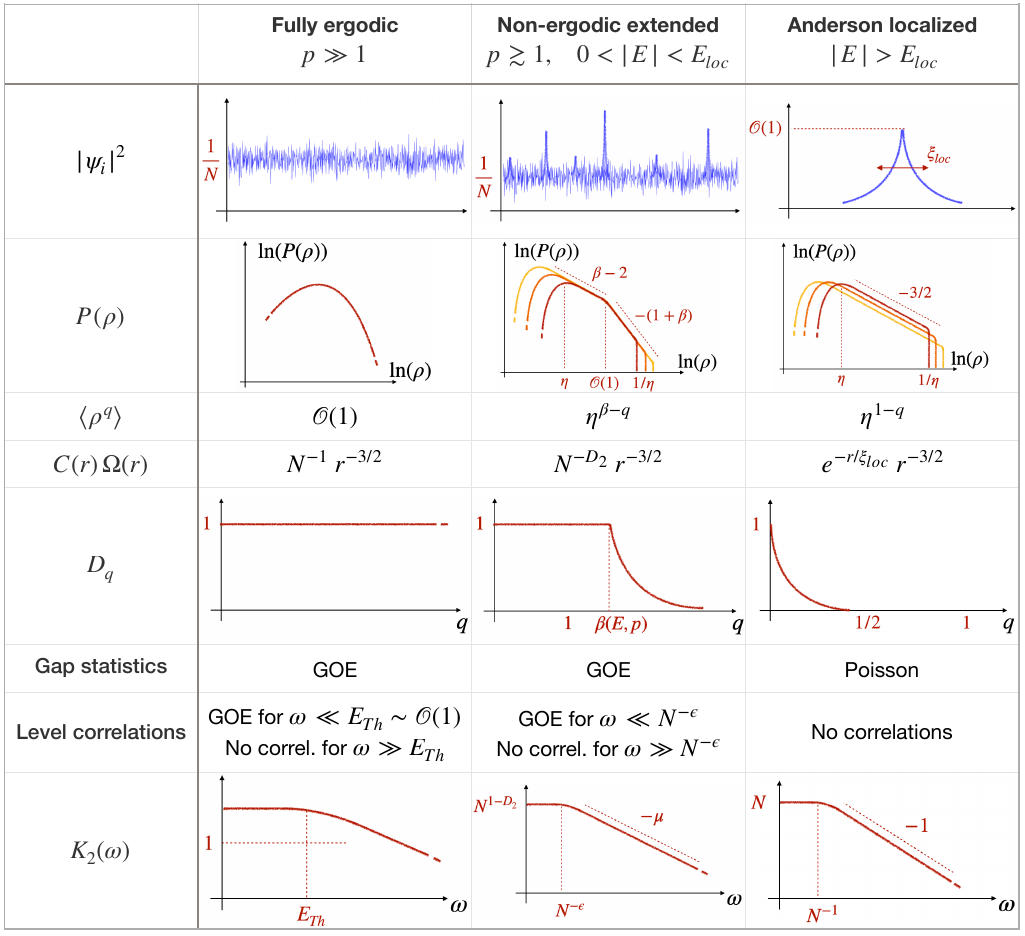} 
    \caption{Graphical summary of some of the observables characterized in this paper, throughout the distinct phases of the model.}
    \label{fig:table}
\end{figure*}

\section{Summary and perspectives} 
\label{sec:conclusions}

In this work we discussed the emergence of delocalized multifractal states within a broad region of the parameter space ($p$--$E$) of the (giant connected component of the) weighted \ER~random graph, where $p$ is the average node 
\rev{degree}
and $E$ is the eigenstates' energy. 
Our analysis revealed an unusual behavior of the fractal dimensions $D_q$ governing the scaling with the system size of the moments of the eigenfunctions' amplitudes (see Eq.~\eqref{eq:Dqbeta} and Fig.~\ref{fig:Dq}). In particular, the support set of the \rev{multifractal} 
extended states is 
\rev{extensive in $N$} 
and occupies a finite fraction of the nodes of the graph (\ie~$D_1=1$), while higher moments of the wavefunctions' coefficients are anomalously large on a vanishingly small fraction of nodes. Upon increasing the average degree $p$ this exotic behavior shifts to larger and larger moments, and the multifractal states progressively recover fully-\rev{extended} 
features, as expected~\cite{mirlin1991localization}.

The phase diagram we presented in Fig.~\ref{fig:PD}(a) was obtained by choosing Gaussian-distributed hopping amplitudes, but we believe its qualitative features to be generic: indeed, as detailed in Sec.~\ref{sec:mechanism}, the origin of the partially-delocalized states has to be traced back to the pronounced heterogeneity in the topology of the graph. In particular, strong fluctuations in local degrees produce many leaves and grafted trees (\ie~groups of nodes that are connected to the bulk via a single edge), while small values of the hopping on these edges cause an effective fragmentation of the graph (as depicted in Fig.~\ref{fig:PD}(b)). This mechanism might be particularly pertinent in the context of MBL and quantum kinetically constrained models, where the fragmentation of the Hilbert space has been argued to play a pivotal role~\cite{pietracaprina2021hilbert,roy2019exact,roy2019percolation,bernien2017probing,serbyn2021quantum,pancotti2020quantum}.

We supported our findings by analyzing the level statistics in the delocalized multifractal phase. Several observables such as the average gap statistics (Sec.~\ref{sec:gap}), the overlap correlation function (Sec.~\ref{sec:overlap}), and the level compressibility (Sec.~\ref{sec:level_comp}) indicate that correlations among the energy levels establish in a narrow energy band that shrinks to zero in the limit of large system sizes, but remains
much larger than the mean level spacing. Yet, in contrast to the RP model and its generalizations,
pronounced finite-$N$ deviations from the Wigner-Dyson statistics are observed already shortly beyond the scale of the average gap.
The table in Fig.~\ref{fig:table} pictorially summarizes the behavior of the various observables discussed in this work throughout the distinct phases of the model.

Furthermore, we demonstrated that the presence of delocalized \rev{multifractal} 
states exerts a profound influence on the spreading dynamics of wave packets under unitary quantum evolution. Specifically, the return probability exhibits a robust power-law decay at small enough $p$ which extends to larger and larger times upon increasing the system size. 

Future extensions of this study could delve into the sparse complex Gaussian Unitary Ensemble case, where a transition in the characteristic polynomial has recently been established rigorously in Ref.~\cite{Afanasiev2016}. Additionally, gaining analytical insights into the structure of density-density correlations would be valuable, with a 
potential avenue being the extension of the diagrammatic expansion developed in Ref.~\cite{baron2023path} towards the low-connectivity regime. 
Another intriguing question is whether the critical properties of Anderson localization when approaching the critical point from the extended (but multifractal) phase are the same as those observed in the Anderson model on the Bethe lattice when the transition is approached from a \rev{standard metal with fully-extended eigenstates}. 

Exploring the impact of on-site random diagonal energies in the Hamiltonian is another interesting direction. These random energies are expected to lift the degeneracy of the states at $E=0$, dispersing them across the energy band, and concomitantly favouring Anderson localization. However, their influence on the properties of multifractal delocalized states remains uncharted territory.

Furthermore,
recent works~\cite{zirnbauer2023wegner,arenz2023wegner} have 
demonstrated the potential
existence of a novel saddle-point solution of the field-theoretical action found within the framework of the supersymmetric approach to the $n$-orbital Wegner model. 
This
new saddle point corresponds to a solution in which supersymmetry is only partially broken, 
in contrast to
the standard fully-extended phase where supersymmetry is 
broken, and 
to
the standard Anderson-localized phase in which supersymmetry is unbroken. 
This new saddle point has been argued to be relevant for a certain parameter range of Wegner's model. Attempting to relate the multifractal extended phase found in the WER ensemble to this novel fixed point is undeniably fascinating, and it certainly provides an intriguing perspective for future investigations.

The peculiar form of weak multifractality discovered within the bulk spectrum of the WER ensemble, as described by Eqs.~\eqref{eq:Dqbeta} and~\eqref{eq:falpha}, closely resembles the phenomenon observed in standard $\beta$-ensembles for $\beta>2$~\cite{Breuer2007,Das_2023,Das_2024}. Recently, suitable one-dimensional random matrix models have been introduced~\cite{Das_2023,Das_2024} so that their eigenvalue distribution follows the $\beta$-ensemble. 
In these models, multifractality arises from the power-law decay $|\psi (x)|^2 \sim 1/|x-x_0|^{1/q_\star}$, with $q_\star > 1$,
observed at large distances $|x-x_0|$ from a specific position $x_0$ where the wavefunction's amplitude is maximal. 
This yields $D_{q<q_\star} = 1$, and $D_{q>q_\star} < 1$, as in Eq.~\eqref{eq:Dqbeta}. Considering that on \ER~graphs the number of nodes at a distance $r$ from a given node grows exponentially with the distance ($\Omega(r) \sim p^r$), it seems plausible that the power-law decay of the wavefunction's amplitudes around a peak on a given node $r_0$ would be replaced by an exponential decay of the form $|\psi(r)|^2 \sim p^{-|r - r_0|/q_\star}$, with $q_\star = \beta$. Investigating the validity of this expectation is of great interest, as it could establish a direct link between the multifractality of the WER ensemble and that of the $\beta$ ensemble.

Finally, the results presented here are obtained for graphs of large but finite size. Although they are consistent and appear to be robust, the possibility that the extended \rev{multifractal} 
phase is just a finite-size phenomenon --- and that 
\rev{standard delocalization} 
is eventually restored on graphs of huge sizes --- cannot be completely ruled out. In fact, a crossover from an apparent delocalized multifractal phase to a fully-\rev{extended} 
phase does indeed take place in the Anderson model on the RRG~\cite{Garcia-Mata_2022,tikhonov2016anderson,garcia2017scaling,biroli2018delocalization,tikhonov2019statistics,baroni2023corrections}. 
However, in that case
such a crossover is observed in the flow of {\it all} the relevant observables concomitantly: the gap statistics and the flowing fractal dimensions related to the anomalous scaling of the wavefunction's amplitudes exhibit a crossover from \rev{an anomalous to a standard metallic} 
behavior all on the 
\textit{same}
characteristic scale (which grows very fast and diverges exponentially at the critical disorder). 
In the WER ensemble considered here, instead, 
should
such a crossover exist, it would be very peculiar. 
In fact, some observables (\eg~$\avg{r}$, $\avg{\kappa}$, $D_1$) converge to their expected asymptotic behavior on some characteristic scale which is smaller than the largest accessible size, while anomalous scaling dimensions associated to higher moments of the wavefunctions' amplitudes, \ie~$D_q$ for $q>\beta$,  would crossover on
a distinct
(and much larger) scale.

The answer to this question is ultimately related to another important unresolved issue:
determining the scaling with the system size of the minimal value of $\nu$ --- which prevents excessively small hopping amplitudes, see Sec.~\ref{sec:mechanism} --- sufficient to restore \rev{a standard metallic behavior}. 
In other words, 
it remains to understand
whether a sharp transition from a \rev{multifractal} 
extended phase to a fully-\rev{extended} 
phase occurs at a {\it finite} value of $\nu$ in the thermodynamic limit.

\begin{acknowledgments}
We thank J.~W.~Baron, G.~Biroli, C.~Bordenave, 
I.~Khaymovich, R.~K\"uhn, and M.~Pino for illuminating discussions.
We especially thank Y.~Fyodorov for drawing our attention to the symmetry of $P(\rho)$ described in Ref.~\cite{Mirlin_2006}, which led us to a better understanding of the multifractal spectrum.
DV acknowledges partial financial support from Erasmus+ 2022-1-IT02-KA131-HED-000067727, and thanks the LPTMC for the warm hospitality during the early stages of this work.
LFC acknowledges financial support from ANR-19-CE30-0014. 
GS acknowledges support from ANR, Grant No.~ANR-23-CE30-0020-01 EDIPS.
\end{acknowledgments}

\appendix

\section{Average spectral density with bimodal weights}
\label{app:delta_peaks}

\begin{figure}
    \centering
    \includegraphics[width=\columnwidth]{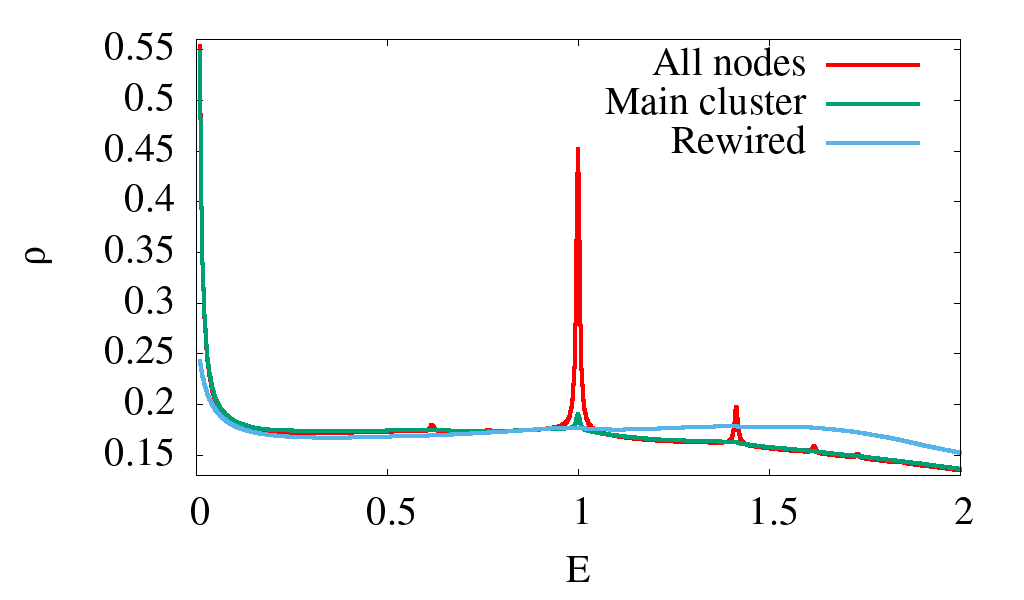}
    \vspace{-0.5cm}
    \caption{Average spectral density in the case of bimodal edge disorder (see Eq.~\eqref{eq:bimodal}), with $p=3$, $\eta=0.005$, and $N=2^{10}$. The plot compares the results of the cavity calculation on the original graph with those following the removal of the isolated clusters, and finally after rewiring the leaves. The first operation reduces the $\delta$-peaks in the bulk of the spectrum, while the second removes them together with the peak at $E=0$. }
    \label{fig:peaks}
\end{figure}

In this appendix we analyze the case in which the weight distribution is chosen to be bimodal, so that
\begin{equation} 
    \mathcal P({H}_{ij}) = \left(1 - \frac{p}{N} \right ) \delta ({H}_{ij}) + \frac{p}{N} \, \pi \left ({H}_{ij}\right) \, , 
\end{equation}
with 
\begin{equation}
    \pi(x) \equiv \frac12 \delta(x+1) + \frac12 \delta(x-1) \, .
    \label{eq:bimodal}
\end{equation}
Inspecting the recursion relations~\eqref{eq:cavity}~and~\eqref{eq:green} reveals that the cavity Green's functions are completely insensitive to the \textit{sign} of the chosen weights. We thus note from the onset that the results of this section would remain unchanged if we replaced the bimodal distribution in Eq.~\eqref{eq:bimodal} by constant weights on the links, \eg~$\pi(x)=\delta(x-1)$.

With this choice, the $\delta$-peak in the average spectral density observed in the Gaussian case for $E=0$ (see Sec.~\ref{sec:spectral_density} and Fig.~\ref{fig:DoS})
is known to be accompanied by several other peaks within the bulk of the spectrum~\cite{Kirkpatrick_1972}, distinct from the ones that appear in its Lifshitz tails~\cite{Rodgers_1988,Semerjian_2002}. 
These peaks are present for small values of $p>1$ above the percolation threshold, while they disappear in the large-connectivity limit.
Their origin was attributed to the eigenstates localized on small isolated structures in Refs.~\cite{Bauer2001,golinelli2003statistics}, where additionally their positions were fully characterized.
As it was later noted in Refs.~\cite{Rogers_2008,Kuhn_2008}, these peaks can be recovered within the cavity approximation only provided that the imaginary regulator $\eta$ is kept sufficiently large --- while they are not captured by other previously introduced approximation schemes (see \eg~Ref.~\cite{Semerjian_2002} and Figs.~5--7 therein).

It is natural to ask how our choice of removing all the small isolated clusters (see Sec.~\ref{sec:model}) affects the peaks at $E\neq 0$. To address this point, we first show in Fig.~\ref{fig:peaks} the average spectral density computed within the cavity approximation \textit{before} the disconnected clusters are removed. The parameters $p$ and $\eta$ were chosen so as to make contact with Fig.~4 in Ref.~\cite{Rogers_2008}, and indeed the same peaks are recovered. Next, solving the cavity equations on the main percolating cluster shows the persistence of the peak at $E=0$, while the height of the other peaks is sensibly reduced. Finally, these peaks completely disappear following the rewiring procedure delineated in Sec.~\ref{sec:mechanism}, which removes the leaves of the graph. As mentioned in Sec.~\ref{sec:mechanism}, such procedure necessarily produces a small modification of the degree distribution, which motivates the change in the shape of the average spectral density 
in Fig.~\ref{fig:peaks}.

\section{Multifractal spectrum}
\label{app:multifractal}

\begin{figure}
\includegraphics[width=0.482\textwidth]{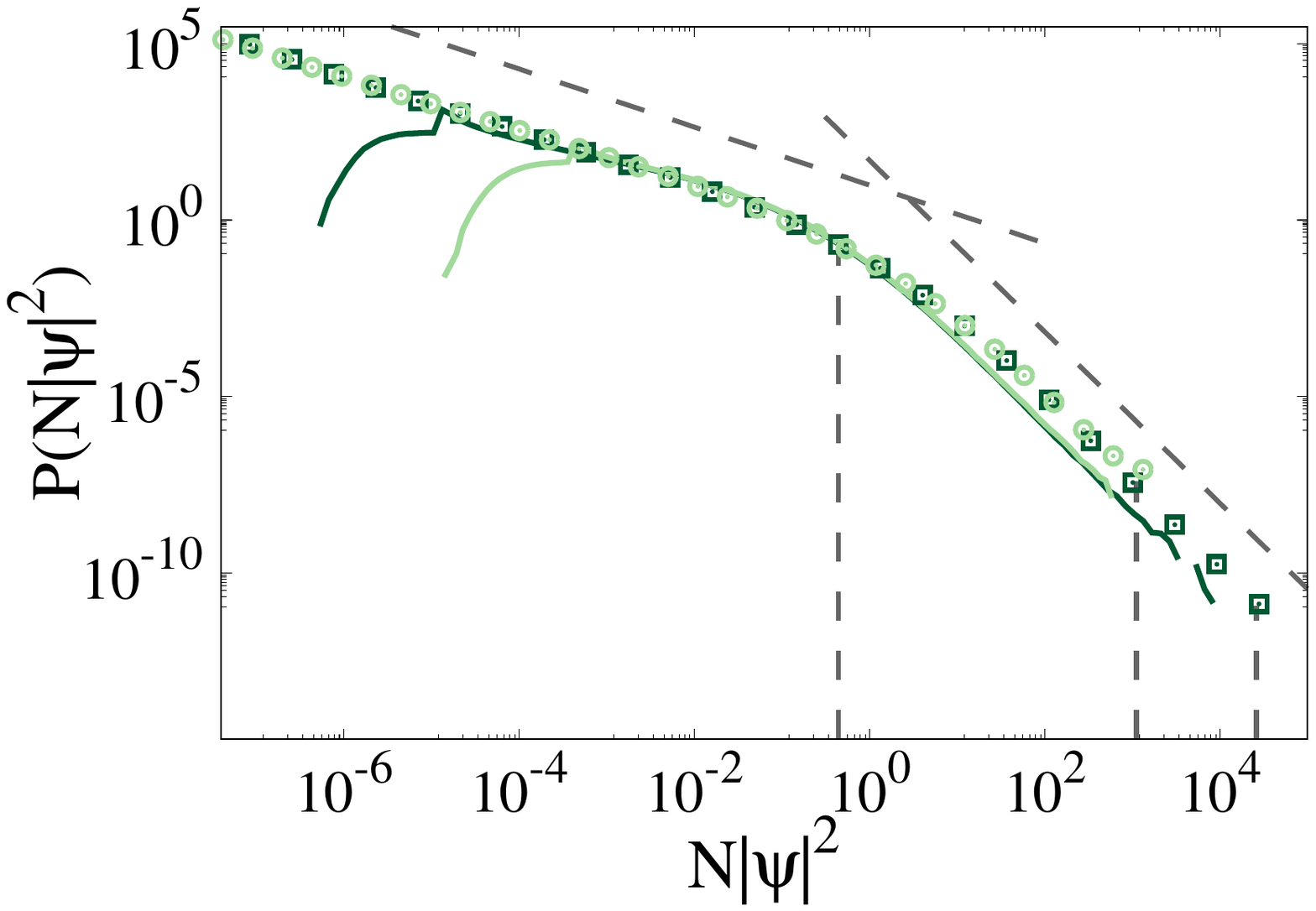} \put(-46,97){\small{$-(1+\beta)$}} \put(-130,150){\small{$\beta-2$}} \put(-114,38){\small{${\cal O}(1)$}} \put(-49,37){\small{$N$}}
\vspace{-0.2cm}
\caption{Probability distribution $P(N|\psi|^2)$ of the wavefunction's amplitudes, here plotted for $N=2^{17}$ (dark green squares) and $N=2^{12}$ (light green circles). Additionally, the plot illustrates the probability distribution $P(\rho)$ of the LDoS calculated using the cavity approximation, as described in Sec.~\ref{sec:LDoS}. By setting $\eta=1/N$, we find that $P(\rho)$ closely reproduces $P(N|\psi|^2)$.}
 \label{fig:sketch_P(x)}
\end{figure}

In this appendix we compute the spectrum of fractal dimensions $f(\alpha)$ reported in Eq.~\eqref{eq:falpha} in the main text, and described \eg~in Refs.~\cite{Kutlin_2024,Mirlin_2006}. 
This is normally defined so that $\sim N^{f(\alpha)}$ nodes have wavefunctions' amplitudes scaling as $N^{-\alpha}$.

To compute $f(\alpha)$, we start by introducing the distribution $P(x)$ of the variable $x\equiv N \abs{\psi}^2$. Such $P(x)$ is closely related to the distribution $P(\rho)$ of the LDoS which we characterized in Sec.~\ref{sec:LDoS}: indeed, as shown in Fig.~\ref{fig:sketch_P(x)}, by setting $\eta=1/N$ one finds that $P(\rho)$ reproduces closely $P(x)$. Specifically,
\begin{eqnarray}
    P(x) \simeq 
    \begin{dcases}
        c_0 x^{\beta-2}, & \text{if}\quad  
        x \ll 1,  \\
        c_0^\prime x^{-1-\beta}, & \text{if}\quad 1 \ll x \ll N, 
    \end{dcases}
    \label{eq:P(x)}
\end{eqnarray}
with $c_0$ and $c_0^\prime$ both of ${\cal O}(1)$, thus ensuring the matching of the two power-law regimes for $x \sim {\cal O}(1)$.
Changing variables as $x\equiv N^{1-\alpha}$ gives 
\begin{align}
    P(x) \, \de x &= P(N^{1-\alpha}) \, N^{1-\alpha} \, \ln N \, \de \alpha 
    \propto N^{f(\alpha)-1},
\end{align}
where the last equivalence is valid by definition in the large-$N$ limit~\cite{Kutlin_2024}. Implementing this change of variables into Eq.~\eqref{eq:P(x)} and using the definition above, one finds the expression of $f(\alpha)$ reported in Eq.~\eqref{eq:falpha} in the main text.

\begin{figure*}
\includegraphics[width=0.42\textwidth]{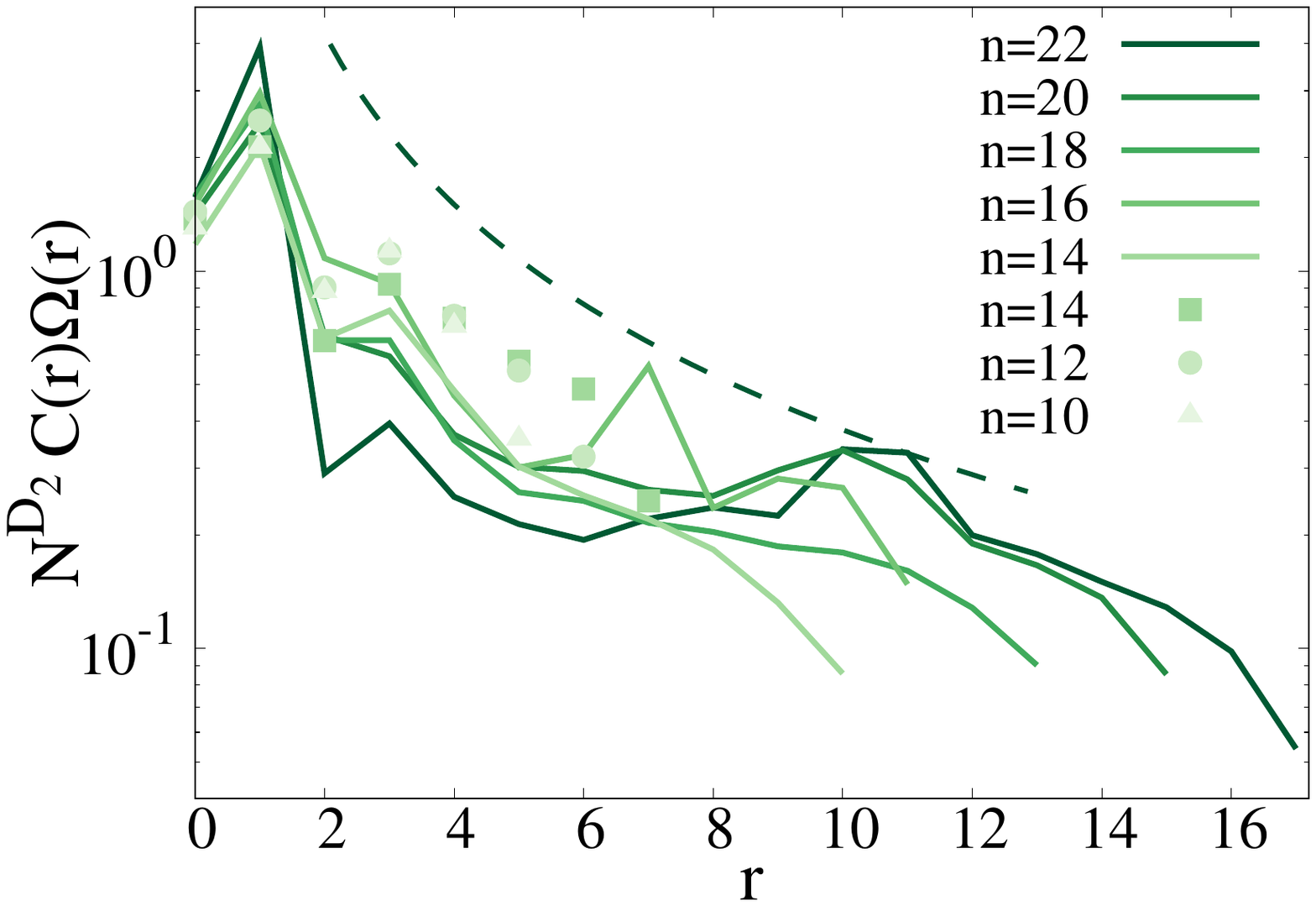} 
\put(-205,133){(a)}
\hspace{0.5cm}
\includegraphics[width=0.42\textwidth]{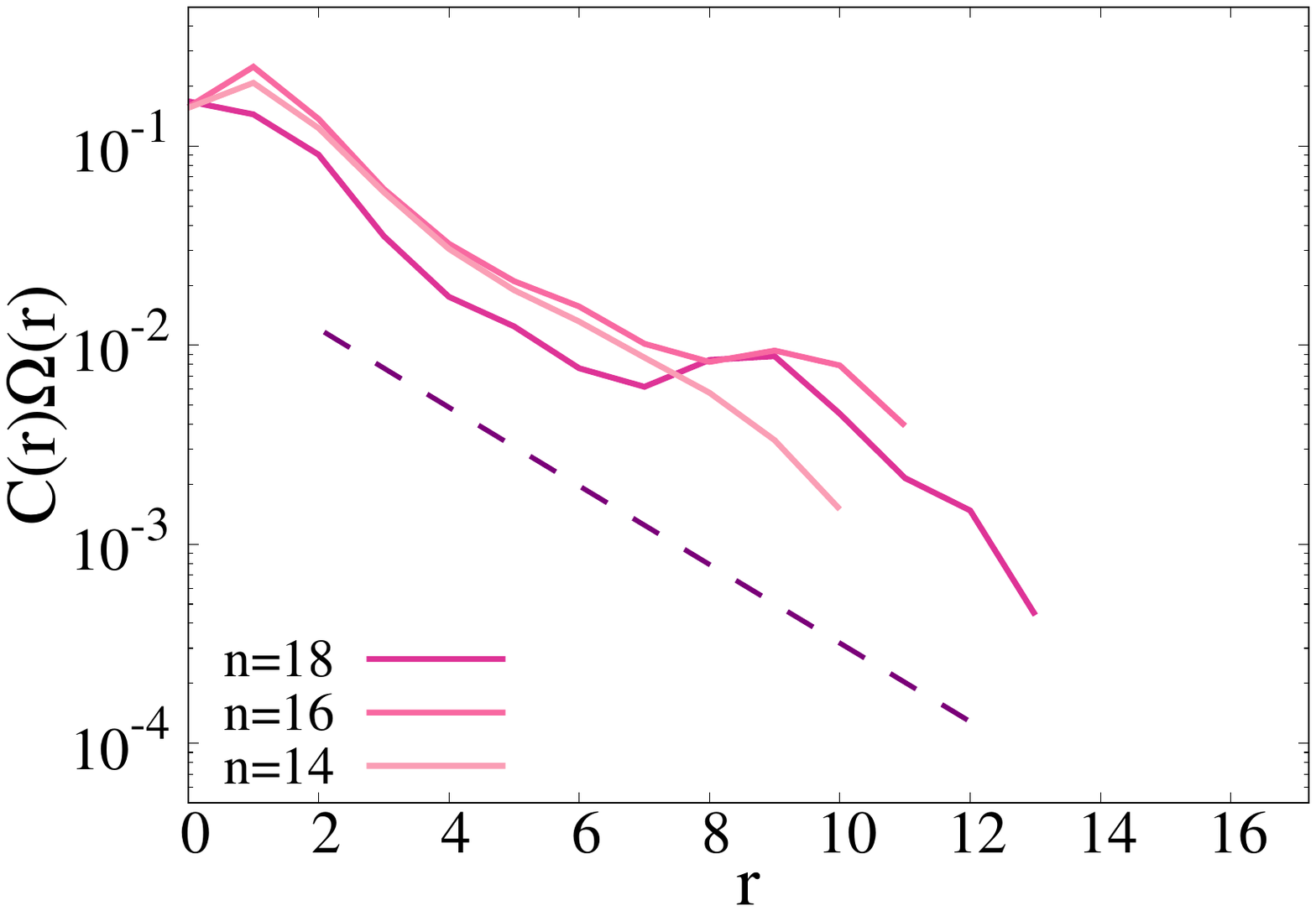} 
\put(-206,133){(b)}
\caption{Two-point spatial correlation function, see Eq.~\eqref{eq:Cr}, multiplied by the number of nodes $\Omega(r)$ at distance $r$ from a given site, for $p=2.4$ and
(a) $E=0.4$
(within the multifractal 
extended regime), or (b) $E=1.9$ (within the Anderson-localized phase). Solid lines correspond to the results found within the cavity approximation (see Eq.~\eqref{eq:Crcav}) for several values of the system size $N=2^n$, with $n$ ranging from $14$ to $22$ (and setting $\eta=\delta \propto 1/N$). Symbols show the exact diagonalization calculations for $n$ varying between $10$ and $14$.  In panel (a) $C(r)\Omega(r)$ is multiplied by $N^{D_2}$
to obtain a collapse of the data corresponding to different system sizes,
with $D_2=0.44$ being
the fractal dimension value already estimated from the analysis of the IPR (see Fig.~\ref{fig:Dq_ED}(a)).
The dashed curve 
represents the power-law decay $r^{-3/2}$ expected in the extended regime, see Eq.~\eqref{eq:Crdeloc}. In panel (b) the dashed line corresponds to the exponential decay $e^{-r/\xi_{\rm loc}}$ expected in the Anderson-localized regime, see Eq.~\eqref{eq:Crloc}, with $\xi_{\rm loc} \simeq 2.2$.
\label{fig:2P}}
\end{figure*}

Note that $f(\alpha)$ is related to the anomalous dimensions $D_q$ via a Legendre transformation. To see this, one can introduce the exponents $\tau_q=(q-1)D_q$, 
which can be obtained as~\cite{Kutlin_2024}
\begin{equation}\label{eq_tauq}
\tau_q = {\rm min}_\alpha [\alpha q - f(\alpha)].
\end{equation}
Using $f(\alpha)$ found in Eq.~\eqref{eq:falpha}, one finds that the minimum is in $\alpha=1$ when $q\leq \beta$, and in 
$\alpha=0$ when $q>\beta$, which correctly renders 
\begin{align}\label{appeq:tauq}
    \tau_q 
    =  
    \begin{dcases}
        q-1, & \text{if}\quad q \leq \beta,\\
         \beta - 1, & \text{if}\quad q > \beta,
    \end{dcases} 
\end{align}
in agreement with the exponents $D_q$ given in Eq.~\eqref{eq:Dqbeta}.

\section{Two-point spatial correlations}
\label{app:two-point}

In this appendix we focus on the spatial decay of the correlations of the wavefunctions' amplitudes $\avgg{\abs{ \psi_\alpha (i) }^2 \abs{\psi_\alpha (i+r) }^2}$ between two points at distance $r$ on the graph. For $r=0$ this object gives the IPR of the eigenstate $\alpha$ divided by $N$, while for $r \gg 1$ one has that
$\avgg{\abs{ \psi_\alpha (i) }^2 \abs{\psi_\alpha (i+r) }^2} \to \avgg{\abs{ \psi_\alpha (i) }^2} \avgg{\abs{\psi_\alpha (i+r) }^2} = N^{-2}$. 
We thus define the connected 
two-point 
correlation function
\begin{equation} \label{eq:Cr}
    C(r;E) = \frac{N \sum\limits_\alpha \avg{\abs{ \psi_\alpha (i) }^2 \abs{\psi_\alpha (i+r)}^2} \delta \left(E-\lambda_\alpha \right) } {\sum\limits_\alpha \delta \left(E-\lambda_\alpha \right) } - \frac{1}{N}  \, ,
\end{equation}
so that $C(r=0;E) = I_2(E)$ and $C(r \gg 1;E) \to 0$.

We then define $\Omega(r)$ as the number of nodes of the graph at distance $r$ from a given node. For distances smaller than the diameter of the graph, $r \ll \ln \tilde{N} / \ln (\tilde{p} - 1)$, $\Omega(r)$ is expected to scale as $\Omega(r) \simeq \tilde{p} (\tilde{p} - 1)^{r-1}$.
In the standard metallic phase of the Anderson tight-binding model on the Bethe lattice, corresponding to  eigenstates uniformly spread over the whole volume, the correlation function behaves as~\cite{tikhonov2019statistics}
\begin{equation} \label{eq:Crdeloc}
    C(r) \Omega(r) \propto N^{-1} \,r^{-3/2}
\end{equation}
(we omitted the $E$-dependence from $C(r;E)$ for simplicity).
Conversely, in the Anderson-localized phase the correlation function reflects the exponential decay of the eigenstates' amplitudes and behaves as~\cite{tikhonov2019statistics}
\begin{equation} \label{eq:Crloc}
    C(r) \Omega(r) \propto e^{-r/\xi_{\rm loc}} \,r^{-3/2} \, ,
\end{equation}
where the localization length $\xi_{\rm loc}$ diverges at the transition.

Such $C(r;E)$ can be directly computed in moderately small systems from exact diagonalization. It also admits a simple spectral representation in terms of the Green's functions (which generalizes Eq.~\eqref{eq:IqG}) as
\begin{equation} \label{eq:Crcav}
    C(r;E) \simeq \frac{\eta \avg{\left \vert G_{i,i+r} \right \vert^2}}{\avg {\im G}} \simeq \frac{\eta \avg{G_{i,i} G_{i+r,i+r}^\star}_c}{\avg {\im G}} \, ,
\end{equation}
which allows one to compute it within the cavity approximation for much larger system sizes. 
In Fig.~\ref{fig:2P} we show numerical results for the two-point function within the \rev{multifractal} 
extended regime (panel (a)), and within the Anderson-localized one (panel (b)). In the former case $C(r)$
behaves as described by Eq.~\eqref{eq:Crdeloc}, corresponding to delocalized eigenstates, with the only difference being that the $N^{-1}$ prefactor is replaced by $N^{-D_2}$. In the latter case we recover the exponential decay described by Eq.~\eqref{eq:Crloc}. Note that by setting $\eta=\delta$ in Eq.~\eqref{eq:Crcav} one finds a very good agreement between the cavity calculation and the exact diagonalization result at fixed $N$.

\bibliography{references}

\end{document}